\documentclass[12pt]{article}
\usepackage{newinutile}
\usepackage{amsmath,amssymb,amsthm,amsfonts}
\usepackage{epsfig,graphics,color,calc,graphicx}
\usepackage{subfigure}
\usepackage{comment}
\usepackage{pgf,tikz}
\usepackage{mathrsfs}
\usepackage{color}
\usetikzlibrary{arrows}
\usepackage{cancel}
\usepackage{mathrsfs}
\usepackage{hyperref}
\hypersetup{colorlinks=true,linkcolor=blue}

\usepackage{cite}
\usepackage{mathrsfs}

\newtheorem{remark}{Remark}

\def\theequation{\arabic{section}.\arabic{equation}}

\newcommand{\rr}[1]{{\normalfont\textrm{#1}}}

\newcommand{\cc}[1]{{\mathcal{#1}}}
\newcommand{\bb}[1]{{\mathbb{#1}}}

\newcommand{\rexit}{\textup{R.E.}}

\newlength{\pecettawidth}
\begin{document}
\title{Localization of defects via residence time measures}

\author{A.\ Ciallella}
\email{alessandro.ciallella@univaq.it}
\affiliation{Dipartimento di Ingegneria Civile, Edile -- Architettura e
             Ambientale, Centro di Ricerca M\&MoCS\\
             Universit\`a degli Studi dell'Aquila,\\
             via Giovanni Gronchi 18, 67100 L'Aquila, Italy.}

\author{E.N.M.\ Cirillo}
\email{emilio.cirillo@uniroma.it}
\affiliation{Dipartimento di Scienze di Base e Applicate per l'Ingegneria,\\
             Sapienza Universit\`a di Roma, \\
             via A.\ Scarpa 16, 00161, Roma, Italy.}

\author{B.\ Vantaggi}
\email{barbara.vantaggi@uniroma1.it}
\affiliation{Dipartimento Metodi e Modelli per l'Economia, il Territorio
             e la Finanza,\\
             Sapienza Universit\`a di Roma, \\
             via Del Castro Laurenziano 9, 00161, Roma, Italy.}


\begin{abstract}
We show that residence time measure can be used to identify
the geometrical and transmission properties of a
defect along a path.
The model we study is based on a
one--dimensional simple random walk.
The sites of the lattice are regular, i.e.,
the jumping probabilities are the same in each site,
except for a site, called \emph{defect}, where the jumping
probabilities are different.
At each side of the lattice an absorbing site is present.
We show that by measuring the fraction of particles crossing
the channel and/or the typical time they need to cross it,
it is possible to identify the main features of the lattice and
of the defect site, namely, the jumping probabilities at
regular and at the defect sites and the position of the
defect in the lattice.
\end{abstract}




\maketitle

\section{Introduction}
\label{s:intro}
The effect of obstacles in the transport of
moving agents has been widely studied in different contexts \cite{CPamm2017}.
Macromolecules,
playing the role of obstacles for the diffusing smaller molecules
\cite{Sbj1994,HFrpp2013,MHSbj2017,ESMCBjcp2014}, tend to
slow down the dynamics, while in granular systems
\cite{TLPprl2001,ZGMPPpre2005,AMAGTOKpre2012,ZJGLAMprl2011}
and in the framework
of pedestrian dynamics
\cite{Hrmp2001,HFMV2011,CMpA2013,MCKBud2014,CMcrm2012,CCCMm3as2018},
the correct positioning of physical
obstacles can accelerate the exit of the agents from a
bounded region hindering clogging patterns
\cite{ABCKsjap2016,HFVn2000,HBJWts2005,EDLR2003}.

Geometrical and dynamical properties of obstacles, such as position,
shape, and attitude to transfer mass, influence the transport
properties of the whole system. We refer
the reader to some old and more recent studies
where the effect of obstacles has been studied with different
techniques and for many different stochastic dynamics,
such as
the simple exclusion 1D model \cite{JanLeb94},
a Probabilistic Cellular Automata \cite{SLM15}
the 1D and 2D symmetric and asymmetric random walk
\cite{CC19,CCSpre2018},
the linear Boltzmann dynamics \cite{CCkrm2018},
the zero range process \cite{CCpre2017,CCMphysicaA2017,CCMpre2016},
and
the simple exclusion 2D process \cite{CKMSpre2016,CKMSSpA2016}.

As it is clearly demonstrated by those studies,
in the applications it can be of the out-most importance the
knowledge of the geometrical and transport
properties of obstacles which limit and affect the dynamics
of the agents in a region of space, since the currents observed
in the system strongly depend on them.
On the other hand obstacles can be often not easily accessible
to instruments for experimental measures: imagine an obstacle placed
inside a biological channel or in a long pipe.
Hence, it is rather natural to ask ourselves if it is possible
to deduce the properties of the obstacle measuring observable quantities
connected to the motion of agents.

In this paper we shall consider a model of particles flowing
on a lane and the observables that will be used to deduce
the properties of the lane and
of the obstacle are the fraction of particles
crossing the lane and the mean time they take to cross it.

We shall consider two different experiments.
In experiment 1 we imagine to start a known number of particles
at one end point of the lane and compute the number of particles
which will exit the lane through the second end point. Thus,
in this experiment we have
access to the fraction of particles that succeed to cross the lane.
In experiment 2 we imagine to start, all at the same time,
an unknown number of particles at one end point of the lane and to measure
the time that the particles that succeed to cross the lane
take to exit it through the
second end point. Thus, in this experiment we are not able
to measure the fraction of particles that actually cross the lane,
but we have access to the mean value of the time that the
particles that cross the lane need to cross it. 
The typical time needed for a particle to cross the channel
will be called \emph{residence time}.
We mention that the idea of residence time as average crossing
time conditioned to crossing a region
was firstly introduced in
\cite{CKMSpre2016,CKMSSpA2016}
for a 2D simple exclusion walk: a thorough study of its
properties in absence of obstacle is reported in
\cite{CKMSSpA2016}, whereas the presence of obstacles was considered in
\cite{CKMSpre2016}.
We notice that
in the 1D version of the simple exclusion model
with Langmuir kinetics the
idea of residence time was exploited in \cite{MRVMSpre2016}.

The problem introduced above is faced in this paper via
a one--dimensional model based on a simple random walk.
At each side of the lattice an absorbing site is present.
We deal with the problem of identify the features of the lattice, once
we measured some  observable such as the fraction of particles reaching
the right absorbing site, and the average time needed for a walker to
reach it.

More precisely,
the model we consider in this paper is a one--dimensional
simple random walk 
on $\{0,1,\dots,L\}$ possibly with a defect site.
The sites $0$ and $L\ge3$ are absorbing,
so that when the particle reaches one of these two sites the walk is stopped.
All the sites $1,\dots,L-1$ are \emph{regular} excepted for one
site called \emph{defect}
\cite{CC19,CCSpre2018}.
The defect site is the site $d$.
At each unit of time the walker
jumps to a neighboring site according to the following rule:
if it is on a regular site, then it
jumps to the right with probability $p$ or to the left
with probability $q=1-p$, so that it cannot stay still at the site.
If it is at the defect site it jumps
to the right with probability $p'$ or to the left
with probability $q'=1-p'$.

The array $1,\dots,L-1$ will
be called the \emph{lane}. The sites $0$ and $L$ will be,
respectively, called the \emph{left} and \emph{right exit} of the lane.
In this framework the residence time is defined by starting
the walk at site $1$ and computing
the typical time that the particle takes to reach the site $L$ provided
the walker reaches $L$ before $0$.
More precisely, we let $x_t$ be the position of the walker at time $t$ and
denote by $\mathbb{P}$ and $\mathbb{E}$ the probability associated to the
process and the related expected operator
for the walk started at $x_0=1$.
We let $\tau$
be the \emph{first hitting time to} $L$ or \emph{crossing time}, namely,
\begin{equation}
\label{one000}
\tau=\inf\{t>0:\,x_t=L\}
\;\;,
\end{equation}
with the convention that
$\tau=\infty$ if the trajectory reaches $0$ before reaching $L$.
Note that the trajectories cannot be trapped for an infinite time in
the lane because $L$ is finite.
Note also that, since the particle starts at site $1$,
the values that $\tau$ can assume are of the form $L+2k-1$,
for any integer $k\geq0$.
We thus define
the \emph{residence time} as
\begin{equation}
\label{one010}
R
=
\mathbb{E}(\tau|\rexit)
\;\;,
\end{equation}
where we conditioned to the event $\rexit$ meaning that the
particle exits the lane through the right exit in $L$.

In absence of defect, namely, the sole parameters that one wants
to deduce is the jumping probability $p$ which accounts
for the transport properties of the lane.
In case of experiment 1, namely, when the fraction of particles
crossing the lane is available, a standard maximum likelihood argument
provides a correct estimate for the parameter, which is
fully identified. On the other hand, in case of experiment 2,
again an estimate is provided by a maximum likelihood argument,
but a complete identification of the parameter $p$ is lacking since
the residence time is symmetric when $p$ is exchanged with $1-p$.

In presence of defects the situation is much more complicated,
since one wants to deduce the three parameters $p$, $d$, and $p'$,
which respectively account for the transport properties of the lane ($p$),
the geometry of the defect (its position $d$), and the transmission
property of the obstacle ($p'$).
In our discussion we shall assume $p$ to be known and we shall
show that both in case of experiment 1 and 2 it will not possible
to estimate the values of $d$ and $p'$, but it will possible
to find an equation connecting the two. In other words, in the
parameter space $d$--$p'$, a level curve on which the pair $(d,p')$
has to lay will be identified.
If the two experiments, 1 and 2, are performed simultaneously,
namely, provided the fraction of particles crossing
the lane and the residence time are both measured,
then in many case the identifiability problem will be
fully solved and an estimate for the parameters $d$ and $p'$
will be found.

The paper is organized as follows: in Section~\ref{s:nodef}
we consider the case in which the lane does not present any defect.
The presence of a defect site is then discussed in Section~\ref{s:defect}.
Finally, the Section~\ref{s:conc} is devoted to some brief conclusions.

\section{Simple random walk: no defect}
\label{s:nodef}
We first consider the simple random walk case, namely,
here we assume that $p'=p$.
For a given  $p$, considering a single
particle, the joint probability of the event that it exits through
the right side starting from site $1$ and the first hitting time
to $L$ is $n=L+2k-1$ is
\begin{equation}\label{eq:005}
\mathbb{P}(\tau=n , \rexit)
= \alpha_n p^{\frac{L+n-1}{2}}q^{\frac{n-(L-1)}{2}}
=\alpha_{L+2k-1} p^{L+k-1} q^k,
\end{equation}
where $\alpha_n$ is the number of different possible paths that
starts in $1$ and exits the lane reaching the site $L$ for the
first time at time $n$, without hitting the site $0$.
Moreover, in each of these paths the walker
performs $(n-(L-1))/2$ steps to the left and
$(n-(L-1))/2+L-1$ steps to the right.
We stress that $\alpha_n$ is smaller that the total number of paths made of
$(n-(L-1))/2$ steps to the left and
$(n-(L-1))/2+L-1$ steps to the right, since the constraint
that the particles reaches $L$ before touching $0$ must be taken into account.

It is well known, see, e.g., \cite{CC19,Fel,S1975,L09},
that the probability of the event $\rexit$
is given by the classic gambler's ruin probability
\begin{equation}
\label{gamb}
\mathbb{P}(\rexit)=\frac{1-(q/p)}{1-(q/p)^L} \textup{ if } p\neq q
\;\;
\textup{ and }
\;\;
\mathbb{P}(\rexit.)=\frac{1}{L} \textup{ if } p=q.
\end{equation}
Thus, for
$p\neq q$
\begin{equation}
\label{eq:015}
\mathbb{P}(\tau=n|\rexit)
= \alpha_n p^{L-1+k}q^k \frac{(q/p)^L -1}{(q/p)-1}
= \alpha_n p^{k}q^k \frac{q^L -p^L}{q-p}
= \alpha_n p^{k}q^k \sum_{j=0}^{L-1} p^j q^{L-1-j},
\end{equation}
where, we recall, the time $n$ is in the form $L+2k-1$.
Note the last expression in \eqref{eq:015}
is valid also for $p=q$, indeed
\begin{displaymath}
\mathbb{P}(\tau=n|\rexit)= \alpha_n p^{L+2k-1 }  L .
\end{displaymath}

Now, we consider $M$ independent identically distributed walkers
starting at $1$ and denote by $\tau_i$ the crossing time of the
particle $i$, for $i=1,\dots,M$.
The probability
that the $M$ particles exit the lane through the right exit
with crossing times $n_i=L+2k_i-1$ is
\begin{equation}
\label{eq:008}
\mathbb{P}(\tau_1=n_1,\tau_2=n_2,\ldots,\tau_M=n_M,
\rexit_1,\rexit_2,\ldots,\rexit_M)
=\prod_{i=1}^M \alpha_{n_i} p^{{L+k_i-1}}q^{{k_i}}
\;,
\end{equation}
where with $\rexit_i$ we denote the event that the $i$ particle
exits the lane through the right exit.
Using, as above, the gambler's ruin result, we get
\begin{equation}
\label{eq:018}
\begin{split}
&\mathbb{P}(\tau_1=n_1,\tau_2=n_2,\ldots,\tau_M=n_M
    |\rexit_1,\rexit_2,\ldots,\rexit_M)=\\
&\;\;\;\;\;\;
 =\prod_{i=1}^M\left( \alpha_{n_i} p^{k_i}q^{k_i}
    \frac{q^L -p^L}{q-p} \right)
=\left(\prod_{i=1}^M \alpha_{n_i}\right)
 p^{\sum_{i=1}^M k_i} q^{\sum_{i=1}^M k_i}
 \left(\sum_{j=0}^{L-1} p^j q^{L-1-j}\right)^M
\end{split}
\end{equation}

We remark that
conditioning to exiting through the right side do not influence
the independence of the observed crossing times for given $p$.
Consider, for simplicity, the case $M=2$:
\begin{equation}
\label{eq:020}
\mathbb{P}(\tau_1=n_1,\tau_2=n_2|\rexit_1,\rexit_2)
=\frac{\mathbb{P} (\tau_1=n_1, \rexit_1,\tau_2=n_2, \rexit_2)}
{\mathbb{P}(\rexit_1,\rexit_2)}
\end{equation}
\begin{equation}\label{eq:021}
\mathbb{P}(\tau_1=n_1|\rexit_1)
\mathbb{P}(\tau_2=n_2|\rexit_2)
=\frac{\mathbb{P}(\tau_1=n_1, \rexit_1)}
      {\mathbb{P}(\rexit_1)}
  \cdot \frac{\mathbb{P}(\tau_2=n_2, \rexit_2)}{\mathbb{P}(\rexit_2)}
\;.
\end{equation}
Since the crossing times are
independent and identically distributed random variables it follows
immediately that
the left--hand sides of
\eqref{eq:020}--\eqref{eq:021} are equal.

As we discussed in the Introduction, the  problem we address in this Section is
the estimate of the parameter $p$: we assume to
perform an experiment and after we measure some observables
we want to trace back the value of $p$.
We discuss, now, two different experiments: in the first case we
measure the fraction of particles which succeed to cross the
lane, namely, the fraction of particles which exit the
lane through the right exit in $L$.
In the second experiment we shall assume to measure only the
average time spent by the particles in the lane before exiting through the
right end without
having any information on the fraction of successful particles.

In these experiments the role of identifiability of models is discussed. In the following a statistical model is said to be identifiable whenever
the likelihood has  no flat region, so it is theoretically possible to estimate the  underlying parameters. This means that different values of the parameters must generate a different value for the likelihood
over the observable variables, i.e. $$L(\theta, \mathbf(x))\neq L(\theta', \mathbf(x)) \,\qquad \mbox{ for any } \theta\neq\theta'.$$
For non-identifiable  models: two or more values of the parameters  are observationally equivalent. In these cases, it is relevant to determine the non-identifiable regions. Furthermore a model is said locally identifiable 
whenever for any $\theta$ in the parameter space $\Theta$ there 
exists an open neighborhood $N_\theta$ of $\theta$ in $\Theta$  such 
that for any $\theta'\in N_\theta$ it holds 
$L(\theta, \mathbf(x))\neq L(\theta', \mathbf(x))$ for any  
$\theta\neq\theta'$.

\subsection{Experiment 1: measuring the fraction of successful particles}
\label{s:identifp}
The knowledge of the fraction of the number of particles
which exit through the right side is sufficient to find a maximum
likelihood estimate
of $p$ using the probability of success in the gambler's
ruin problem \eqref{gamb}.
This probability as a function of $p$ is strictly monotonic,
hence, if we prepare $N$ particles at site $1$ and denote by
$m$ the number of those particles that
exits the lane through the right exit in $L$,
the equation in the unknown $p$
\begin{equation}
\label{eq:027}
\frac{m}{N}=\bb{P}(\rexit)
\end{equation}
has an unique solution $\hat{p}_m$.
Note that $m/N$ is the maximum likelihood estimate of $\mathbb{P}(\rexit)$.
Indeed,
the random variables
$\chi_i$ taking value $1$ if the particle $i$ 
reaches the
right side of the interval and $0$ otherwise
are i.i.d.\ Bernoulli random variables, so that we
recall here that in this case the maximum likelihood estimator
$\sum_{i=1}^N\chi_i/N$ is not biased.

%

The classical computation follows for completeness:
we can estimate $p$ by finding the
\begin{displaymath}
\mathrm{arg}\max_{p\in[0,1]} {\binom{N} {m}}
\left(\mathbb{P}(\rexit)\right)^m
\left(1-\mathbb{P}(\rexit)\right)^{N-m}
\;.
\end{displaymath}
By considering the log--likelihood we write
\begin{equation}
\begin{split}
\ell(p)
&=\ln {\binom N  m} + m \ln \mathbb{P}(\rexit)
+ (N-m) \ln (1-\mathbb{P}(\rexit))
\;.
\end{split}
\end{equation}
To search for its critical point we derive with respect to $p$:
\begin{equation}
\begin{split}
\frac{\mathrm{d}}{\mathrm{dp}}
\ell(\mathbb{P}(\rexit))
&=m \frac{1}{\mathbb{P}(\rexit)}
\frac{\mathrm{d}}{\mathrm{dp}}\mathbb{P}(\rexit)
-(N-m) \frac{1}{1-\mathbb{P}(\rexit)}
\frac{\mathrm{d}}{\mathrm{dp}} \mathbb{P}(\rexit)=
 \\
&=\frac{\mathrm{d}}{\mathrm{dp}}\mathbb{P}(\rexit)
[\frac{m}{\mathbb{P}(\rexit)}-\frac{N-m}{1-\mathbb{P}(\rexit)}]=0
\end{split}
\end{equation}
Since $\mathrm{d}\mathbb{P}(\rexit)/\mathrm{dp}$ is strictly positive,
the former has a unique solution
$\mathbb{P}(\rexit)=m/N$,
as we wrote in \eqref{eq:027}.

\subsection{Experiment 2: measuring the crossing time}
\label{s:meares}
In case the fraction of particles crossing the lattice is not know,
it is possible to setup an estimate of $p$ by measuring the
average crossing time. Even in this case we will be able to
perform a maximum likelihood estimate.
Suppose we measure the crossing time $n_i$ of $m$ particles,
for $i=1,\dots,m$,
we evaluate the parameter $p$ by comparing the experimental mean crossing
time
$\big(\sum_{i=1}^mn_i\big)/m$
to the theoretical residence time $R$.
The equation providing our estimate for $p$ will be found via
a maximum likelihood argument.

\begin{figure}[ht!]
\begin{picture}(120,130)(-40,5)
\put(60,0){
  \includegraphics[width=0.5\textwidth]{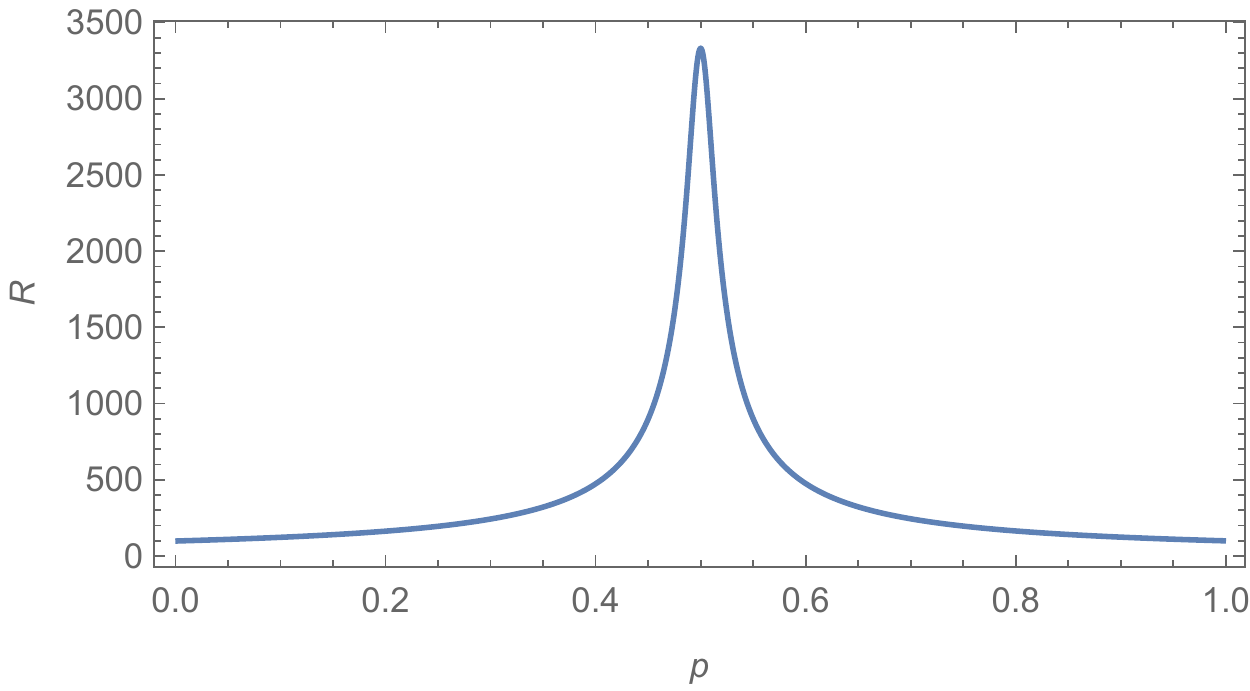}
}
\end{picture}
\caption{Residence time $R$ as a function of $p\in[0,1]$ for
$L=100$ in absence of defect.
}
\label{f:000}
\end{figure}

We note that
the residence time presents a symmetry in the role of
$p$ and $q=1-p$ that suggests the presence of identifiability problems.
That is to say, if the residence time of walkers is the sole information
we have access to, it is not possible to uniquely deduce the
jumping parameter $p$ that characterize the walk.
Indeed, looking at \eqref{eq:018}, we notice that the crossing
time observed for particles that eventually exit through the
site $L$ starting from site $1$,
is symmetric\footnote{We refer the interested reader
to \cite{L09} and \cite{WB77} where
other classical results from Stern and Samuels for walks starting from
the middle point are discussed.}
in $p\in[0,1]$ with respect to $1/2$
(recall that $q=1-p$).
The existence of this symmetry is also evident if one looks at
the dependence of the residence time on $p$, see, e.g.,
Fig.~\ref{f:000}.

In other words, looking at the symmetry residence time, we can expect that the experimental information on
the crossing time won't be sufficient to distinguish different estimates of $p$.
More precisely,
we find two possible estimates $\bar{p}$ and $1-\bar{p}$ for
the parameter $p$.
These two solutions will reduce to a single
one in the case in which $\bar{p}=1-\bar{p}=1/2$. This will be, indeed,
the sole case in which we will have complete identifiability for the
parameter $p$.
Note, however, that,
if the value of $p$ is not very close to $1/2$,
a partial information on the fraction of particles exiting from the
right side (e.g., the knowledge of the order of magnitude of the number of
particles started from 1) is sufficient to distinguish between the two cases and identify the correct estimate $\bar{p}$ or $1-\bar{p}$.

%

To construct the estimate of $p$ based on the residence time measure
we, now, setup a maximum likelihood argument:
we focus on $m$ particles that we know to have reached the site $L$ and
observe their crossing times $n_i$.
We know, from \eqref{eq:015} and \eqref{eq:018} the probability of
observing $\tau_1=n_1$, $\tau_2=n_2$,$\ldots$, $\tau_m=n_m$
conditioned to $\rexit_i$, for $i=1,\dots,m$.
Maximum likelihood gives the estimation $\overline{p}_m$ of $p$ as
\begin{equation}
\label{eq:028}
\begin{split}
\overline{p}_m
&=\rm{arg} \max_{p\in[0,1]}
\left(\prod_{i=1}^m \alpha_{n_i}\right) p^{\sum_i k_i} q^{\sum_i k_i}
\left(\frac{q^L -p^L}{q-p}\right)^m
\\
&=
\rm{arg} \max_{p\in[0,1]}
\left(\prod_{i=1}^m \alpha_{n_i}\right) p^{\sum_i k_i} q^{\sum_i k_i}
\left(\sum_{j=0}^{L-1} p^j q^{L-1-j}\right)^m
,
\end{split}
\end{equation}
where $n_i=L+2k_i-1$.
Notice that the last expression shows that the function
is symmetric in the exchange $p$ into $1-p$.
The log--likelihood is
\begin{equation}
\ell(p)=\sum_{i=1}^m \ln \alpha_{n_i} + \sum_{i=1}^m k_i \ln p + \sum_{i=1}^m k_i \ln (1-p) + m\ln \left(\frac{ (1-p)^L - p^L}{1-2p} \right)
\end{equation}
and, looking for
the $\mathrm{arg}\max$ of $\ell(p)$, we find the
critical points of $\ell(p)$
as the solutions
of the equation
\begin{equation}
\label{eq:029}
\left(\sum_{i=1}^m \frac{k_i}{m} \right)\frac{1-2p}{p-p^2}
= \frac{2}{2p-1} + L\frac{(1-p)^{L-1} + p^{L-1}}{(1-p)^L - p^L}.
\end{equation}

To discuss the structure of the solutions of \eqref{eq:029},
due to the invariance under the exchange $p$ in $1-p$, we
restrict the discussion to the interval $[1/2,1]$ and
denote by $f(p)$ and $g(p)$ its left--hand and right--hand sides.
It is possible to check that the following statements hold: 
\begin{itemize}
\item[--]
${\displaystyle \lim_{p\to1/2+} f(p)=0}$,
${\displaystyle \lim_{p\to1^-} f(p)=-\infty}$,
${\displaystyle \lim_{p\to1/2+} g(p)=0}$,
${\displaystyle\lim_{p\to1^-} g(p)=-L+2}$;
\item[--]
${\displaystyle \lim_{p\to 1/2+}f'(p)
=-8\sum_{i=1}^m\frac{k_i}{m}}$,
${\displaystyle \lim_{p\to 1/2+}g'(p)=-8/3 + 4 L - 4L^2/3}$;
\item[--]
$f(p)$ is a concave monotonically decreasing function in $(1/2,1)$.
\end{itemize}
Moreover, we checekd numerically that 
there exists $p^*\in (1/2,1]$ that depends on $L$ such that
$g(p)$ is convex and strictly monotonically decreasing in $(1/2,p^*)$, 
while it is monotonically increasing in $(p^*,1)$.
Hence,
if $-8\sum_{i=1}^m \frac{k_i}{m}\le-8/3 + 4 L - 4L^2/3$
the sole solution of \eqref{eq:029} is $p=1/2$, and it is a maximum point.
Otherwise,
there exists one more solution of \eqref{eq:029}
in $(1/2,1)$.
It is easy to see that such a solution is a maximum point, while $1/2$ becomes in this case a minimum point.

The above remarks yield the following conclusions:
If $-8\sum_{i=1}^m \frac{k_i}{m}\le-8/3 + 4 L - 4L^2/3$
the model is globally identifiable and the maximum likelihood estimate is
$\bar{p}_m=1/2$.
Otherwise, the model is locally (not globally) identifiable and there
exists  
 two maximum likelihood estimates $\bar{p}_m$ and
$1-\bar{p}_m$.

\begin{figure}[ht!]
\begin{picture}(200,90)(-10,0)
\put(0,0){
  \includegraphics[width=0.2\textwidth]{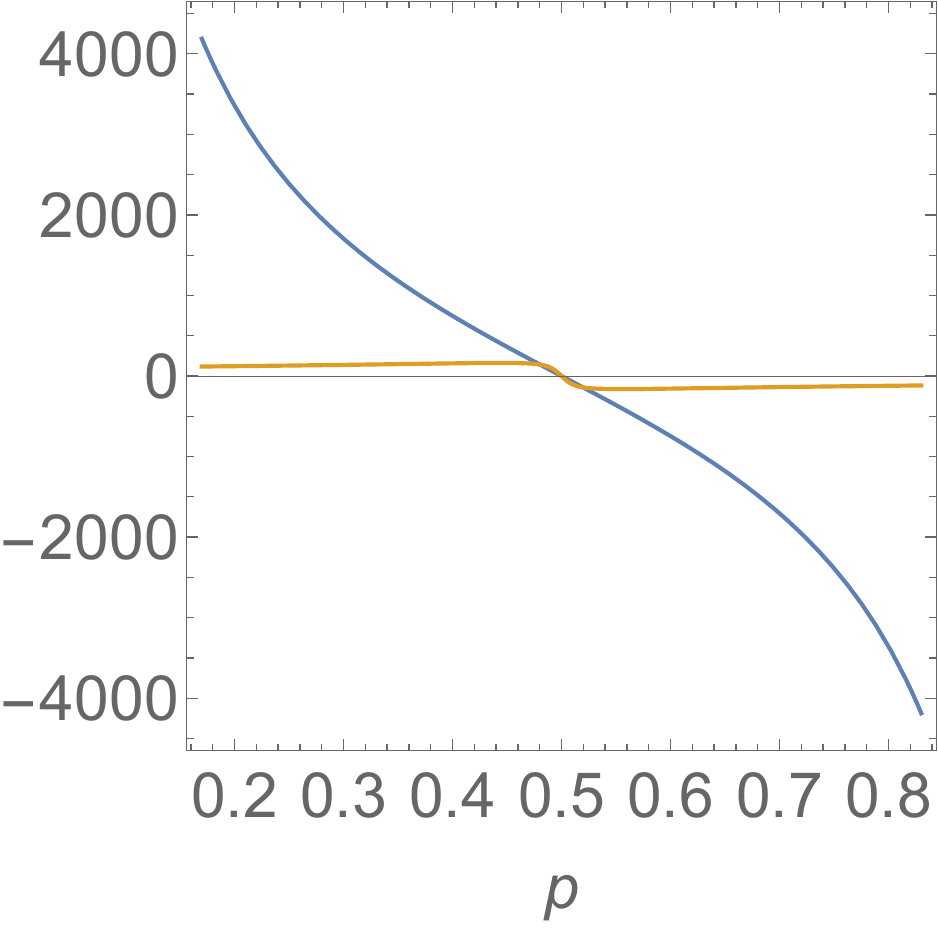}
}
\put(110,0){
  \includegraphics[width=0.2\textwidth]{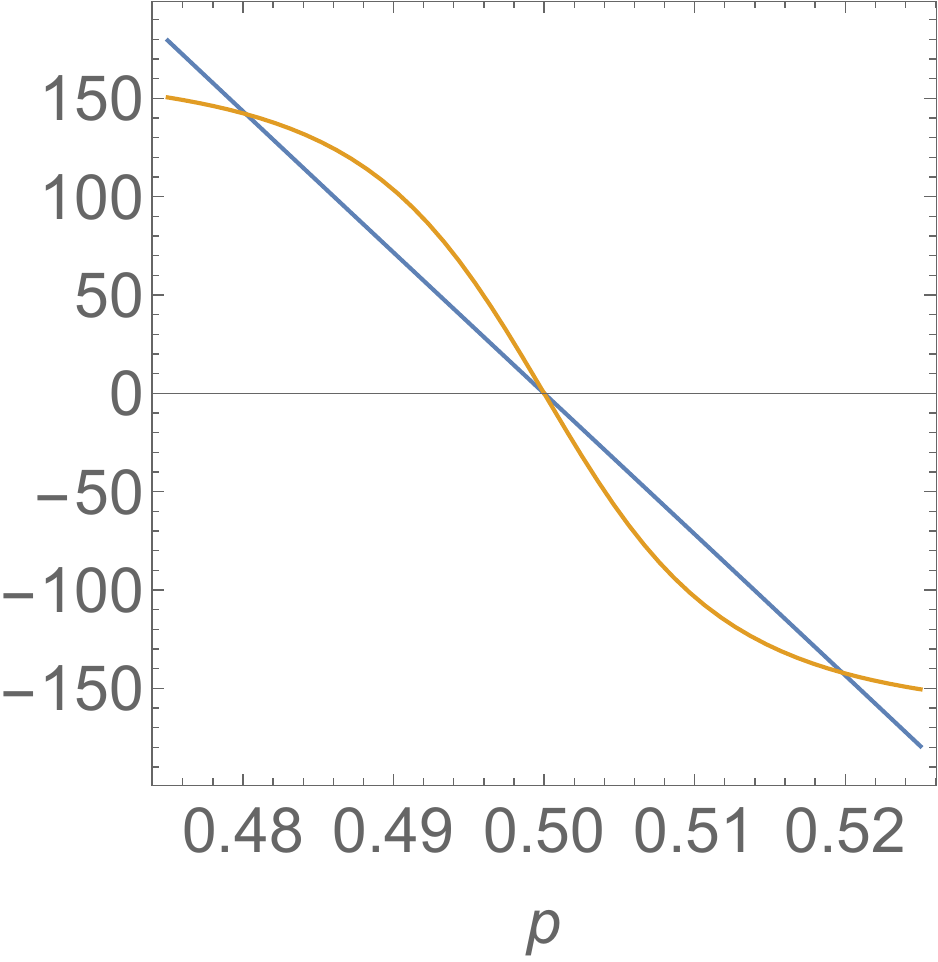}
}
\put(220,0){
  \includegraphics[width=0.2\textwidth]{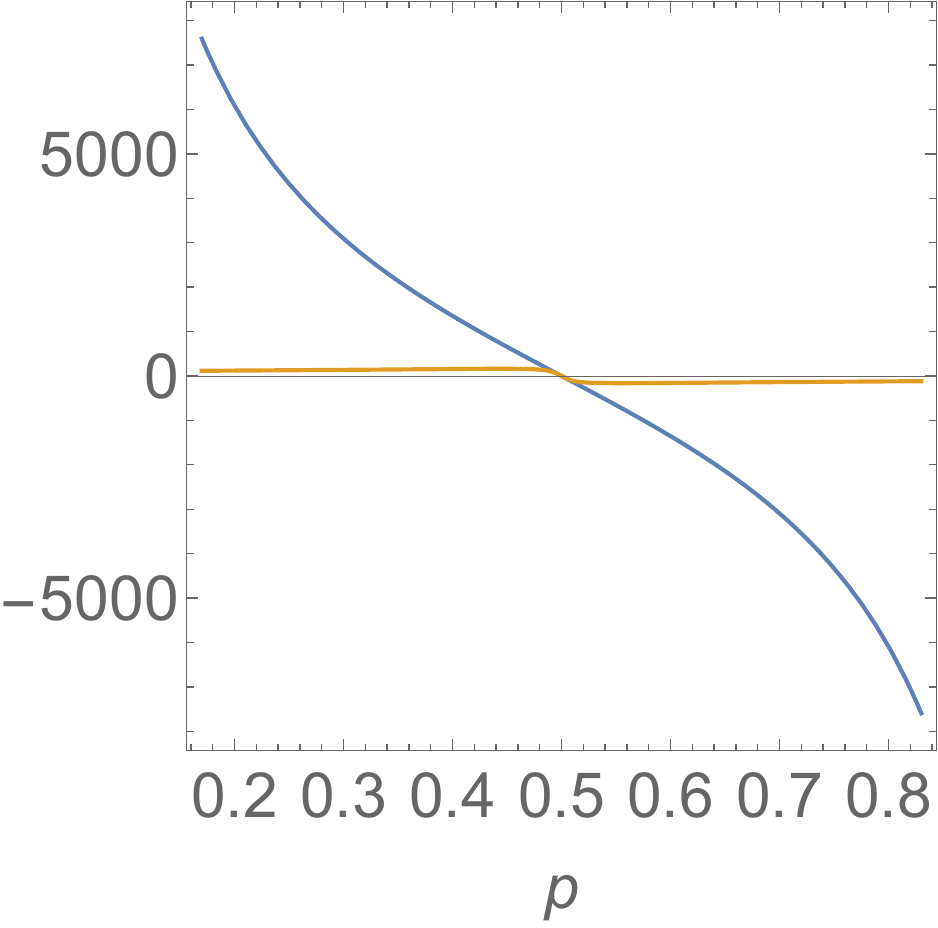}
}
\put(330,0){
  \includegraphics[width=0.2\textwidth]{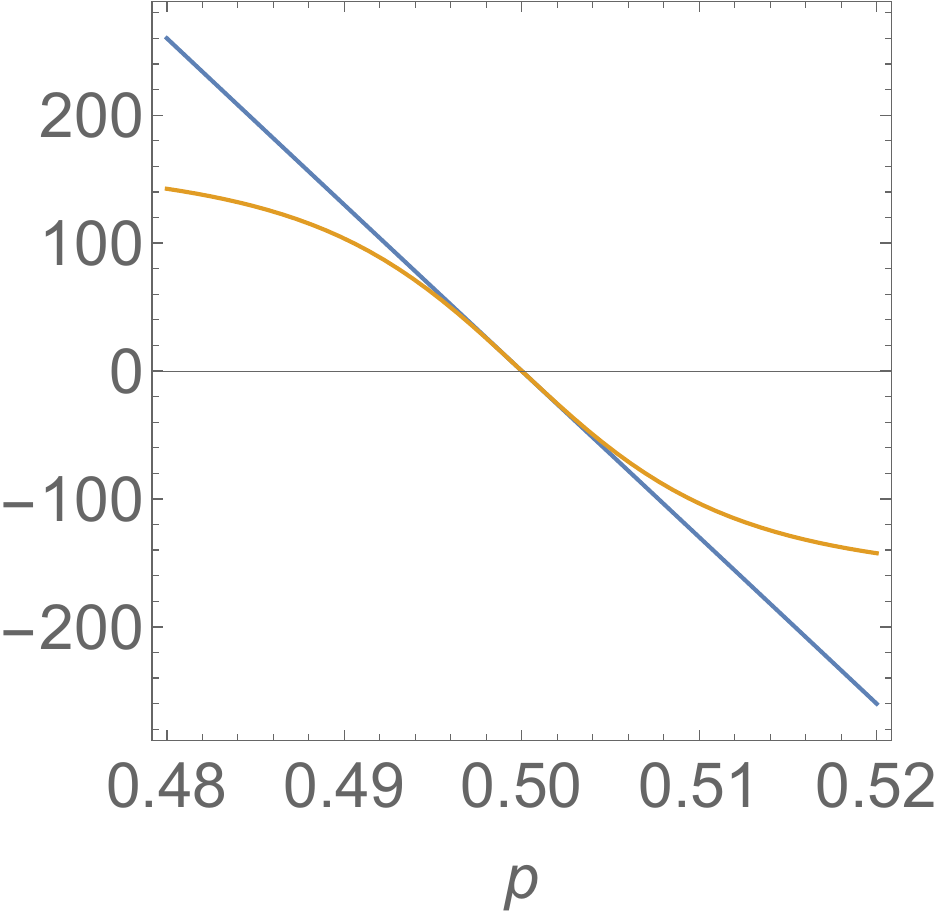}
}
\end{picture}
\caption{Graphical solution of equation \eqref{eq:029} for $L=100$:
the orange and the blue curves (color online) are, respectively,
the plots of the right and left hand sides of the equation.
The second and the fourth panel are a magnification of the first and the
third, respectively.
The experimental measure $\sum_ik_i/m$ is equal to $895$ in the
first two panels and $1623$ in the last two.
The experimental data used in the first two plot have been obtained by
simulating the process with $p=0.48$, whereas for the last two plots
the simulation has been run with $p=0.5$.
The two curves intersect each others at $0.4802$, $0.5$ and $0.5198$
in the first two plots and at $0.5$ in the last two.
}
\label{f:015}
\end{figure}

We conclude this section testing the above procedure
performing a numerical experiment. We first run the model
with $L=100$ and $p=0.48$. We find $1889.17$ as numerical estimate for the
residence time, so that the experimental value of the quantity
$\sum_{i=1}^mk_i/m$ appearing in \eqref{eq:020} is $895.085$.
The graphical solution of \eqref{eq:029} is depicted in
the two leftmost panels of Fig.~\ref{f:015}. The two curves
intersect in $1/2$ and in the two symmetric points
$0.4802$ and $0.5198$. These two last values are the
maximum likelihood estimate for $p$ and, as we discussed above,
the information provided by this experiment is not
sufficient to identify correctly the parameter.
On the other hand, by running the simulation with $p=1/2$,
we found $3344.65$ as numerical estimate for the
residence time, so that the experimental value of the quantity
$\sum_{i=1}^mk_i/m$ appearing in \eqref{eq:020} is $1622.83$.
The graphical solution of \eqref{eq:029} is, now, depicted in
the two rightmost panels of Fig.~\ref{f:015}. The two curves
intersect only in $p=1/2$ which is the maximum likelihood estimate
for $p$ in this case.



%
%

\section{Identification of a defect}
\label{s:defect}
We consider now the case of presence of a defect in the lane.
The defect is the site $d$, where the probabilities are $p'$
to jump to the right and $q'=1-p'$ to the left.
We can have different situations: $p$, $d$, $p'$ can be known or unknown.
In this section\footnote
{In case we suppose that
the geometric and transport properties of the
defect which, in our model, are represented by $d$ and $p'$,
are known, then we face the problem of estimating $p$.
Suppose we know $N$ and we measure the number $m$ of crossing
particles:
since for fixed $p'$ and $d$
the probability of reaching the site $L$ is
monotonically increasing as a function of $p$,
as we did in the case of no defect (see eq.\eqref{eq:027}),
the maximum likelihood estimation gives
$m/N=\mathbb{P}(\rexit)$.
Hence, solving such an equation
we get an estimate for $p$ and the parameter $p$ is completely identifiable.
On the other hand,
in case $N$ is not known, but it is possible to measure the
mean crossing time we find a lack of complete identifiability.
Indeed,
as in the case of no defect, the identification of $p$ as the most likely
value producing the observed residence time in general do not allow
to distinguish between two different possible values of $p$, since the
residence time as a function of $p$ is monotonically increasing up
to a value $p^*$ and then it decreases (see, for instance,
\cite[right panel of Fig.~9]{CC19}).
The effect of the presence of the defect is that of modifying
the numerical value of the measured residence time and to shift the
value of $p^*$, that in the case of no defect is $0.5$.
}
we suppose that $p$ is known, namely, the
transport properties of the lane are known, and we want
to estimate the geometric and transport properties of the
defect which, in our model, are represented by $d$ and $p'$.

Note first of all that this case is more complex than the previous one without
the defect site. The probability for a walker to jump out from the
right--side exit in a time exactly $n$, starting from $1$ is now
\begin{equation}
\label{eqe000}
\mathbb{P}(\tau=n, \rexit)
=
\sum_{j=1}^k \sum_{i=1}^j \alpha_{i,j,d}
p^{L-1+k-i} (p')^i q^{k-(j-i)} (q')^{j-i}
\,,
\end{equation}
where $j$ counts the total number of times that the particles
jumped on the defect $d$ before reaching the site $L$, while
$i$ is the number of times that the particle starting from $d$
jumped  to $d+1$. Moreover,
$\alpha_{i,j,d}$ is the total number of different
paths that the particle can perform in the line for each choice
of $i$, $j$, $d$.
Calculation of $\alpha_{i,j,d} $ is in principle possible
but it is not easy and not necessary for our purpose.

%
%
%
%
%
%
%
%
%

\subsection{Measuring the fraction of successful particle in presence of a defect}
It is possible to produce explicit formulas for the probability
of exiting through the right end point even in presence of a defect (see \cite{CC19}).
Indeed, the probability to cross the lane is given
by \eqref{eq:030} and \eqref{eq:040} and can be considered
(for $p$ fixed and known) as a function of $p'$ and $d$:
\begin{equation}
\label{eq:030}
\mathbb{P}(\rexit)(p',d)
=
\frac{p' (p-q)q}{p(-p'q(-1 + (q/p)^d)+ pq'(-(q/p)^L + (q/p)^d))},
\quad p\neq q;
\end{equation}
\begin{equation}\label{eq:040}
\mathbb{P}(\rexit)(p',d)= \frac{1}{d+ (L-d)q'/p'}
,\quad p= q.
\end{equation}

Using the expressions of the probability to reach the
right side \eqref{eq:030} and \eqref{eq:040},
the observation of how many particles reached the site $L$ allows
to estimate $(p',d)$ through a maximum likelihood procedure.
Indeed,
if $m$ of $N$ independent
particles started at $1$ exit the lane through the
right end point,
similarly to what we did in absence of defect
we can write
\begin{equation}
(\hat p', \hat d)
=
\mathrm{arg}\,\max_{p'\in[0,1],d\in\{2,\ldots,L-2\}}
\binom N m (\mathbb{P}(\rexit))^m(1-\mathbb{P}(\rexit))^{N-m}
\;,
\end{equation}
under the assumption that
the Bernoulli random variables taking values 1 if
$\rexit$ is true and zero otherwise are independent and identically
distributed.

We consider $d$ to be a continuous variable, to simplify the evaluation, but we should remember that only discrete values $d=\{2,3,\ldots,L-2\}$ for the defect are significant in our problem.

Considering the log--likelihood we write
\begin{equation}
\label{eq:e010}
\begin{split}
\ell(d,p')
&
=\ln {N\choose m} + m \ln \mathbb{P}(\rexit)
+ (N-m) \ln (1-\mathbb{P}(\rexit))
\end{split}
\end{equation}
and we derive it with respect to $p'$ and $d$ to search for critical points
\begin{displaymath}
\begin{cases}
&\frac{\partial\ell}{\partial p'}
=
\frac{m}{\mathbb{P}(\rexit)}
 \frac{\partial}{\partial p'}
\mathbb{P}(\rexit)
-\frac{N-m}{1-\mathbb{P}(\rexit)}
\frac{\partial}{\partial p'}\mathbb{P}(\rexit)=
\big [ \frac{m}{\mathbb{P}(\rexit)}- \frac{N-m }{1-\mathbb{P}(\rexit)}\big]
 \frac{\partial}{\partial p'}\mathbb{P}(\rexit)=0\\
&
\frac{\partial\ell}{\partial d}
=
\frac{m}{\mathbb{P}(\rexit)}
\frac{\partial}{\partial d}\mathbb{P}(\rexit)
- \frac{N-m}{1-\mathbb{P}(\rexit)}
\frac{\partial}{\partial d}\mathbb{P}(\rexit)
= \big[ \frac{m}{\mathbb{P}(\rexit)}-\frac{N-m}{1-\mathbb{P}(\rexit)}\big]
 \frac{\partial}{\partial d}\mathbb{P}(\rexit)=0
\;.
\end{cases}
\end{displaymath}
Hence,
all the points $(d,p')$ that verify
\begin{equation}
\label{eq:050}
\mathbb{P}(\rexit)=\frac{m}{N}
\end{equation}
are critical, where, we recall, $\mathbb{P}(\rexit)$ is given
in \eqref{eq:030} and \eqref{eq:040}.

To prove that the solutions of \eqref{eq:050} are the sole critical point
for the maximum likelihood function, we show that
$\partial\mathbb{P}(\rexit)/\partial p'$ and
$\partial\mathbb{P}(\rexit)/\partial d$ are different from zero.

We first discuss in detail the case $p=1/2$:
from \eqref{eq:040} we have
\begin{displaymath}
\frac{\partial}{\partial p'}\mathbb{P}(\rexit)
=
\frac{(L-d)/p'^2}{\big[ d+(L-d)(1-p')/p'\big]^2}
\;\;\textrm{ and }\;\;
\frac{\partial}{\partial d}\mathbb{P}(\rexit)
=
\frac{1-(1-p')/p'}{\big[d+(L-d)(1-p')/p'\big]^2}
\;.
\end{displaymath}
The quantity on the left
is always positive for $d<L$ and
it is equal to $0$ if and only if $d=L$,
while
the quantity on the right
is positive for $p'>1/2$ and negative for $p'<1/2$,
and it is  equal to $0$ if and only if $p'=1/2$,
that is to say in absence of defect ($p=q=p'=1/2$).

The case $p\neq q$ can be discussed analogously and, again, we find that
the partial derivative with respect to $p'$ is equal to zero
if and only if $d=L$, while
the partial derivative with respect to $d$ is equal to zero
if and only if $p'=p$, that is to say in the case where no defect is present.

\begin{figure}[ht!]
\begin{picture}(200,350)(0,0)
\put(-5,180){
  \includegraphics[width=0.35\textwidth]{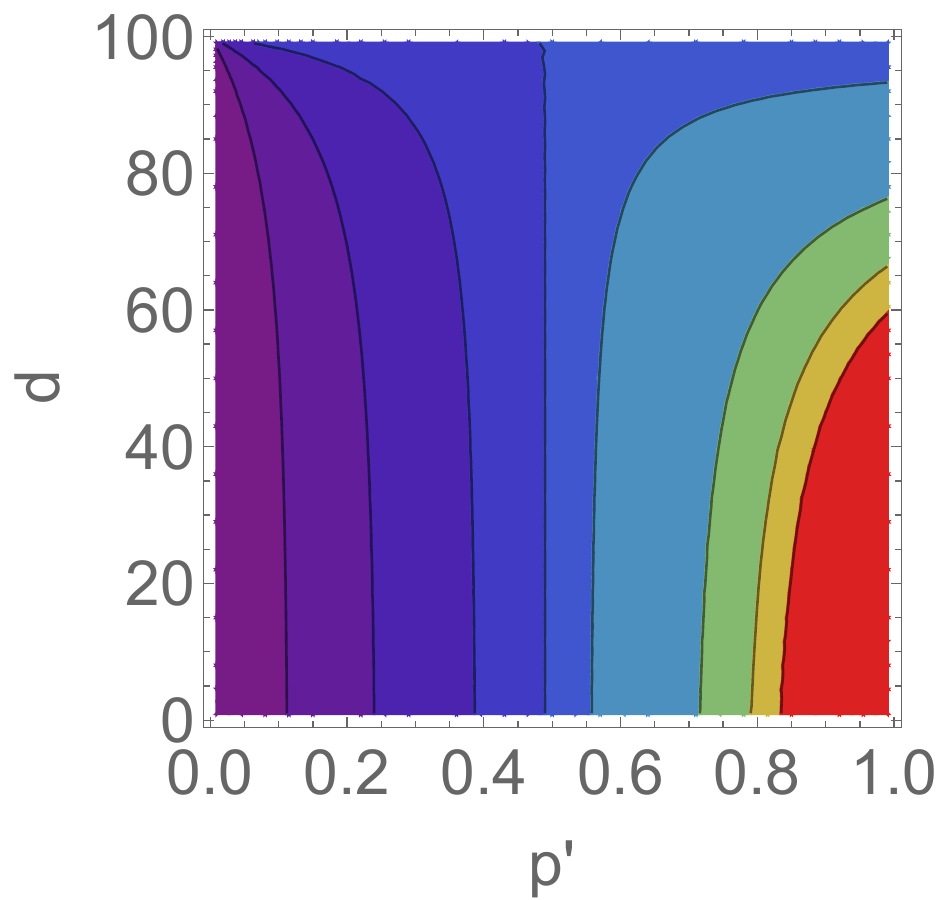}
}
\put(154,195){
  \includegraphics[width=0.15\textwidth]{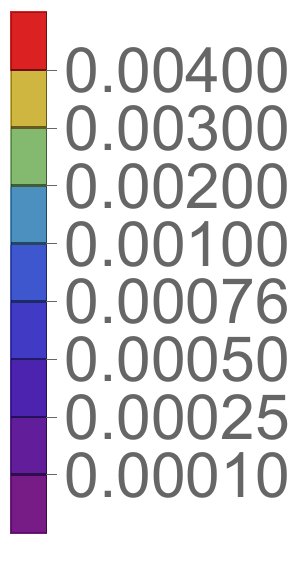}
}
\put(238,180){
  \includegraphics[width=0.35\textwidth]{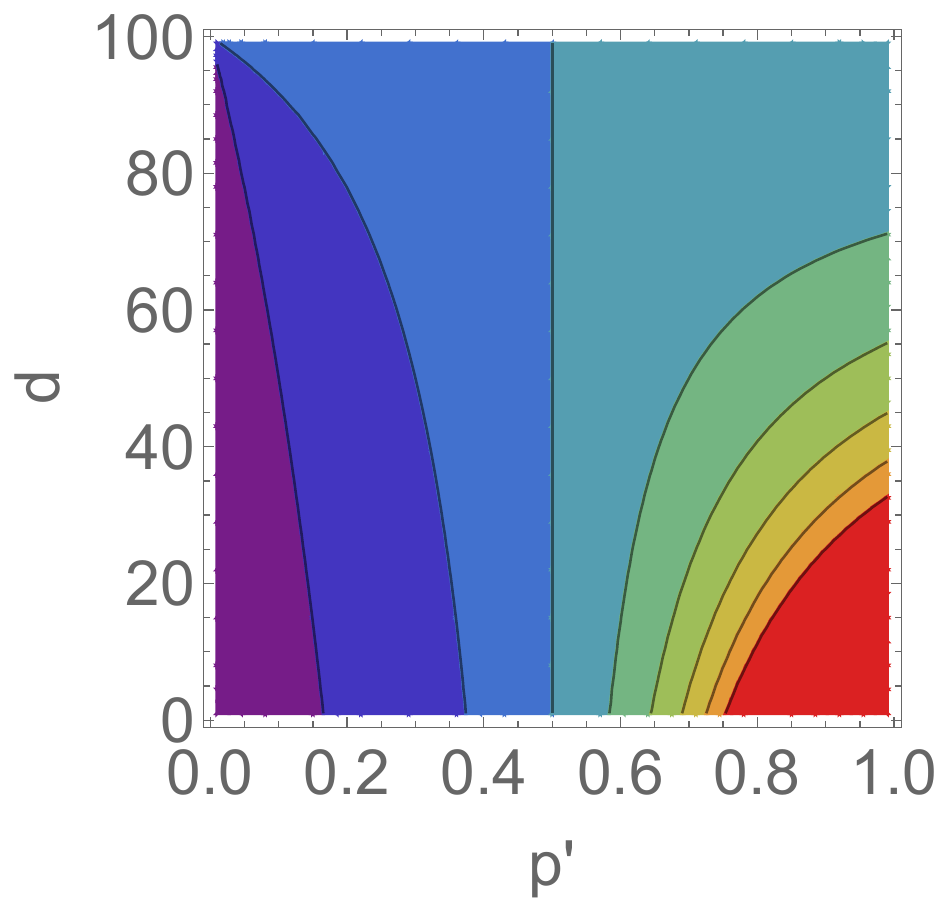}
}
\put(397,196){
  \includegraphics[width=0.113\textwidth]{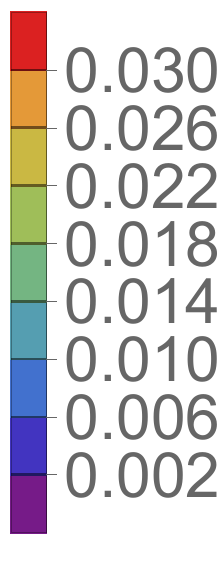}
}
\put(-5,0){
  \includegraphics[width=0.35\textwidth]{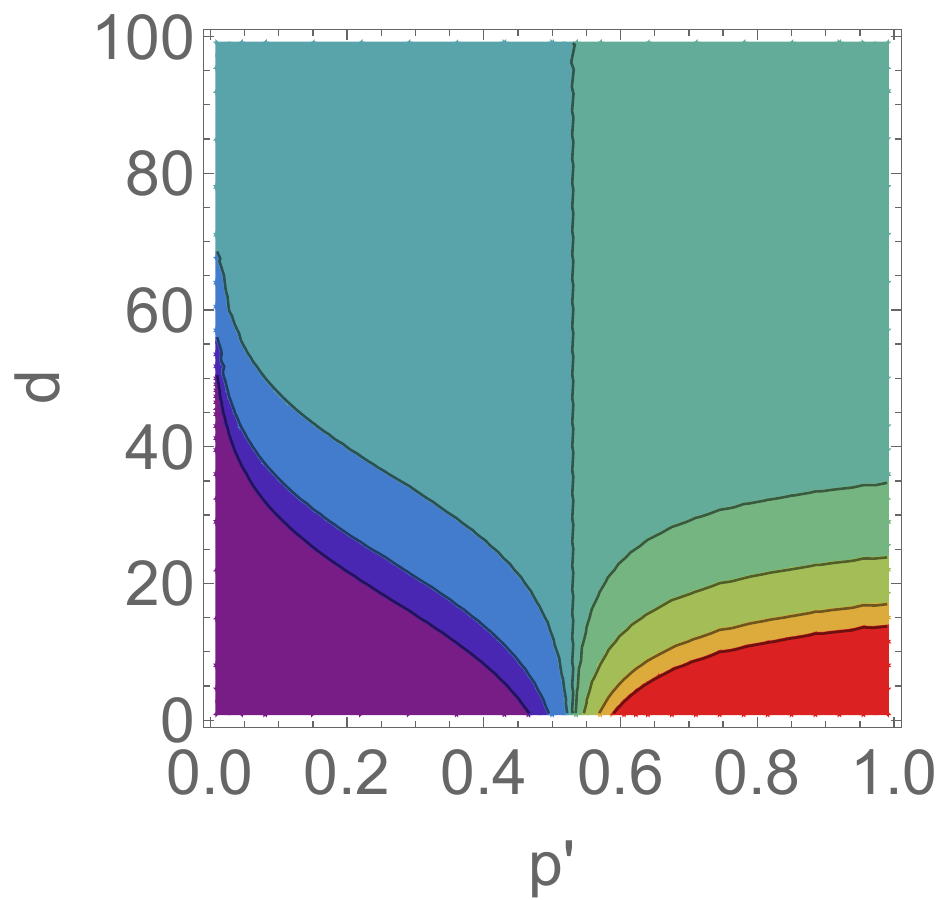}
}
\put(154,15){
  \includegraphics[width=0.15\textwidth]{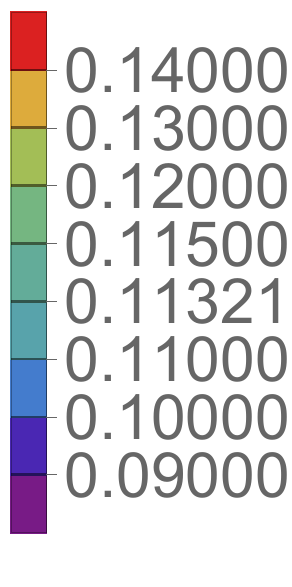}
}
\put(238,0){
  \includegraphics[width=0.35\textwidth]{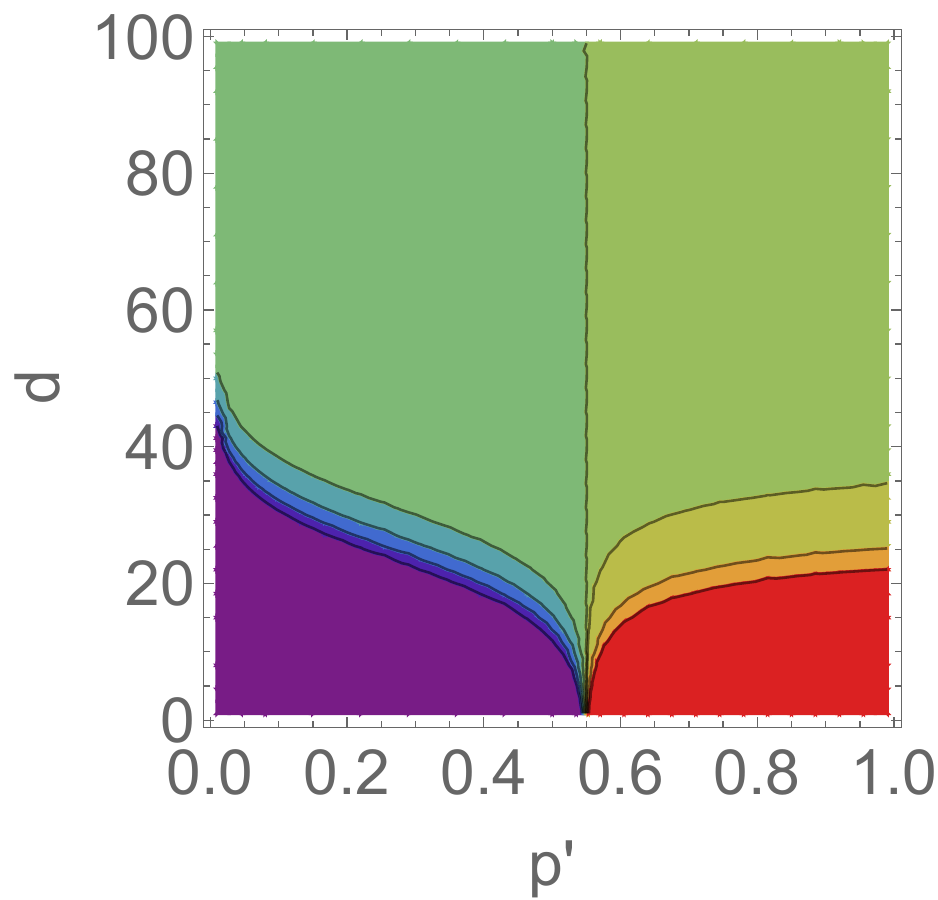}
}
\put(397,15){
  \includegraphics[width=0.15\textwidth]{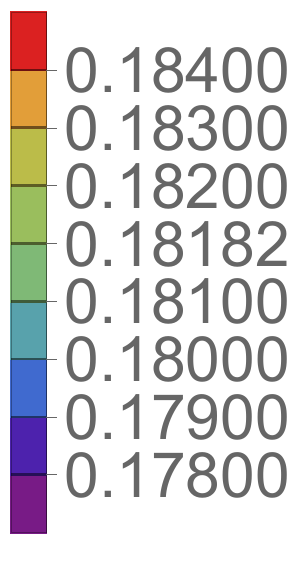}
}
\end{picture}
\caption{Contour plot of $\mathbb{P}(\rexit)$ in the plane
$p'$--$d$ for $L=100$.
Top left $p=0.49$, top right $p=0.5$,
bottom left $p=0.53$, bottom right $p=0.55$.
}
\label{f:501}
\end{figure}

Given $m$ and $N$, the solutions of the equation \eqref{eq:050}
determine a curve in the plane $p'$--$d$.
These curves are depicted in Fig.~\ref{f:501} for different
values of $p$.

The contour plots in Fig.~\ref{f:501} show that,
given $m$ and $N$, the geometric locus made of the solutions
of \eqref{eq:050} provides $d$ as a function of $p'$
which is decreasing if $p'<p$,
increasing for $p'>p$,
and constant for $p'=p$.
Note that equation \eqref{eq:050} is easily explicitly solvable,
finding $p'$ as a function of $d$ and taking into account only the values
of $d$ that produce $p'\in[0,1]$.

Summarizing, dealing with the experiment 1  we are not able to
fully estimate the transport and geometric properties
of the obstacle, $p'$ and $d$. Indeed, we are able to
find only a relation between the two parameters, that is to say
we can determine a curve in the plane $p'$--$d$ where the parameters
have to lay.
Hence, the model is not anymore locally identifiable.

\subsection{Residence time in presence of a defect}
\label{s:drt}
In presence of a defect we have just shown that experiment 1,
namely, the measure of the fraction of particle crossing the lane,
is not sufficient to achieve an estimate for the model
parameters $p'$ and $d$. In this section we approach the
problem also from the point of view of the residence time estimate: we shall
see that using both the fraction and the residence time measure, in
some cases, it will be possible to identify completely the two
parameters.
Due to the complicated structure of the problem, we do not rely on
a pure maximum likelihood estimate, but we simply identify the
theoretical residence time with the experimental
mean crossing time by referring the method of moments. 

In order to derive a theoretical expression for the
residence time we follow the strategy outlined in
\cite{Fel,S1975,L09,CC19} and based on a generating function
computation.
We let $u_{n}$ be the probability that the walk started at $1$
exits through the right end after $n$ steps.
We construct the generating function of the probability of exiting from the
right side starting from $1$,
$U(s)=\sum_{n\geq 0} u_{n} s^{n}$.
The generating function can be explicitly calculated but has not a
simple and concise expression; we report the computation
in the Appendix~\ref{appendice} for completeness, see also \cite{CC19}.

Now, we note that $\mathbb{P}(\rexit)=U(1)$.
Since the derivative of the generating function is
$U'(s)=\sum_{n\geq1} n u_{n} s^{n-1}$,
following \cite{S1975}
we have that
\begin{equation}
\label{res000}
\lim_{s\to 1^-} U'(s)
=
\sum_{n\geq1} n u_{n}
=\sum_{n\geq 1} n \mathbb{P}(\tau=n,\rexit)
=\sum_{n\geq 1} n \mathbb{P}(\tau=n|\rexit)\mathbb{P}(\rexit)
=R\, \mathbb{P}(\rexit) ,
\end{equation}
where $R$ is the residence time as defined in \eqref{one010}.
We note that $R$ is finite for any fixed
$p$, $q$, $p'$, and $q'$ in $(0,1)$.

By deriving two times $U(s)$ we find the following expression, that
is useful to calculate the variance of the residence time:
\begin{displaymath}
\begin{split}
\lim_{s\to 1-} U''(s)&
=\sum_{n\geq 2} n(n-1) u_n=\sum_{n\geq 1} n^2 u_n
- \sum_{n\geq 1} n u_n= \mathbb{P}(\rexit) \left[\sum_{n\geq 1} n^2
\mathbb{P}(\tau =n|\rexit)- R \right] \\
&=\left[\mathbb{E}(\tau^2|\rexit) - R\right] \mathbb{P}(\rexit)
\;.
\end{split}
\end{displaymath}
Thus, it is possible to calculate the variance of the residence time as
\begin{displaymath}
\begin{split}
\mathrm{Var}(\tau|\rexit)
=
\mathbb{E}(\tau^2|\rexit)
- \left[\mathbb{E}(\tau|\rexit)\right]^2
= \frac{1}{\mathbb{P}(\rexit)}
\left[ U''(1) + U'(1) \right]-R^2
\;.
\end{split}
\end{displaymath}

\begin{figure}[ht!]
\begin{picture}(200,350)(0,0)
\put(-80,170){
  \includegraphics[width=1\textwidth]{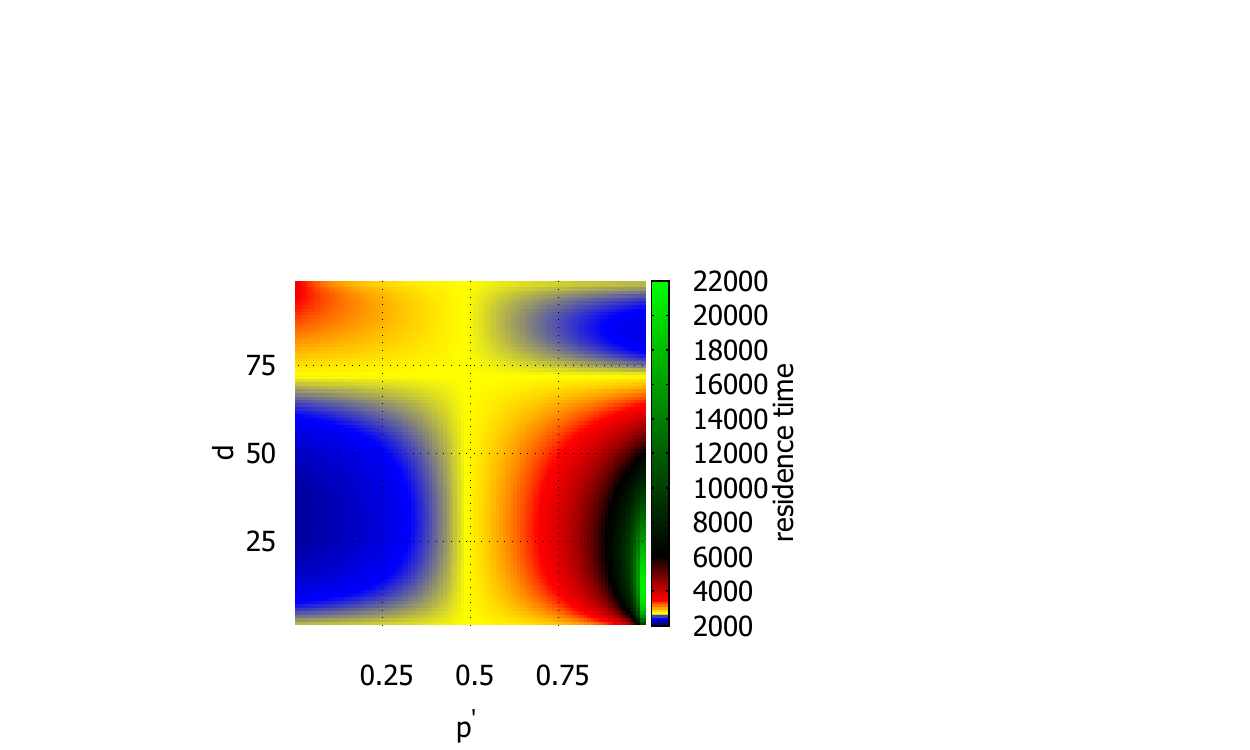}
}
\put(165,170){
  \includegraphics[width=1\textwidth]{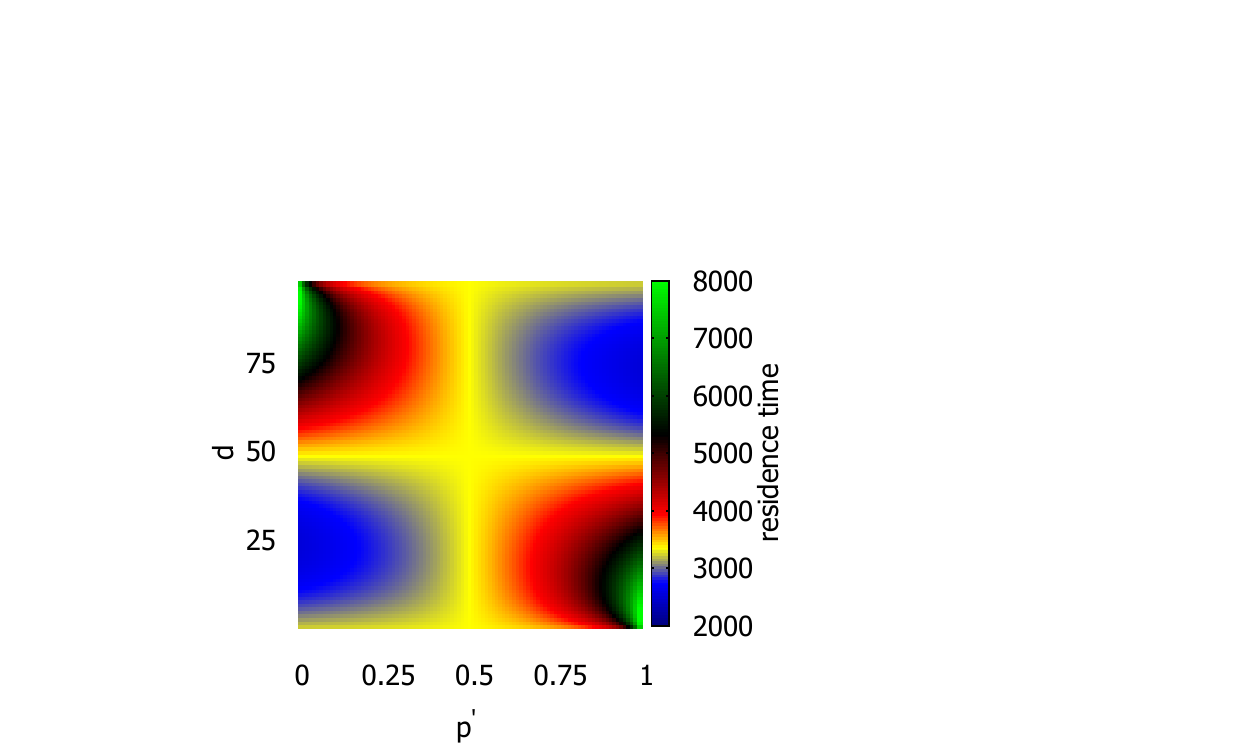}
}
\put(-80,-10){
  \includegraphics[width=1\textwidth]{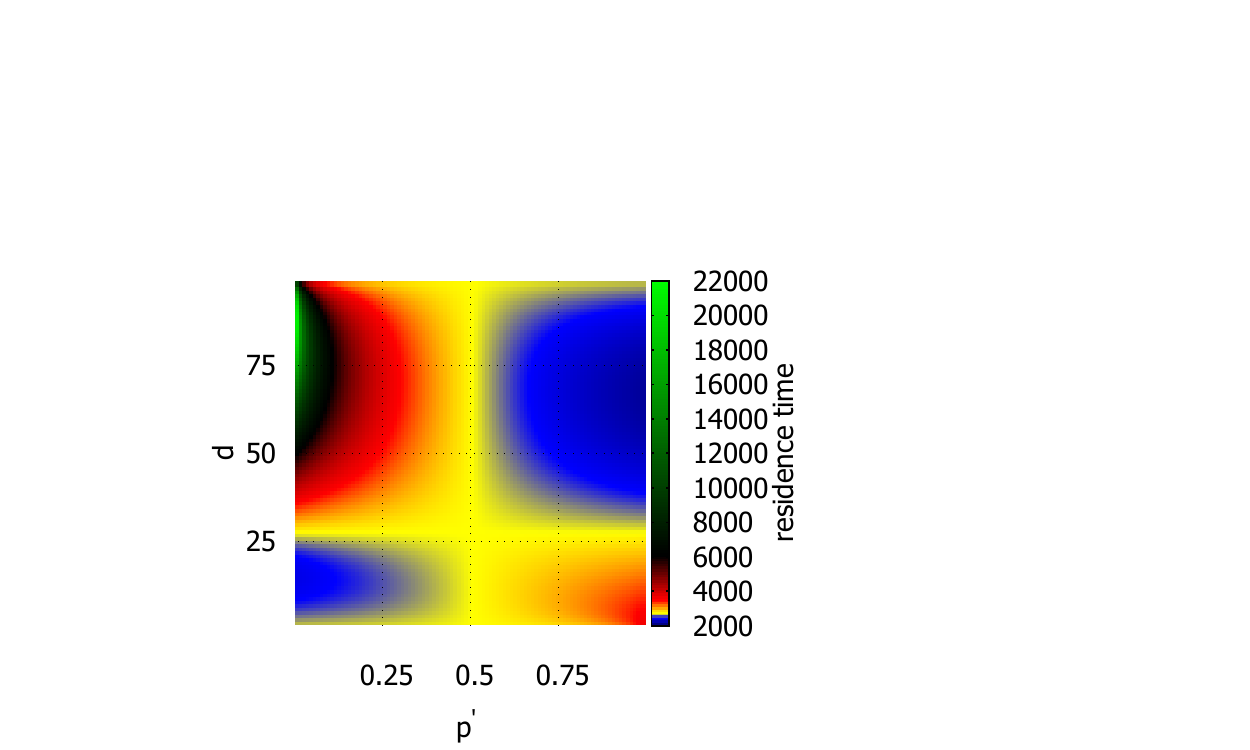}
}
\put(165,-10){
  \includegraphics[width=1\textwidth]{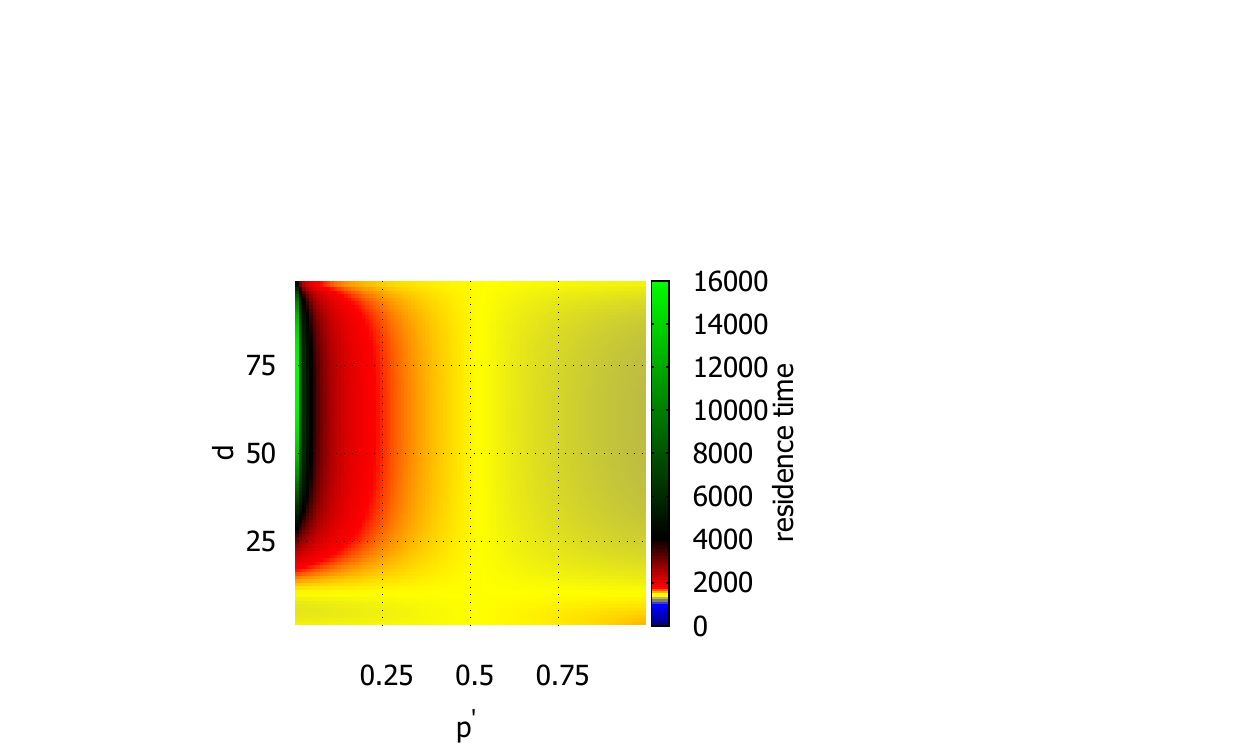}
}
\end{picture}
\caption{Contour plot of the residence time $R$
for $L=100$,
$p=0.49$ (top left),
$p=0.50$ (top right),
$p=0.51$ (bottom left),
$p=0.53$ (bottom right).
}
\label{f:505}
\end{figure}

The  residence time as a function of $p'$ and $d$ can be computed
using the equations shown above. Data for $L=100$ are
reported in Fig.~\ref{f:505} as a scatter plot.

We first notice that the picture in the case $p=0.49$ is specular to that one
for $p=0.51$, and this is always true for any choice of $p$ and $1-p$.

In the symmetric case $p=1/2$ the residence time has a peculiar behavior:
for $p'=p=1/2$ (no real defect is indeed present) and in the case
$d=L/2$ the residence time in presence of the defect is not influenced by
the defect, it is the same as the symmetric case with no defect, see the
yellow region in the plot (panel top right).
These two straight lines divide the
rectangle $[0,1]\times \mathbb{Z}\cap\{2,3,\ldots,L-2\}$ into
four zones symmetric with respect to the center:
two specular regions where the residence time is smaller with respect
to the case of no defect (blue zone in the plot) and two specular regions
where the presence of the defect produce a larger residence time (from red
to green in the plot).
In particular, a defect with $p'<p=1/2$ close to the entrance favors
a faster egress of (a smaller number of) particles, as well as the
presence of a defect with $p'>p=1/2$ close to the exit.
The defects close to the exit with $p'<p=1/2$ tend to trap particles
that reached the second half of the interval, as well as defects with
$p'>p=1/2$ close to the entrance that tend to favor a long permanence
of the walkers in the first half of the interval.
We remark that every observation about the residence time takes into
account only the particles that eventually reach the right exit site $L$, 
ignoring those that come out of the interval at the $0$ site.

In the case $p>1/2$, the behavior is a perturbation of the
case $p=1/2$: the four zones are still present if $p$ is close
to $1/2$, but the two regions in the top of the plot tend to invade the
whole rectangle $[0,1]\times \mathbb{Z}\cap\{2,3,\ldots,L-2\}$
when $p$ is increased.
It is still present a cross made of points (yellow in the plots
in Fig.~\ref{f:505}) where the residence time is the same as in the
case of no defect, the vertical arm of the cross is at $p'=p$,
while the horizontal one moves downward when p is increased
(in the plot is visible at about $d=27$ for $p=0.51$,
at about $d=10$ for $p=0.53$).
The regions below the horizontal arm
are not the specularly symmetric of the top regions, indeed, the values
of the residence time vary more consistently in the top regions.
When $p$ is close to 1
in the residence time plot only two regions are left:
for $p'<p$ the residence time is larger than in the no defect case,
while for $p'>p$ it is the opposite.

The case $p<1/2$ is specular to the case $p>1/2$.

\subsection{Numerical Experiments}
\label{s:numesp}
We consider, now, a numerical experiment in which we melt
both experiments 1 and 2, namely, considered $N$ particles started
at site $1$ of the lane, we count the number of particles
$m$ which exit the lane through the right end 
and we measure the crossing time $n_i$
that each of them takes to reach the exit.
Hence, we get both a numerical measure of the crossing fraction
$m/N$ and of the mean crossing time $\sum_{i=1}^mn_i/m$.

The question we pose is the following: supposing to know the
transport properties of the lane expressed by the
parameter $p$, can we estimate the geometry and the transport
property of the obstacle represented respectively by $d$ and $p'$?

Recalling the definition of the i.i.d.\ Bernoulli random variables
$\chi_i$ introduced in Section~\ref{s:identifp}, we note that
$m$ is the value assumed in the experiment by the random
variable $(\chi_1+\ldots+\chi_N)/N$ and remark that, by the law of large
numbers, such a variable
converges to $\mathbb{P}(\rexit)$ as $N\to\infty$.
We now estimate the rate of convergence:
for any given $\varepsilon>0$, thanks to the Central Limit Theorem,
we have that
\begin{equation}
\label{eq:123stella}
P\Big(\Big|\frac{(\chi_1+\ldots+\chi_N)}{N} - \mathbb{P}(\rexit))\Big|
\leq\varepsilon\Big)
\simeq
2\Phi\Big(
\frac{\sqrt{N}\varepsilon}{\mathbb{P}(\rexit)(1-\mathbb{P}(\rexit))}\Big)-1
\end{equation}
where $\Phi$ is the cumulative Gaussian distribution. 
Given $\varepsilon>0$, for
$\sqrt{N}\varepsilon/[\mathbb{P}(\rexit)(1-\mathbb{P}(\rexit))]\geq 3$
we can estimate this probability to be very close to $1$ (about $0.9974$).
Note that we do not know the correct value of $\mathbb{P}(\rexit)$ since we
do not know $(p',d)$ that identify the defect.
By equation \eqref{eq:050} the probability $\mathbb{P}(\rexit)$ 
is estimated by $m/N$.

Considering the unbiased sample variance estimator of
the i.i.d.\ $\chi_i$
\begin{displaymath}
\hat{s}^2
=
\frac{1}{N-1} \sum_{i=1}^N \Big(\chi_i -\frac{m}{N}\Big)^2
= \frac{N}{N-1}\left(\frac{1}{N}\sum_{i=1}^N \chi_i^2
- \Big(\frac{m}{N}\Big)^2\right)
\;,
\end{displaymath}
we estimate $\hat{s}^2$ as
\begin{displaymath}
 \frac{N}{N-1} \left(\frac{m}{N} -\frac{m^2}{N^2}\right)
=\frac{m}{N-1} \frac{N-m}{N}
\;.
\end{displaymath}
The standard error of the mean is now computed as the observed
realization of the sample standard deviation divided by the square
root of the sample size:
\begin{displaymath}
\hat{\sigma}_\chi
=\frac{\hat{s}}{\sqrt{N}}
= \sqrt{ \frac{m}{N-1} \frac{N-m}{N^2}   }
\;.
\end{displaymath}

Given an experimental observation of the fraction of the particles
exiting the lane through $L$, we 
can consider the interval 
$[m/N-3\hat{\sigma}_\chi,m/N+3\hat{\sigma}_\chi]$ 
where 
$\mathbb{P}(\rexit)$ lies with very high probability. 

\begin{table}
\centering
\begin{tabular}{c|c|c|c|c|c|c}
\hline\hline
label & $p$ & $p'$ & $d$ & $N$ & $m$ & $r_\textup{ex}$ \\
\hline\hline
1 & 0.51 & 0.130 & 19 & $1.0\cdot10^8$ & 1069462 & 2441.266 \\
\hline
2 & 0.51 & 0.130 &  8 & $2.0\cdot10^8$ & 1505795 & 2460.133 \\
\hline
3 & 0.51 & 0.065 & 92 & $1.0\cdot10^6$ & 36345 &  6366.541\\
\hline
4 & 0.51 & 0.130 & 70 & $2.0\cdot10^7$ & 634906 &  5440.954 \\
\hline
5 & 0.51 & 0.750 & 10 & $2.0\cdot10^7$ & 1412009 & 2930.654\\
\hline
6 & 0.51 & 0.800 & 85 & $1.0\cdot10^6$ & 40614 & 2358.224\\
\hline
7 & 0.51 & 0.800 & 85 & $1.0\cdot10^7$ & 404194 & 2349.55\\
\hline
8 & 0.51 & 0.800 & 85 & $1.0\cdot10^8$ & 4040196 & 2349.153 \\
\hline
9 & 0.53 & 0.050 & 43 & $5.0\cdot10^6$ & 507651 & 3751.445 \\
\hline
10 & 0.53 & 0.800 & 50 & $1.0\cdot10^7$ & 1134138 & 1292.222\\
\hline
11 & 0.53 & 0.250 & 75 & $1.0\cdot10^7$ & 1132454 & 1701.426\\
\hline
12 & 0.53 & 0.250 & 75 & $5.0\cdot10^7$ & 5660505 & 1702.267\\
\hline
13 & 0.53 & 0.250 & 75 & $2.0\cdot10^8$ & 22631148 & 1701.527\\
\hline
14 & 0.53 & 0.800 & 15 & $1.0\cdot10^7$ & 1285723 & 1361.100 \\
\hline
15 & 0.55 & 0.250 & 78 & $1.0\cdot10^7$ & 1817491 & 1030.056\\
\hline
16 & 0.55 & 0.100 & 43 & $5.0\cdot10^6$ & 908079 &  1392.074 \\
\hline
17 & 0.55 & 0.700 & 25 & $5.0\cdot10^6$ & 914097 & 877.913\\
\hline\hline
\end{tabular}
\caption{List of experiments discussed in Section~\ref{s:numesp}.
A progressive number is associated to each experiment and the
parameters $p$, $p'$, $d$, and $N$ are listed.
In the last two columns we list the measured values for $m$ and
$r_\textup{ex}$.
}
\label{tab:lista}
\end{table}

In the same way we study the standard error of the residence time,
i.e., of the average crossing time, producing the confidence
interval at level $0.9974$.
Then, we can find the regions of points of the plane $p'$--$d$
to which corresponds
a residence time in that interval and plot those regions.
Note that in this second case the particles producing the
experimental estimate of the residence time, and so producing the
corresponding confidence interval, are only
those $m$ which exit the lane through the right end point at $L$.

We can thus conclude that, given the numerical measures
$f_\textup{ex}=m/N$ for the fraction of crossing particles
and $r_\textup{ex}=\sum_{i=1}^mn_i/m$ for the mean crossing time,
the estimate for the parameters $d$ and $p'$ will be
constructed by intersecting the regions of the plane
$p'$--$d$ corresponding to the confidence intervals
found for the crossing fraction and for the crossing time.
In the sequel of this section we discuss several experiments
performed with different values of the jump probability $p=0.51,0.53,0.55$ 
(see the list of the experiments in Table~\ref{tab:lista}).

\begin{figure}[ht!]
\begin{picture}(200,145)(0,0)
\put(0,0){
  \includegraphics[width=0.25\textwidth]{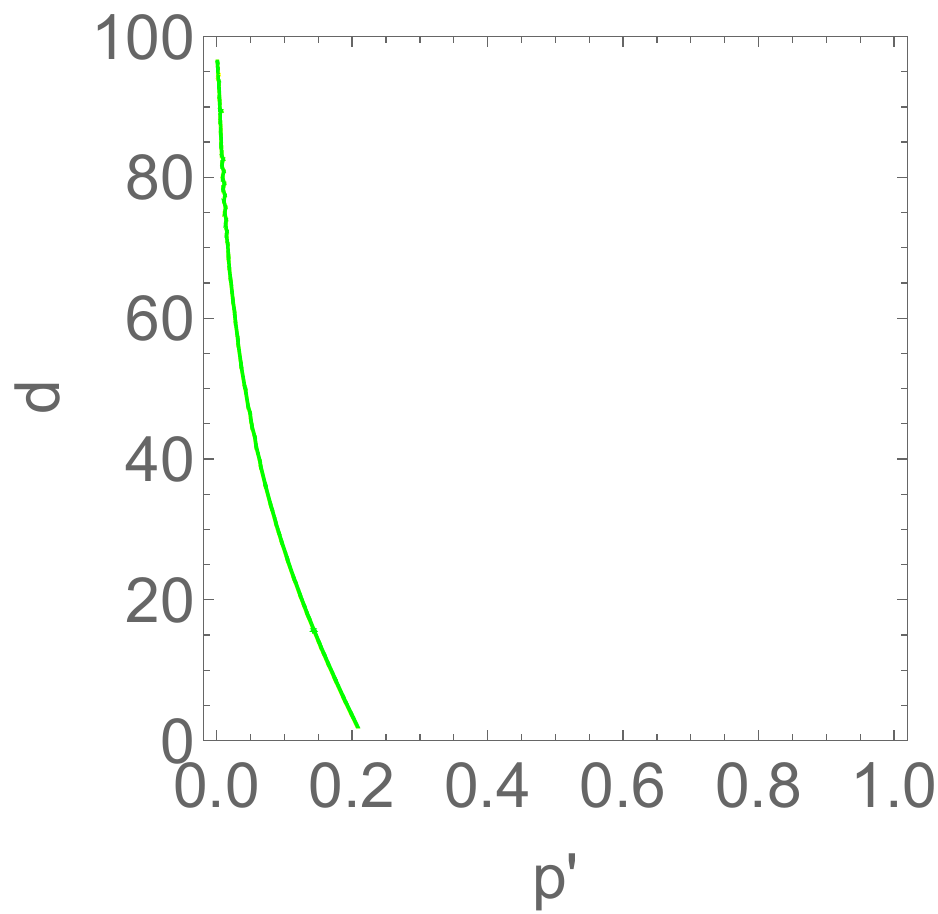}
}
\put(115,20){
  \includegraphics[width=0.12\textwidth]{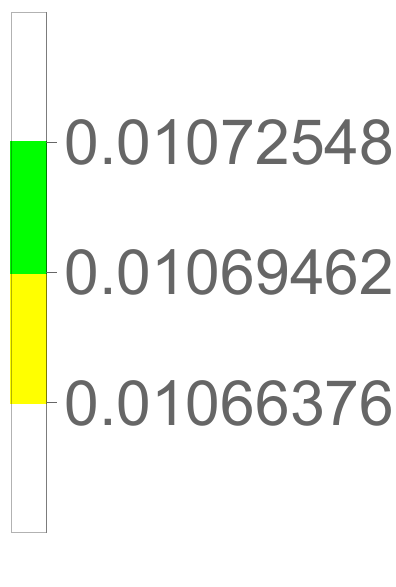}
}
\put(180,0){
  \includegraphics[width=0.25\textwidth]{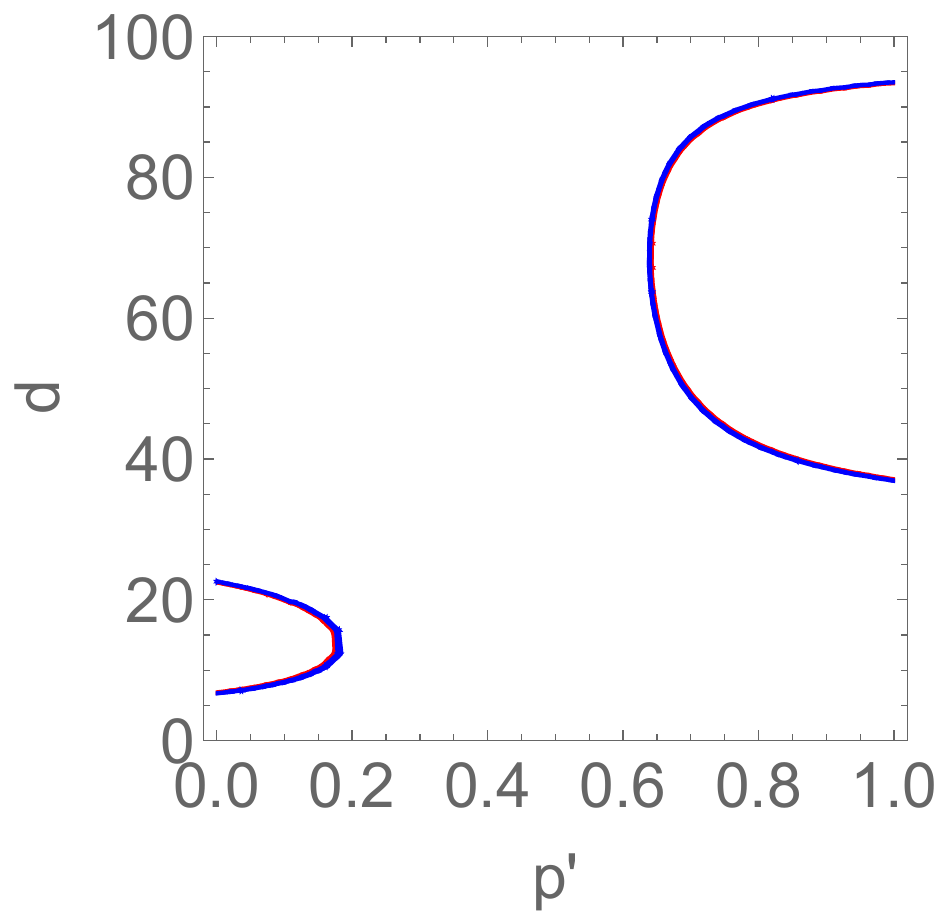}
}
\put(295,20){
  \includegraphics[width=0.10\textwidth]{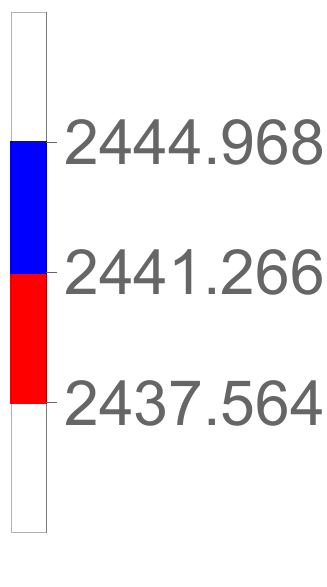}
}
\put(345,0){
  \includegraphics[width=0.25\textwidth]{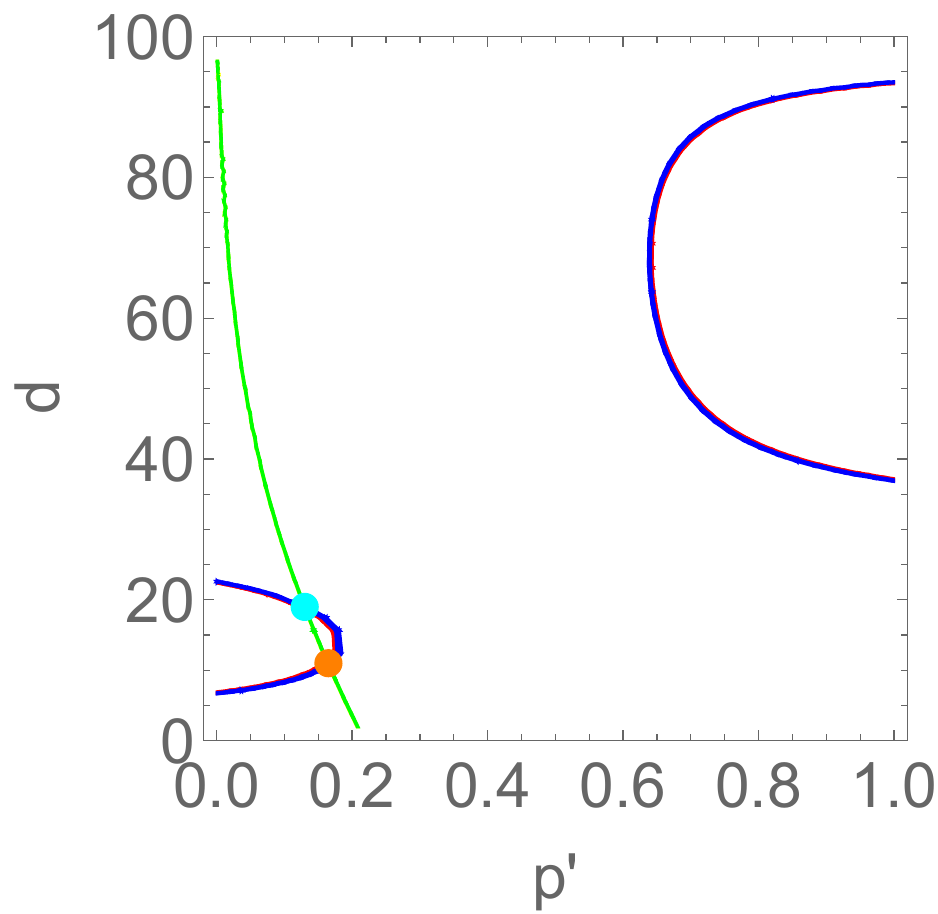}
}
\end{picture}
\caption{Experiment~1.
Left: region associated to confidence interval for $f_\textup{ex}$.
Center: region associated to confidence interval for $r_\textup{ex}$.
Right: intersection of the two regions.
Beside the real value of the obstacle parameters  identified by the cyan point, a second intersection	region around
$(0.1645,11)$ is found (orange point).
}
\label{f:510}
\end{figure}

In Fig.~\ref{f:510} we show results for experiment~1:
the regions associated to the confidence interval for
$f_\textup{ex}$ and $r_\textup{ex}$
are shown together with their intersection.
The confidence intervals for $f_\textup{ex}$ and $r_\textup{ex}$
are, respectively,
$[0.01066376,0.01072548]$ and $[2437.564,2444.968]$.
The intersection is made by two 
small regions, from which we select as candidates only
pairs $(p',d)$ with integer $d$.
The candidates we find are  $p'\in[0.129445,0.130208]$ for integer values of $d=19$, and  $p'\in[0.163927,0.164804]$ for $d=11$. For these values of the parameters, the residence time lies in the interval $[2441.69,2442.21]$ when $p'\in[0.129445,0.130208]$ and $d=19$ and it lies in the interval $[2440.63,2441.15]$ for $p'\in[0.163927,0.164804]$ and $d=11$. Thus, all these candidate pairs $(p',d)$ give a residence time in the confidence interval for $r_\textup{ex}$, so it is not possible to reject any of them.

Hence, a complete identification is not possible.
Nevertheless, it is clear that
all candidate pairs $(p',d)$ correspond to
a defect close to the entrance which tends to obstruct the passage
of walkers ($p'<p$).
No more pairs $(p',d)$ with integer $d$ lie on the intersection region.
 
\begin{figure}[ht!]
\begin{picture}(200,145)(0,0)
\put(0,0){
  \includegraphics[width=0.25\textwidth]{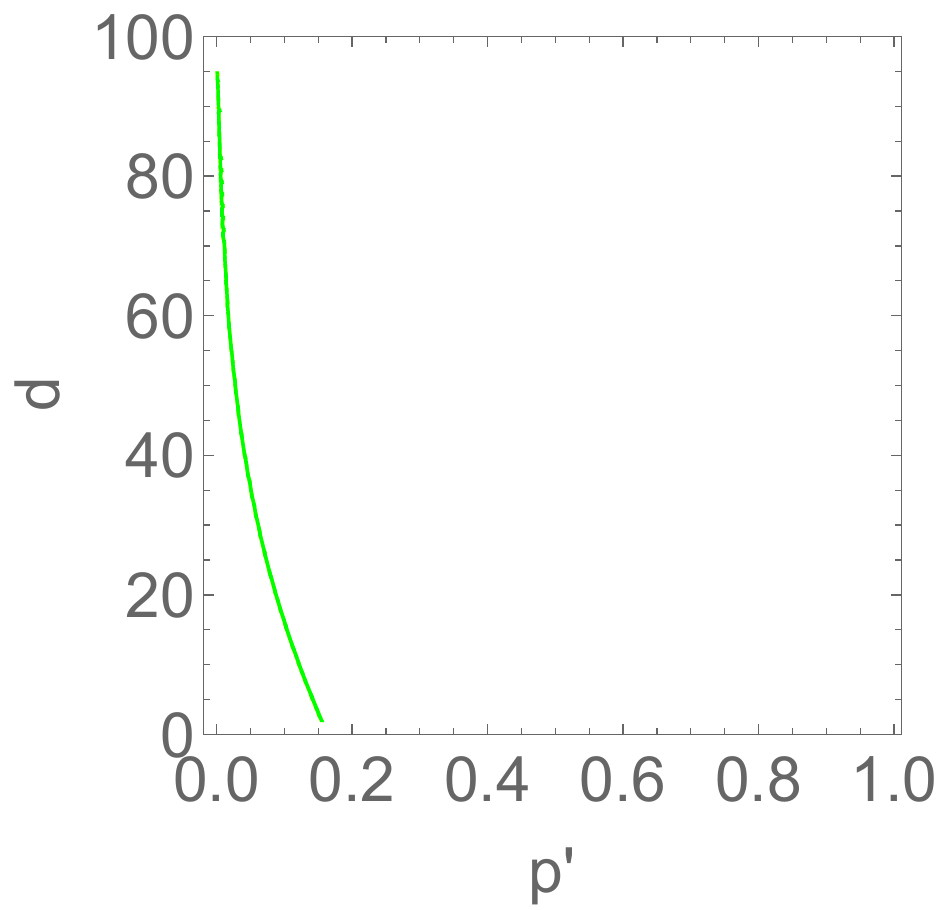}
}
\put(115,20){
  \includegraphics[width=0.12\textwidth]{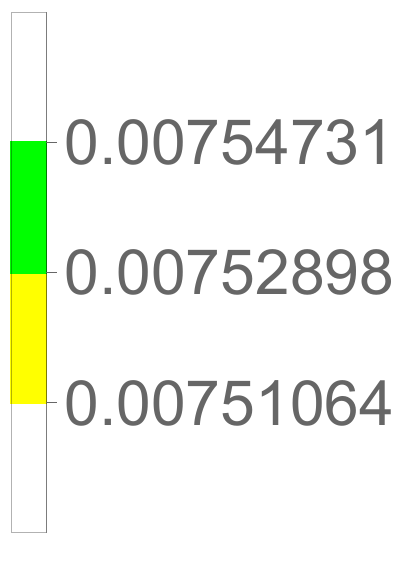}
}
\put(180,0){
  \includegraphics[width=0.25\textwidth]{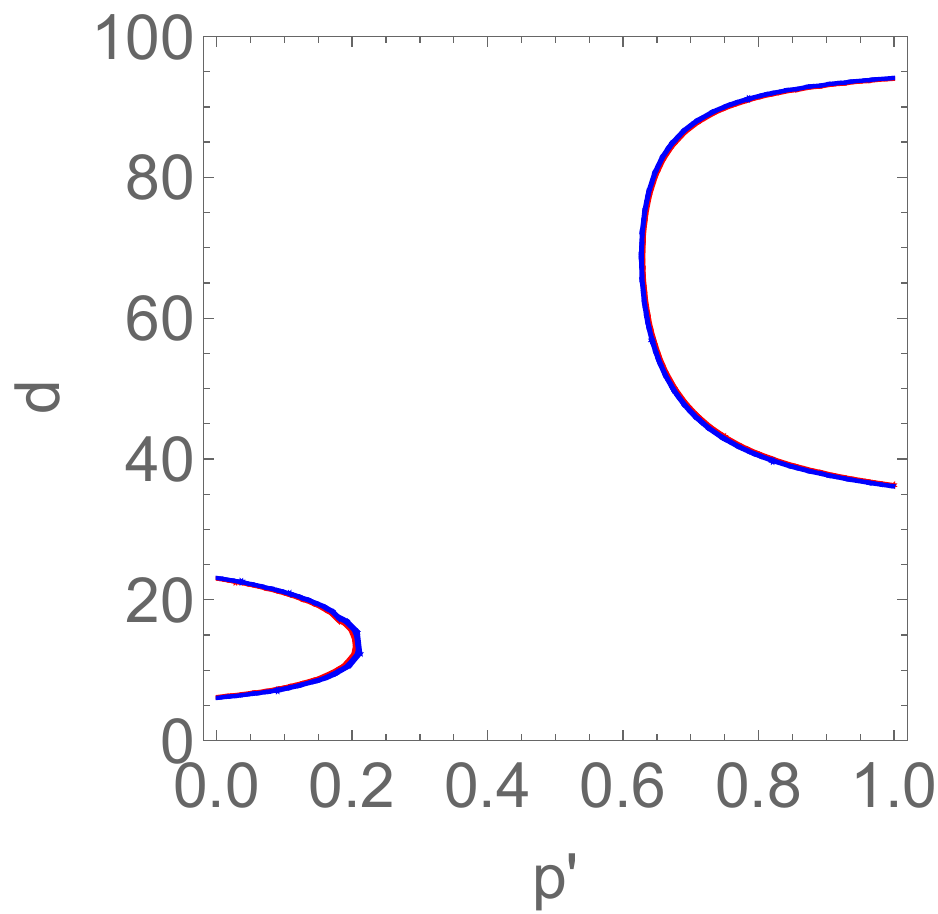}
}
\put(295,20){
  \includegraphics[width=0.10\textwidth]{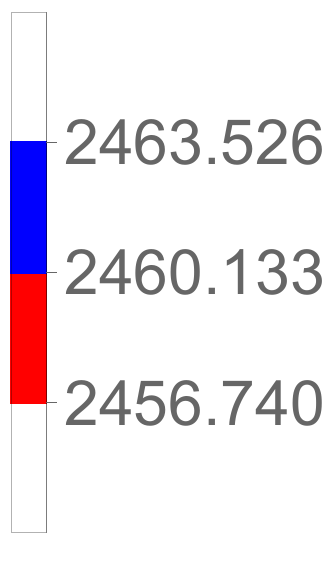}
}
\put(345,0){
  \includegraphics[width=0.23\textwidth]{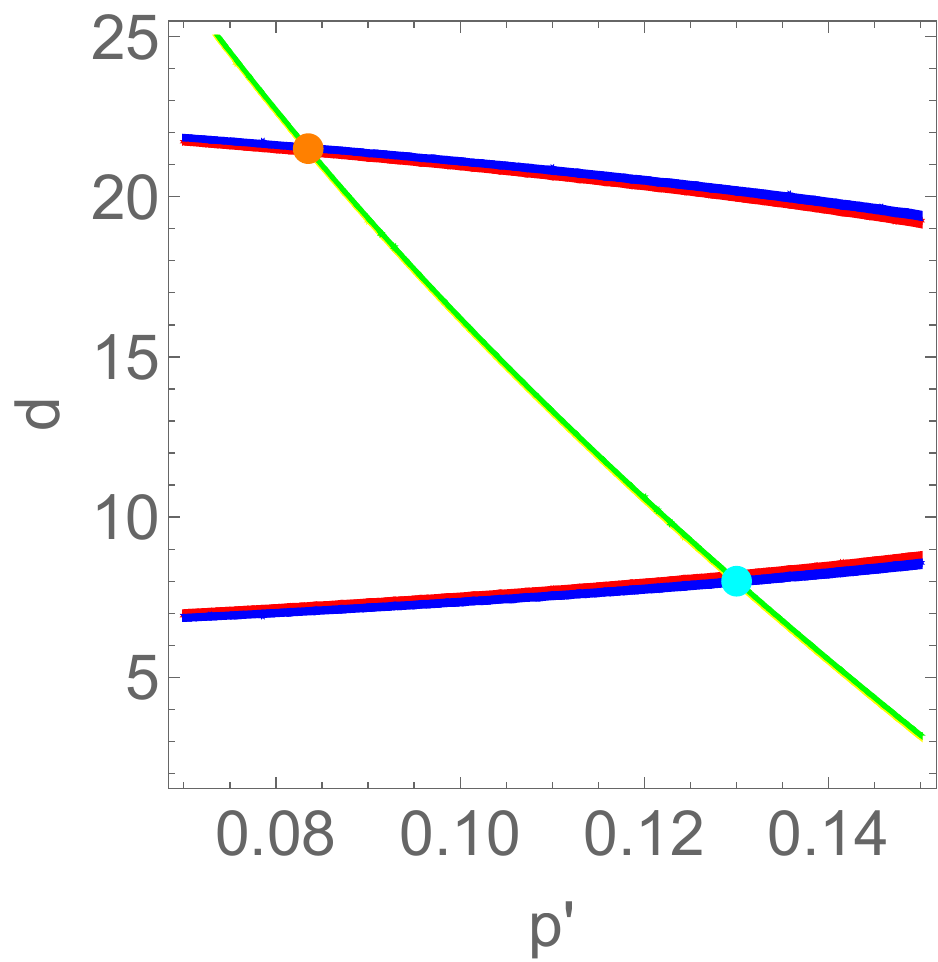}
}\end{picture}
\caption{
Experiment~2.
Left: region associated to confidence interval for $f_\textup{ex}$.
Center: region associated to confidence interval for $r_\textup{ex}$.
Right: magnification of the intersection of the two regions. Points identify the two intersections: the first around the actual defect (in cyan) and a second intersection at about $(0.0835,21.5)$ (in orange).
}
\label{f:520}
\end{figure}

In Fig.~\ref{f:520} we show the results of experiment 2. The setting is similar to the experiment 1, but we are here able to identify the defect parameter $d$ and restrict the possible values of $p'$ to a small interval. 
The regions associated to the confidence interval for
$f_\textup{ex}$ and $r_\textup{ex}$
are shown together with a magnification of their intersection.
The confidence intervals for $f_\textup{ex}$ and $r_\textup{ex}$
are, respectively,
$[0.00751064,0.0754731]$ and $[2456.74,2465.526]$.

The intersection is made by two small regions.
As we did in the case of experiment $1$, we look for the pairs $(p',d)$ with integer $d$ which lie in this regions.
The first region let us identify as possible parameters of the defect $p'\in[0.129606,0.130195]$ for the integer value of $d=8$.
No more admissible pairs can be found.
In fact, considering here the two closest integer values ($d_1=21$ and $d_2=22$) to points in the second region.
Taking $d=21$, one can verify that the values of $p'$ that give a crossing probability laying in the confidence interval for $f_\textup{ex}$ are $p'\in[0.0846366,0.0850611]$ but the residence times associated to these pairs are in the interval $[2449.79,2450.06]$, out of the confidence interval for $r_\textup{ex}$.
In the same way, taking $d=22$, the candidate pairs associated to the confidence interval for $f_\textup{ex}$ are $p'\in[0.0817259,0.0821386]$ whose associated residence times lie in the interval 
$[2472.9,2473.15]$, out of the confidence interval for $r_\textup{ex}$.
Hence, we identify the defect to have $p'\in[0.129606,0.130195]$ and $d=8$.


\begin{figure}[ht!]
\begin{picture}(200,145)(0,0)
\put(0,0){
  \includegraphics[width=0.25\textwidth]{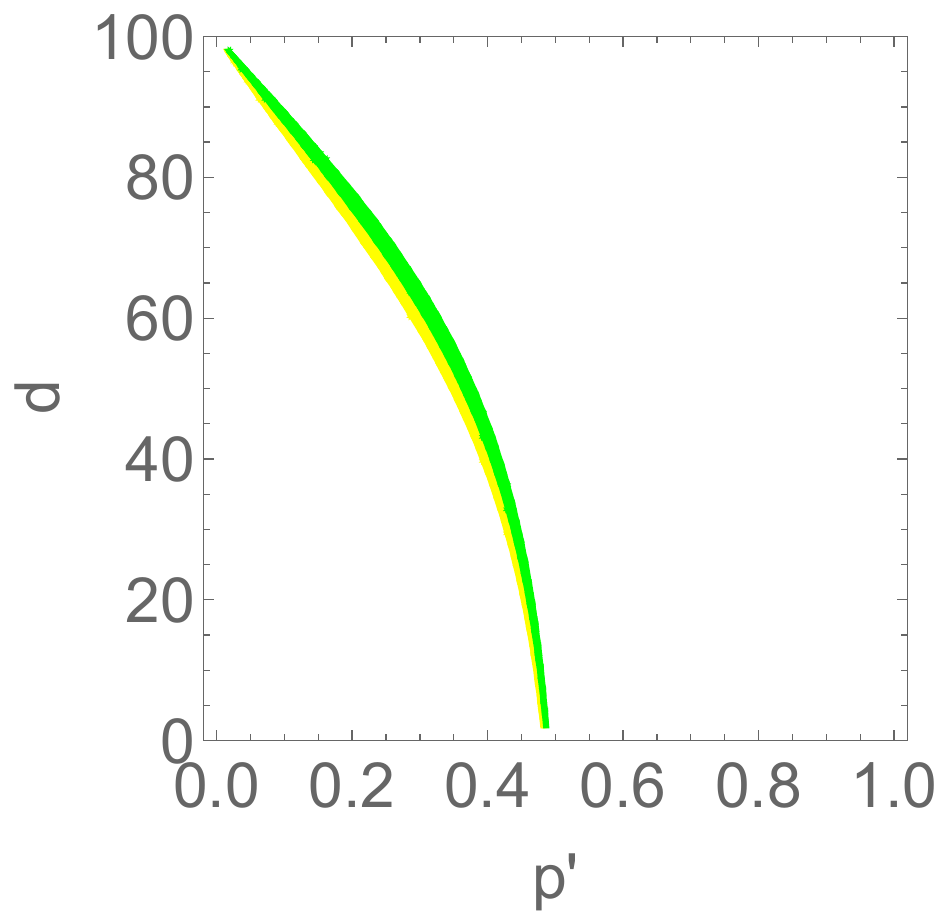}
}
\put(115,20){
  \includegraphics[width=0.11\textwidth]{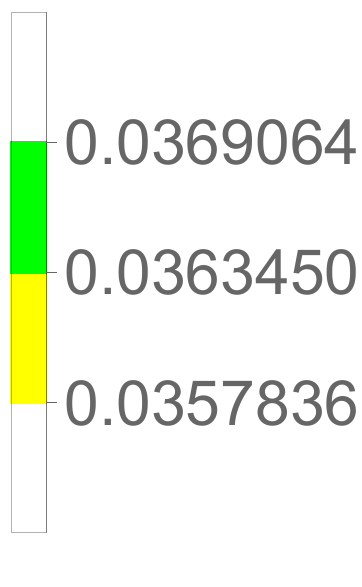}
}
\put(180,0){
  \includegraphics[width=0.25\textwidth]{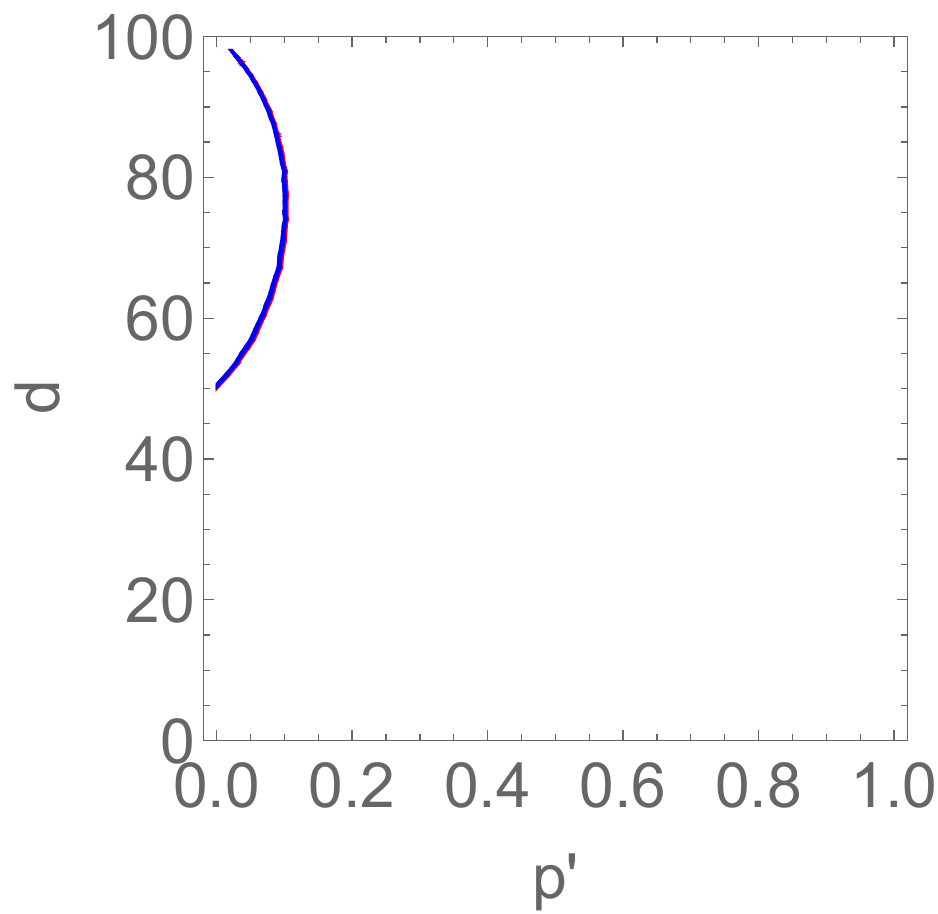}
}
\put(295,20){
  \includegraphics[width=0.085\textwidth]{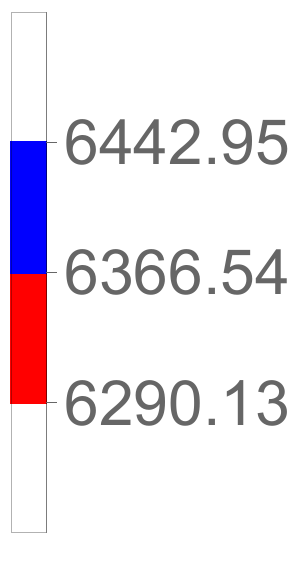}
}
\put(345,0){
  \includegraphics[width=0.25\textwidth]{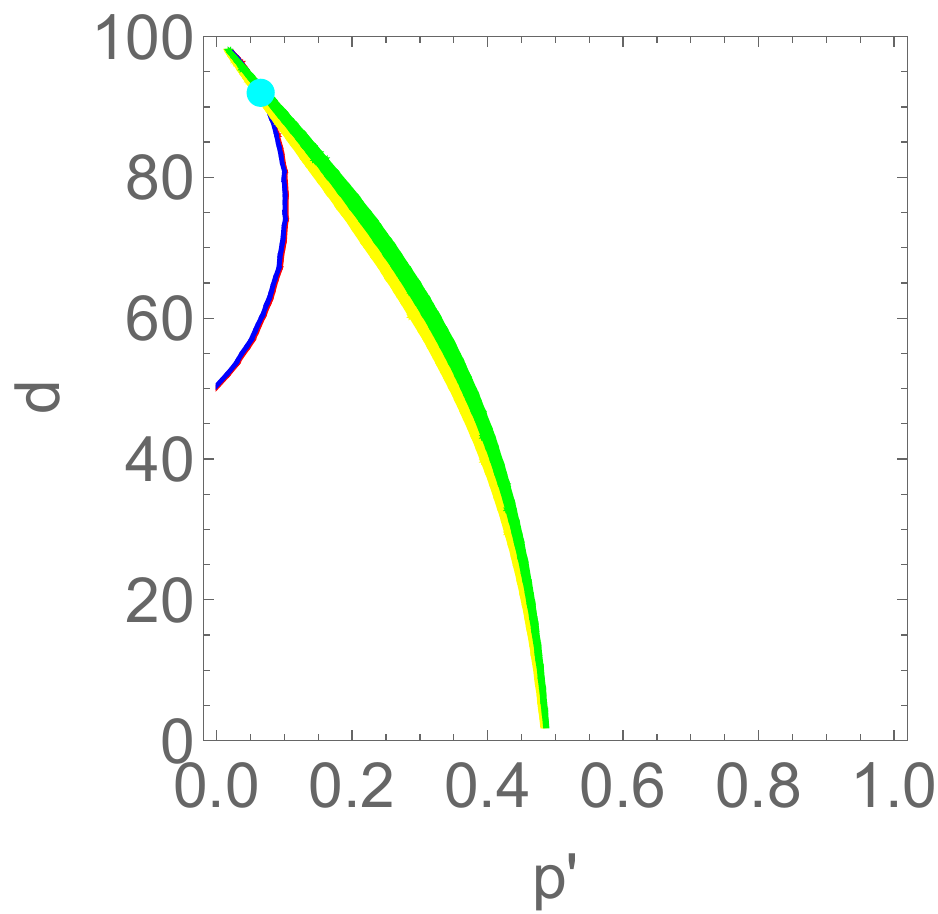}
}\end{picture}
\caption{Experiment~3.
Left, center and right panel as in Fig.~\ref{f:510}.
The cyan point identifies the defect.
}
\label{f:51-3}
\end{figure}

\begin{figure}[ht!]
\begin{picture}(200,145)(0,0)
\put(0,0){
  \includegraphics[width=0.25\textwidth]{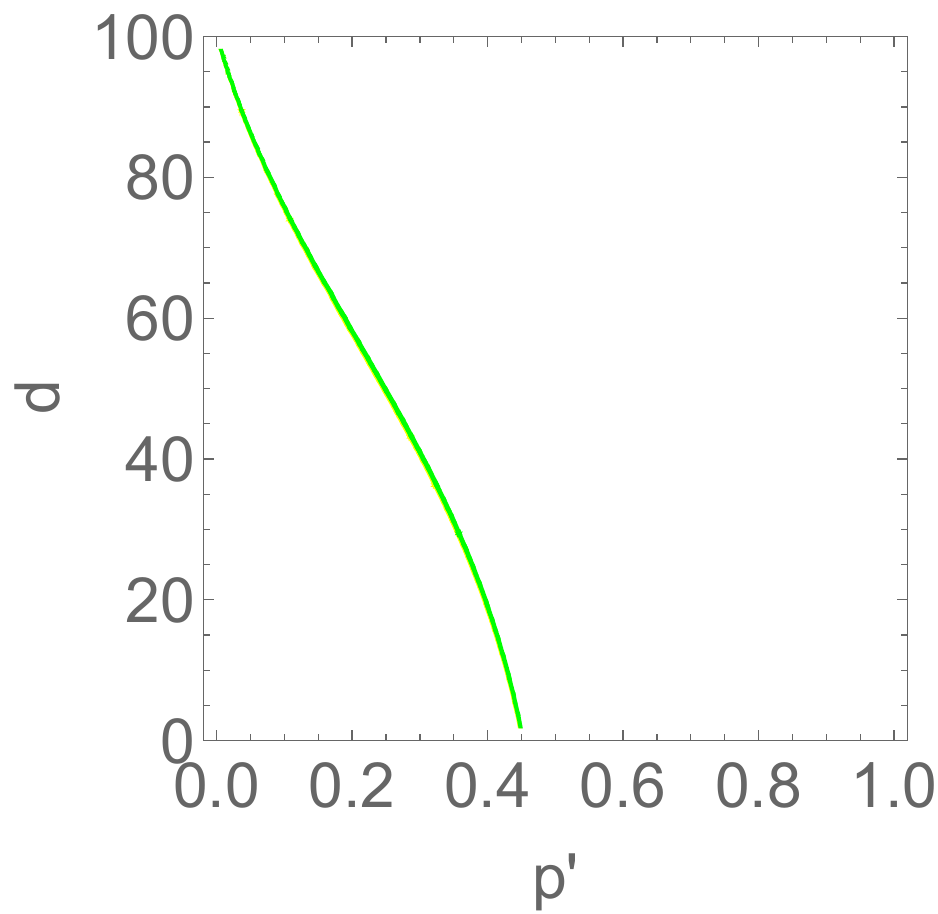}
}
\put(115,20){
  \includegraphics[width=0.11\textwidth]{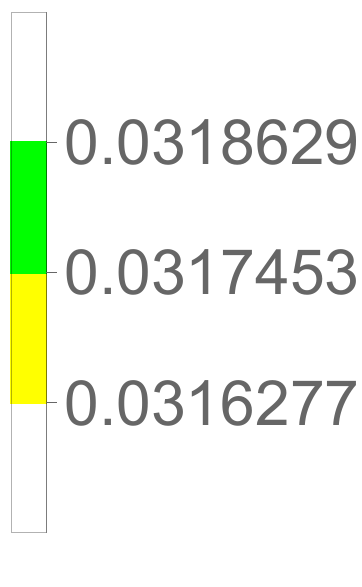}
}
\put(180,0){
  \includegraphics[width=0.25\textwidth]{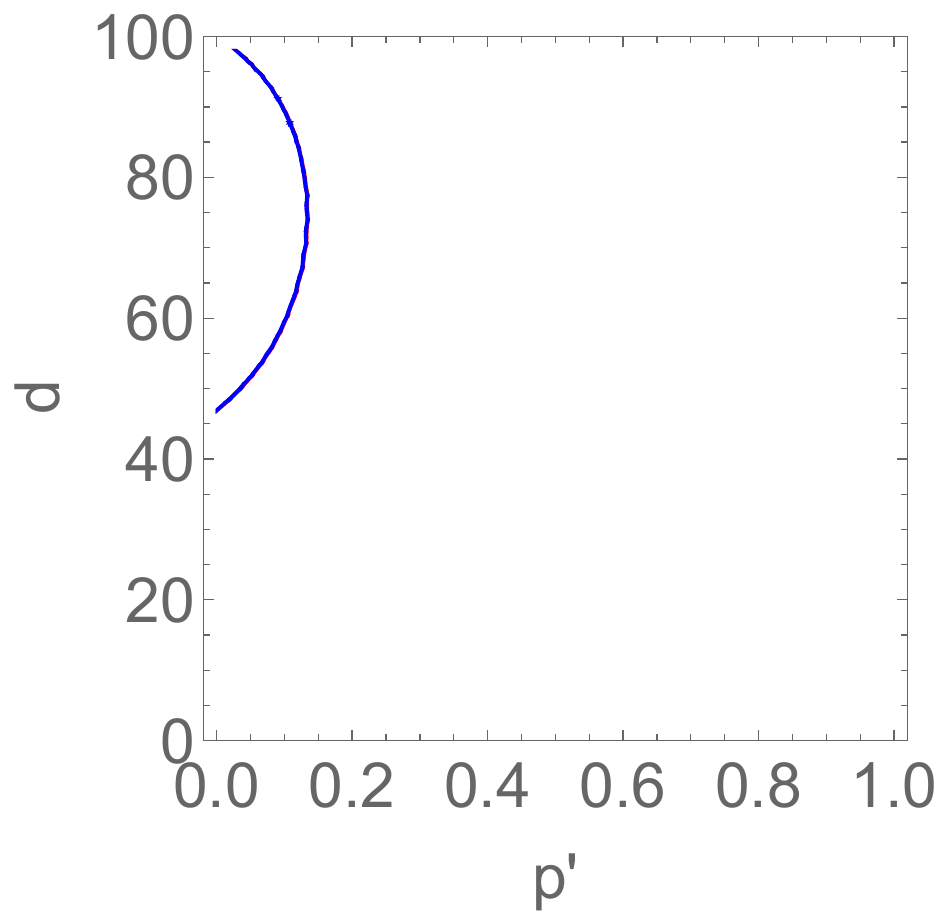}
}
\put(295,20){
  \includegraphics[width=0.085\textwidth]{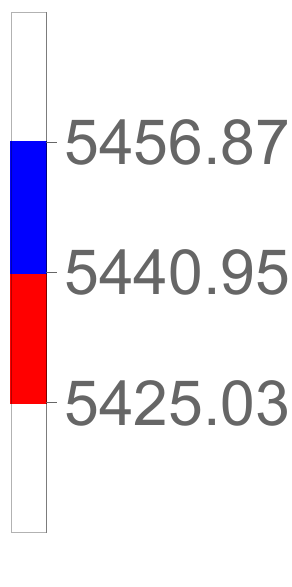}
}
\put(345,0){
  \includegraphics[width=0.25\textwidth]{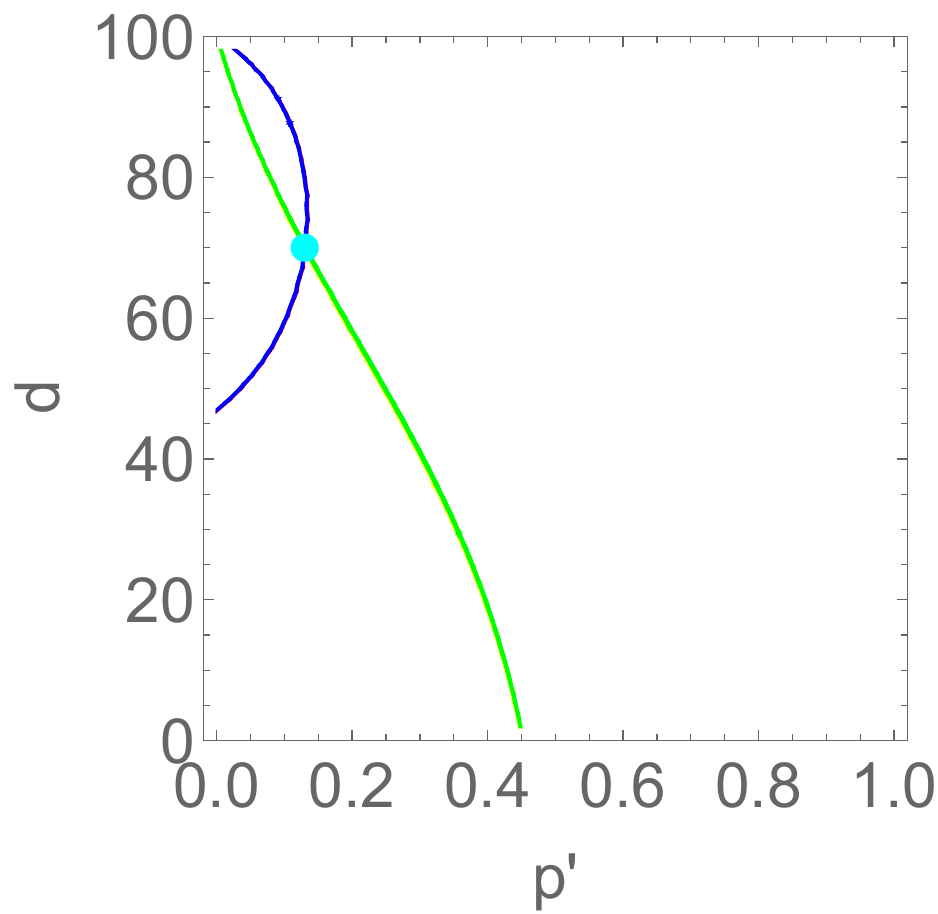}
}
\end{picture}
\caption{
Experiment~4.
See caption of Fig.~\ref{f:51-3}.
}
\label{f:51-4}
\end{figure}

In experiments $3$ and $4$, a defect with $p'<p$, close to the exit site, can be identified as shown in Figs.~\ref{f:51-3} and \ref{f:51-4}. The intersection produces a unique connected region whose dimension is related to the number of starting particles $N$. 
In experiment $3$, see Fig.~\ref{f:51-3}, 
the regions associated to the confidence interval for
$f_\textup{ex}$ and $r_\textup{ex}$
are shown together with their intersection.
The confidence intervals for $f_\textup{ex}$ and $r_\textup{ex}$
are, respectively,
$[0.0357836,0.0369064]$ and $[6290.13,6442.95]$.
The number of starting particles is $N=10^6$, that is sufficient to produce an intersection region that clearly identifies some peculiarities of the defect ($d$ is close to $L$, so the defect is close to the exit site, and $p'$ is small, so that the defect hinder the passage) but not to identify a single possible value of $d$.

In the case of experiment $4$, instead,
the confidence intervals for $f_\textup{ex}$ and $r_\textup{ex}$
are, respectively,
$[0.0316277,0.0318629]$ and $[5425.03,5456.87]$.
The large value of $N=2\cdot 10^7$ allow to individuate a smaller intersection region, so that the defect is completely identifiable in this case. The only pairs with integer $d$ laying in the intersection region are those with $p'\in[0.129184,0.130592]$ and $d=70$.

\begin{figure}[ht!]
\begin{picture}(200,145)(0,0)
\put(0,0){
  \includegraphics[width=0.25\textwidth]{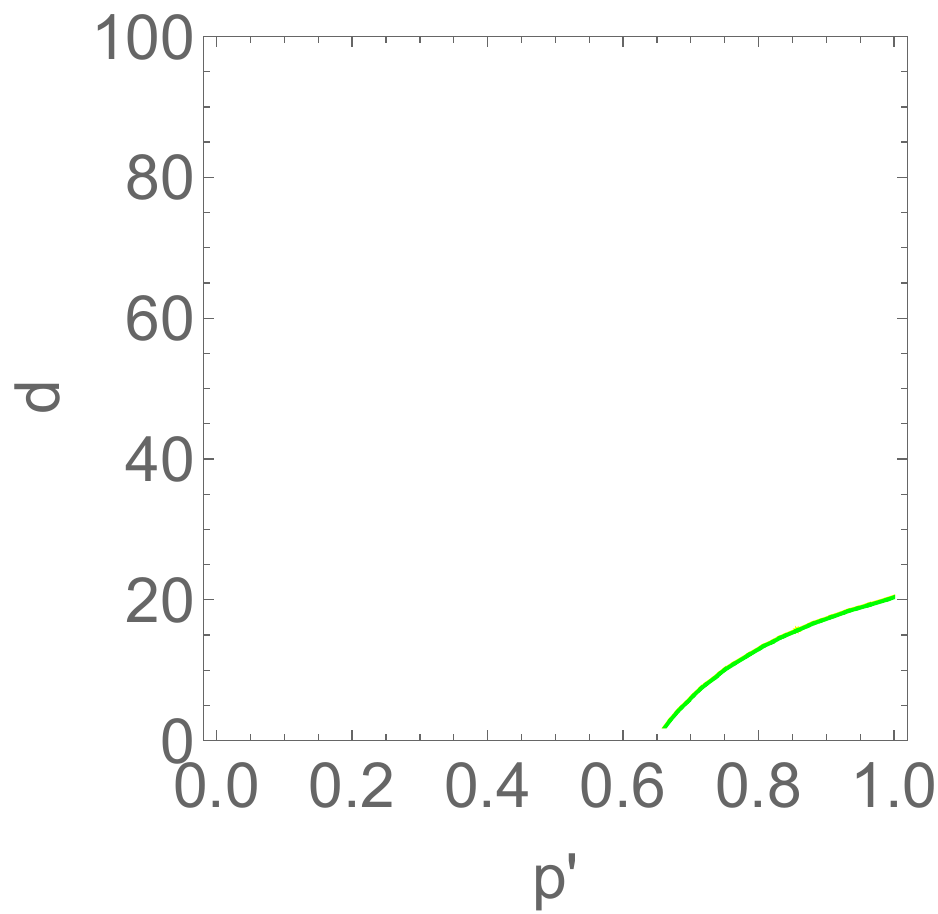}
}
\put(115,20){
  \includegraphics[width=0.11\textwidth]{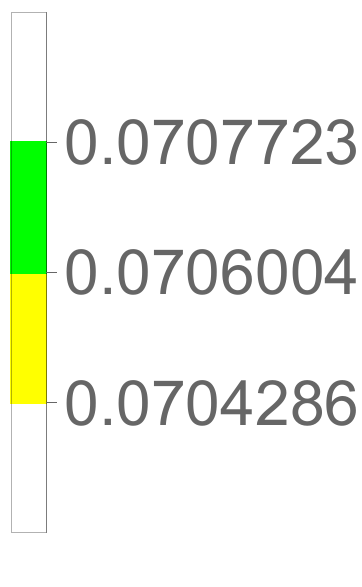}
}
\put(180,0){
  \includegraphics[width=0.25\textwidth]{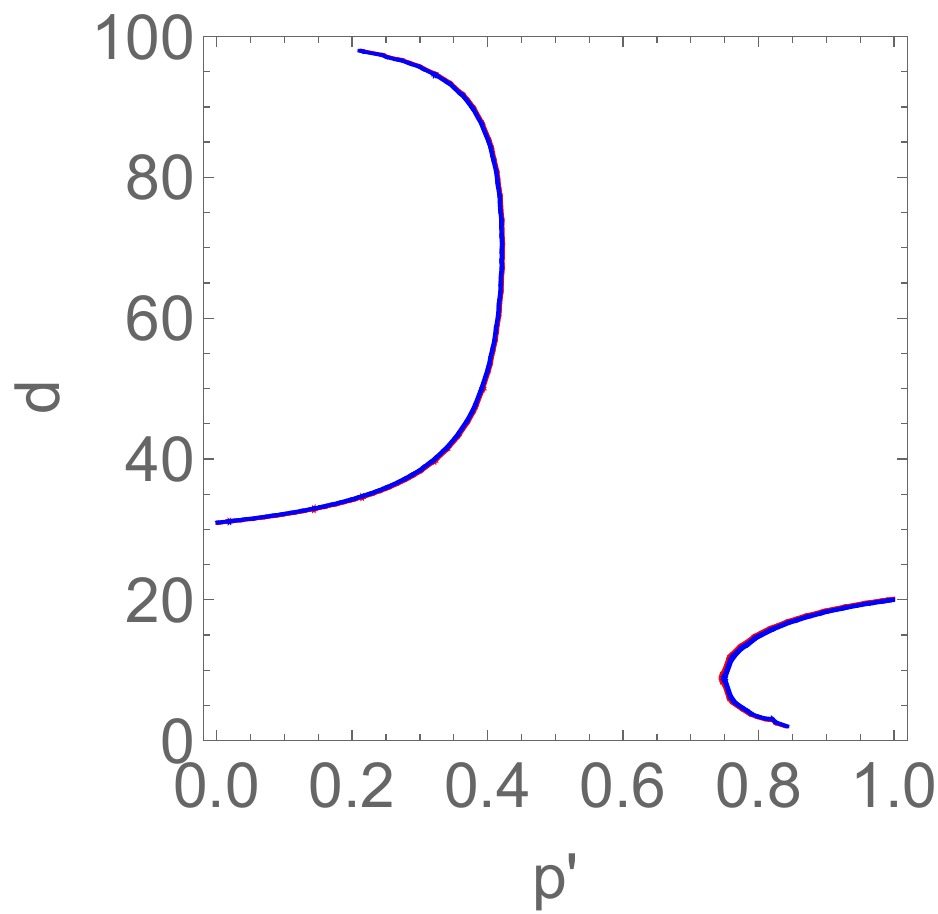}
}
\put(295,20){
  \includegraphics[width=0.10\textwidth]{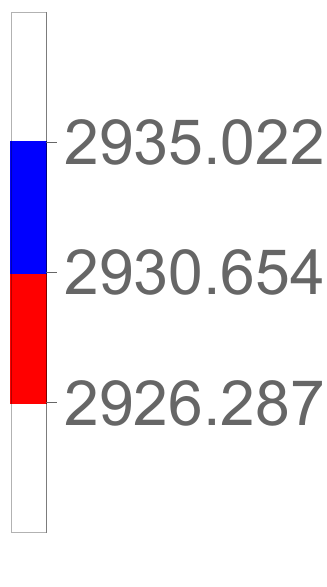}
}
\put(345,0){
  \includegraphics[width=0.25\textwidth]{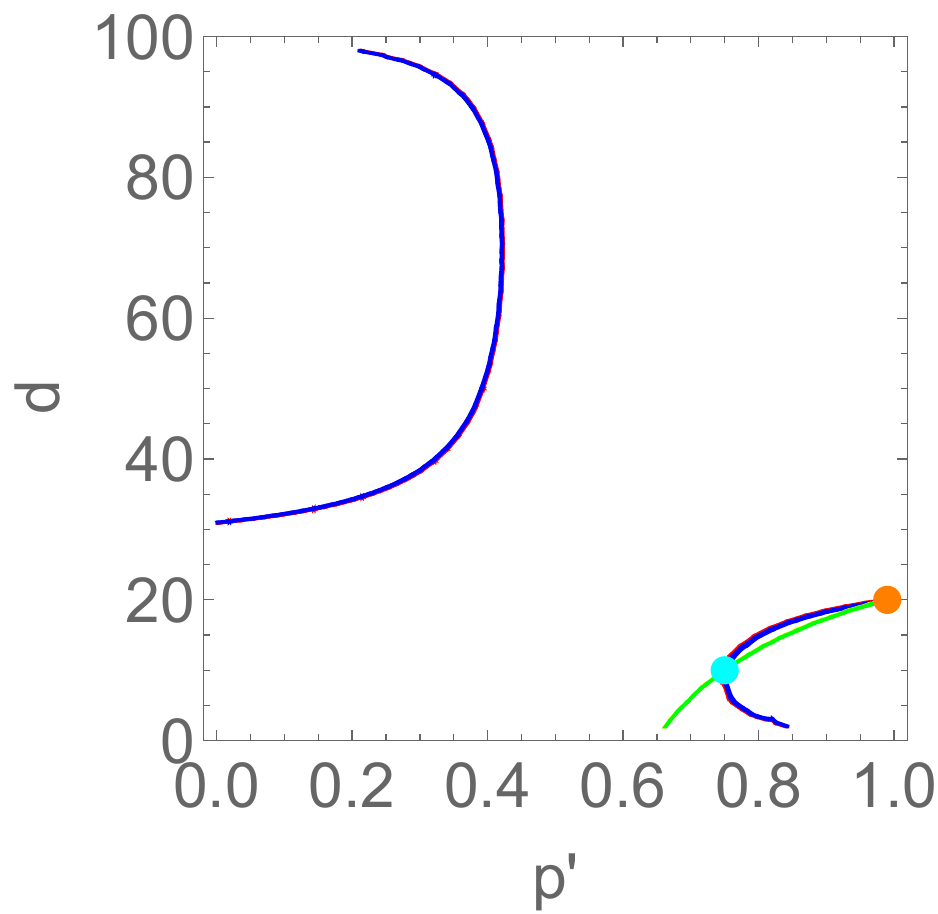}
}
\end{picture}
\caption{Experiment~5.
Left, center and right panel as in Fig.~\ref{f:510}.
Beside the real value of the obstacle parameters  identified by the cyan point, a second intersection	region around
$(0.99,20)$ is found (orange point).
}
\label{f:550}
\end{figure}

Defects that favor the passage ($p'>p$) and close to the entrance site $1$ can produce a lack of complete identifiability as shown in Fig.~\ref{f:550}.
There, experiment $5$ is depicted.
The intersections between the two
regions associated to the confidence interval for
$f_\textup{ex}$ and $r_\textup{ex}$
produce two connected regions.
The confidence intervals for $f_\textup{ex}$ and $r_\textup{ex}$
are, respectively,
$[0.0704286,0.0707723]$ and $[2926.287,2935.022]$.
 In both regions of intersection pairs with integer $d$ are present, so that a complete identification is not possible.
 The estimated values of the defect parameters are $p'\in[0.749253,0.751496]$ when $d=10$, or $p'\in[0.986593,0.992496]$ when $d=20$.


\begin{figure}[ht!]
\begin{picture}(200,145)(0,0)
\put(0,0)
{
  \includegraphics[width=0.25\textwidth]{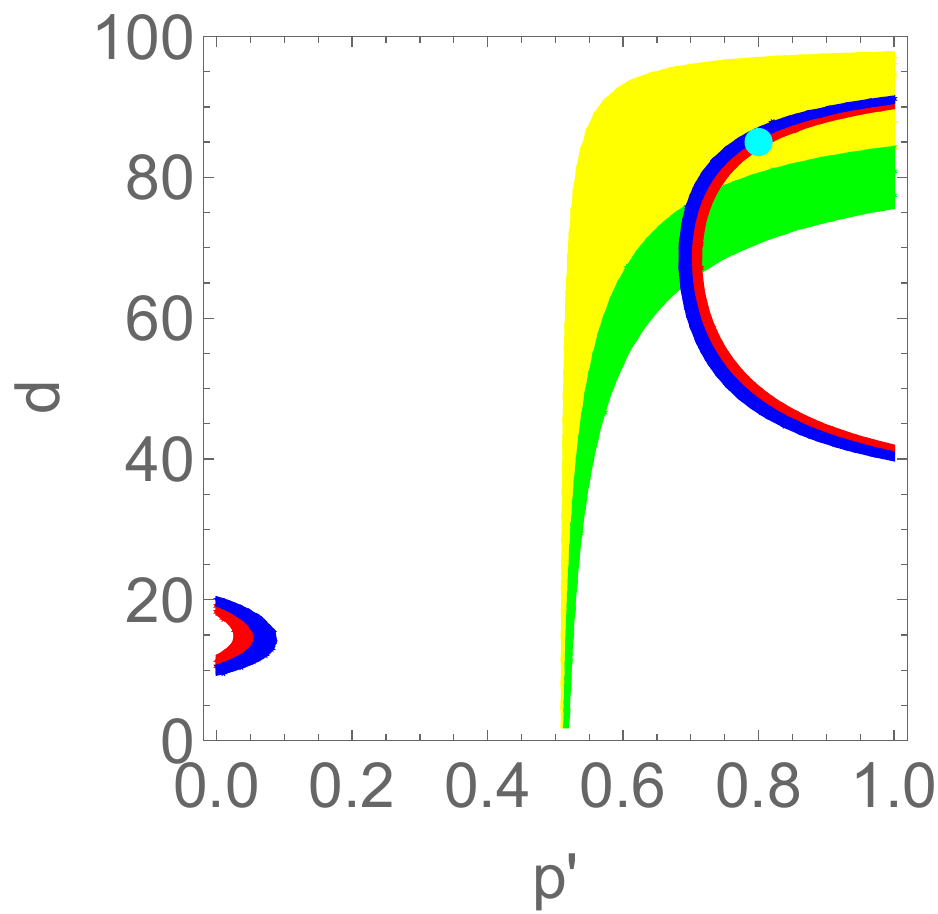}
}
\put(120,15)
{
  \includegraphics[width=0.06\textwidth]{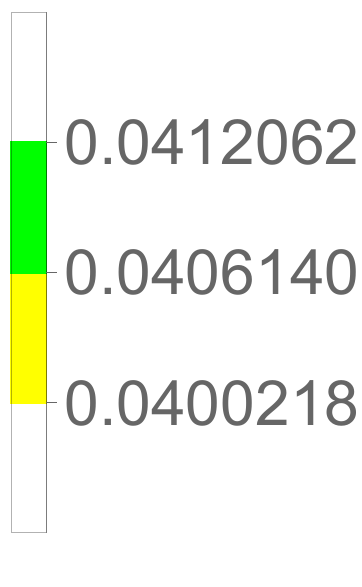}
}
\put(120,60)
{
  \includegraphics[width=0.06\textwidth]{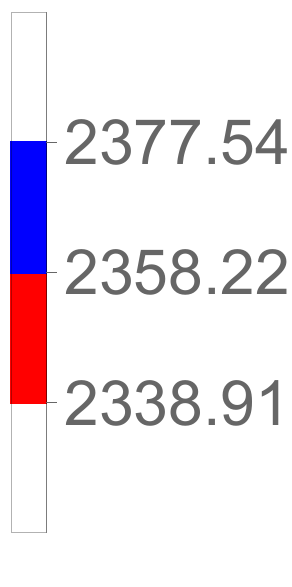}
}
\put(150,0){
  \includegraphics[width=0.25\textwidth]{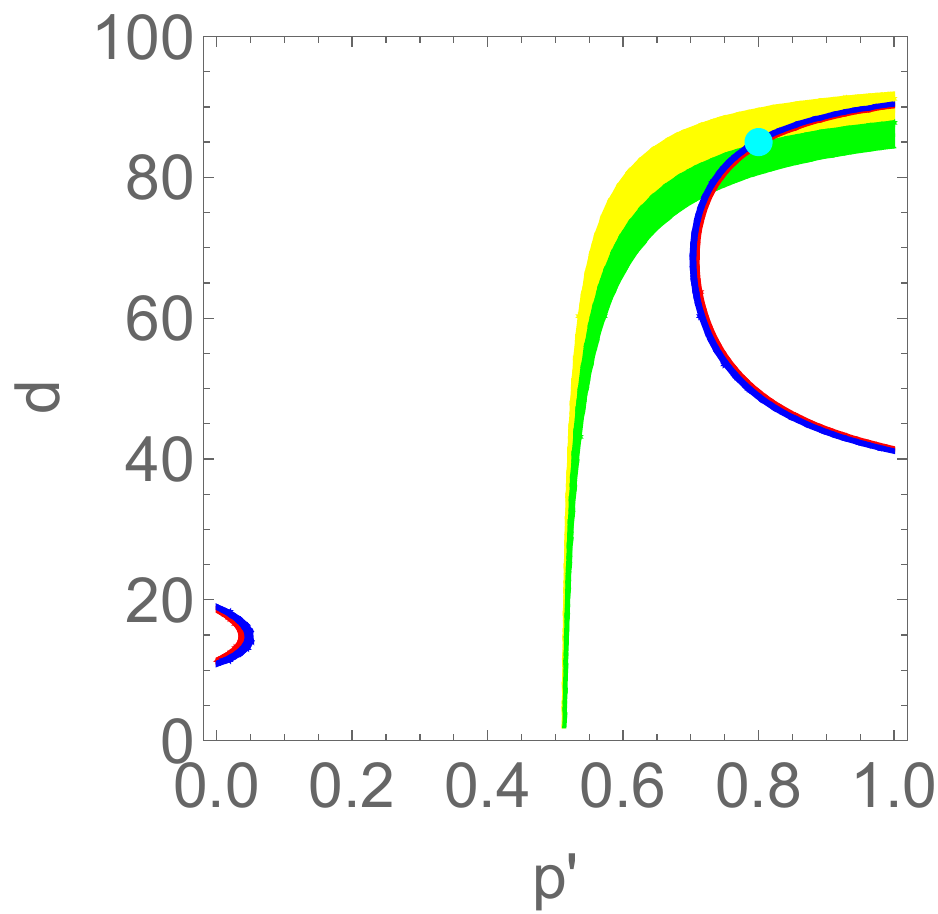}
}
\put(270,15){
  \includegraphics[width=0.06\textwidth]{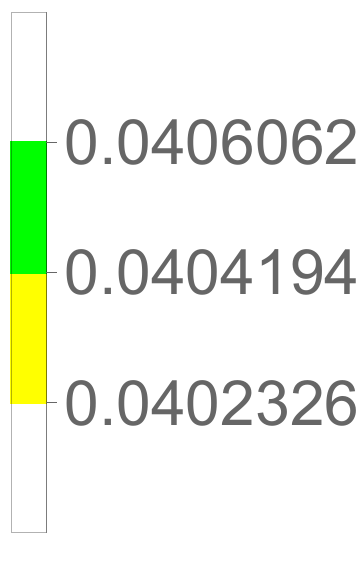}
}
\put(270,60){
  \includegraphics[width=0.06\textwidth]{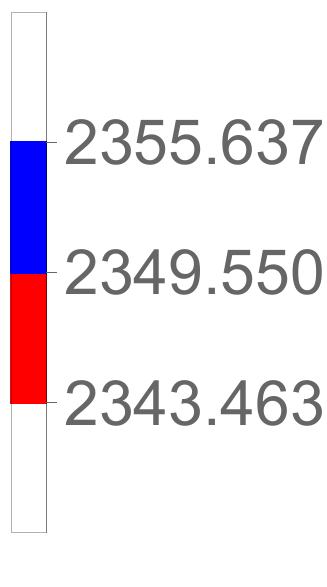}
}
\put(300,0){
  \includegraphics[width=0.25\textwidth]{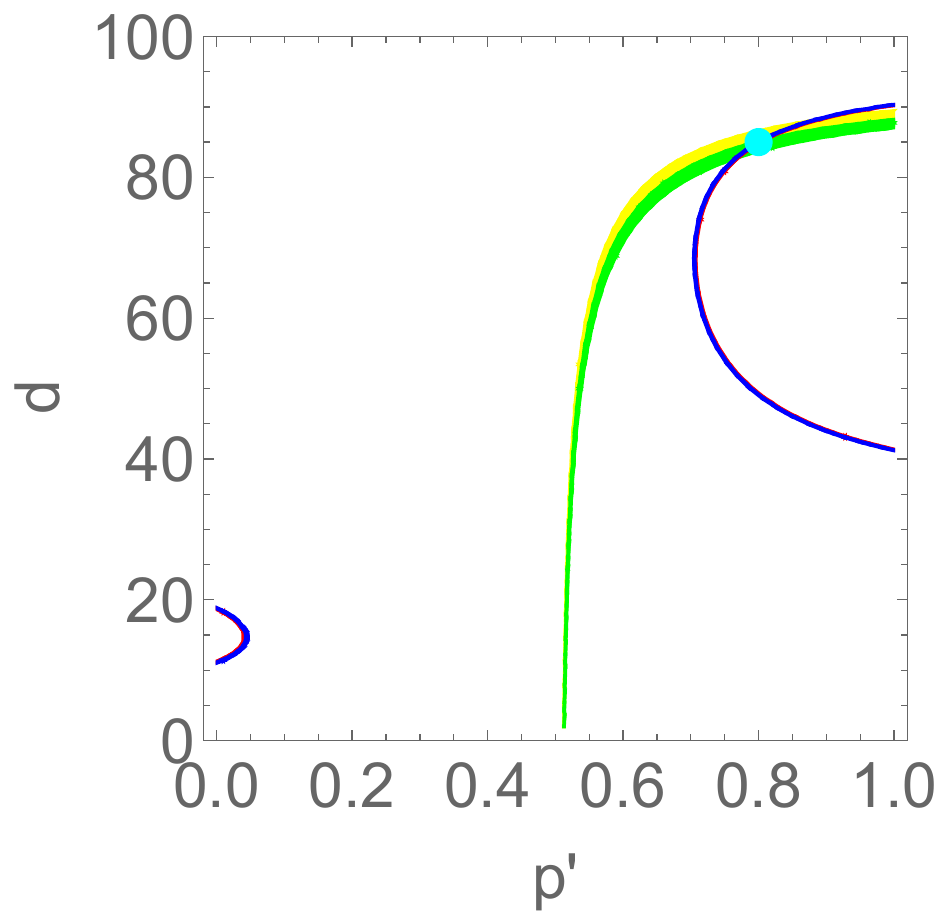}
}
\put(420,15){
  \includegraphics[width=0.06\textwidth]{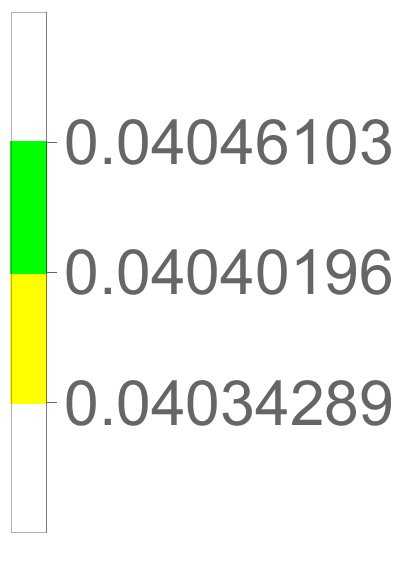}
}
\put(420,60){
  \includegraphics[width=0.06\textwidth]{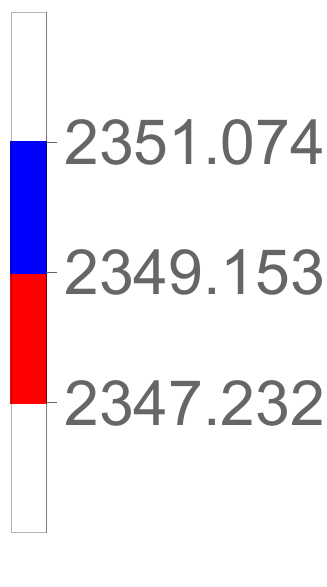}
}
\end{picture}
\caption{
From left to right: experiments 6 -- 7 -- 8 respectively. The cyan point identifies the actual defect, that is the same in the three experiments. The number of starting particles is different:  $N=10^6$ in experiment $6$, $N=10^7$ in experiment $7$, $N=10^8$ in experiment $8$.
}
\label{f:560}
\end{figure}

In Fig.~\ref{f:560} the results of experiments $6$, $7$ and $8$ are presented.
Here, the defect parameters are still the same in the three cases, but the different number of starting particles $N$ allows to find different intersection regions. In this case, it is interesting to show that even for a relatively small number of starting particles ($N=10^6$, experiment $6$) a unique connected region is found. Nevertheless, there are many different possible values of $d$.
When $N$ considered is larger, the intersection region dimension decreases, and the defect is identified to be in the last quarter of the lane, and to favor the passage of the particle ($p'>p$), but still a large number $N=10^8$ is not sufficient to identifies exactly the position of the defect. A further increasing of the number of starting particles $N$ is needed to enhance the estimate.

It is interesting to notice that in all the considered experiments the knowledge of the numbers of crossing particles $n$ and starting particles $N$ allows to deduce if $p'$ is larger or smaller than $p$, but on its own it does  not give any clue where the defect is, except for the experiment $5$.
On the other hand, one can note that the only knowledge of the residence time is generally not sufficient 
by itself to deduce if the defect 
is pushing forward or backward the walkers, since two unconnected regions with both possible behaviors are identified by the confidence interval found (see experiments $1$--$2$ and $5$--$8$).

The lack of identifiability tends to disappear when $p$ is larger, i.e. a single connected region is usually found as the intersection of the regions associated to the confidence interval for
$f_\textup{ex}$ and $r_\textup{ex}$.
However, it is common that the intersection identifies a unique, connected but large region, so that many different values of $d$ are possible. 
This happens 
due to the fact that in presence of a consistent drift ($p$ far enough from  $1/2$) a single defect do not modify very consistently the probability to cross the interval if it is far from the entrance site $1$ or if $p'$ is very close to $p$. 
Hence, this phenomenon was more rare in the case of $p=0.51$ previously considered since the drift was sufficiently small, so that the region associated to the confidence interval of $f_\textup{ex}$ was usually quite small.
 The cases in which the defect is close to site $1$ while $p'$ is not close to $p$ are instead the cases in which it is easier to fully identify the defect.
Note that in principle, increasing sufficiently the number of starting particle $N$, it is possible to let the intersection region become as small as wanted, but the required $N$ can be very large.

However, the larger the drift is, the more is common to deduce some information about the defect from the knowledge of the residence time alone.
For instance, in many of the following experiments we can notice that the residence time allows to individuate if $p'$ is smaller or larger than $p$.
 Indeed, we recall that as shown in Fig.~\ref{f:505} the scatter plot of the residence time tends to reduce,  when $p$ move away from $1/2$, to a first region where $p'<p$ and a second region ($p'>p$), one of them in which  the residence time is larger than the case of no defect, and the opposite in the other one.

To clarify what we said we present now some experiments in which the different situations are encountered.
Let us now consider the case of $p=0.53$.

\begin{figure}[ht!]
\begin{picture}(200,145)(0,0)
\put(0,0){
  \includegraphics[width=0.25\textwidth]{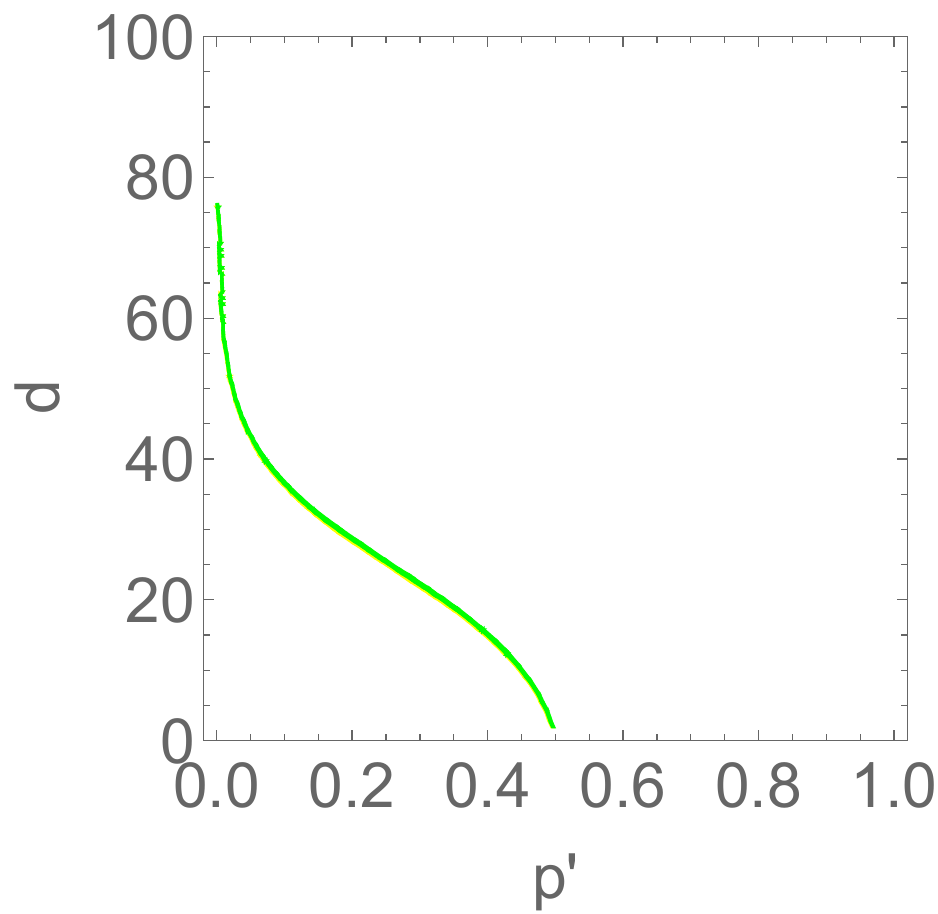}
}
\put(115,20){
  \includegraphics[width=0.11\textwidth]{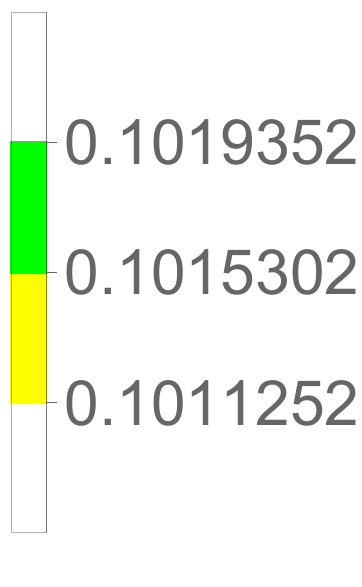}
}
\put(180,0){
  \includegraphics[width=0.25\textwidth]{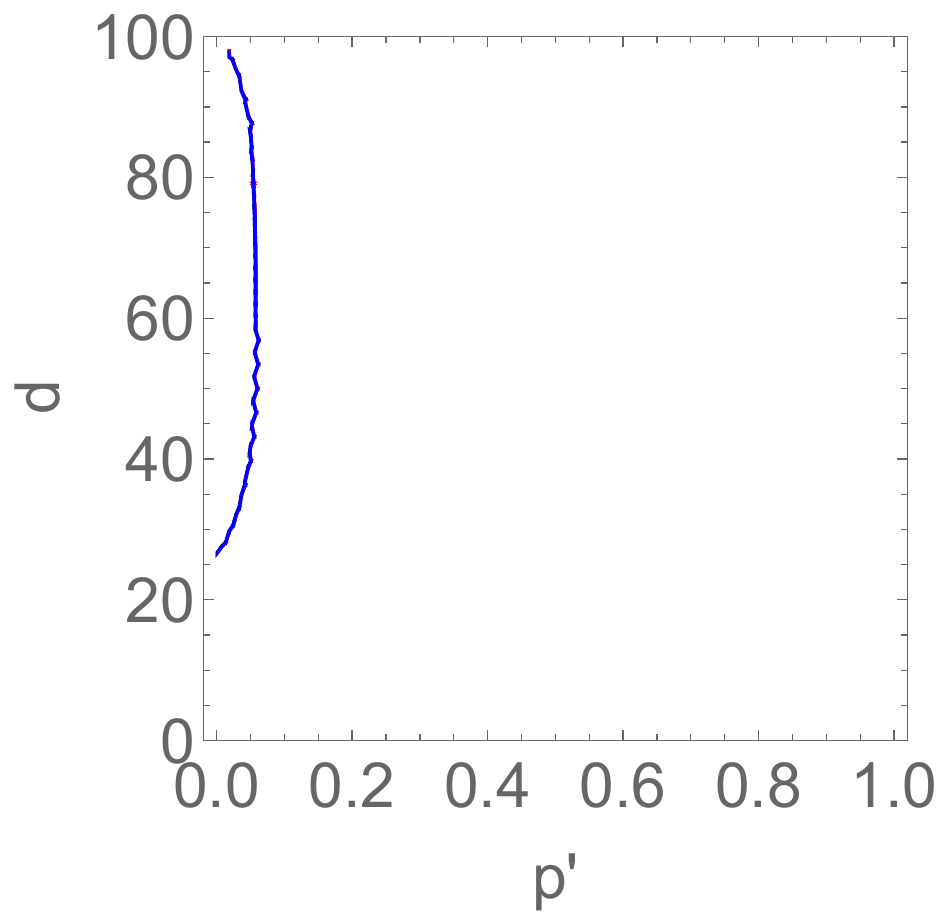}
}
\put(295,20){
  \includegraphics[width=0.10\textwidth]{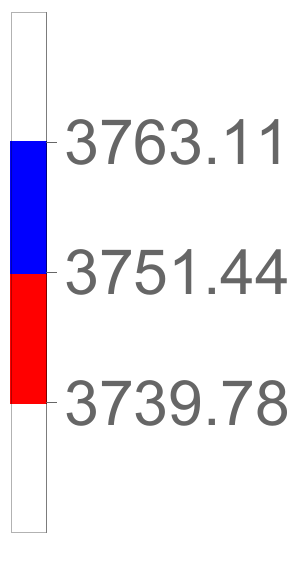}
}
\put(345,0){
  \includegraphics[width=0.25\textwidth]{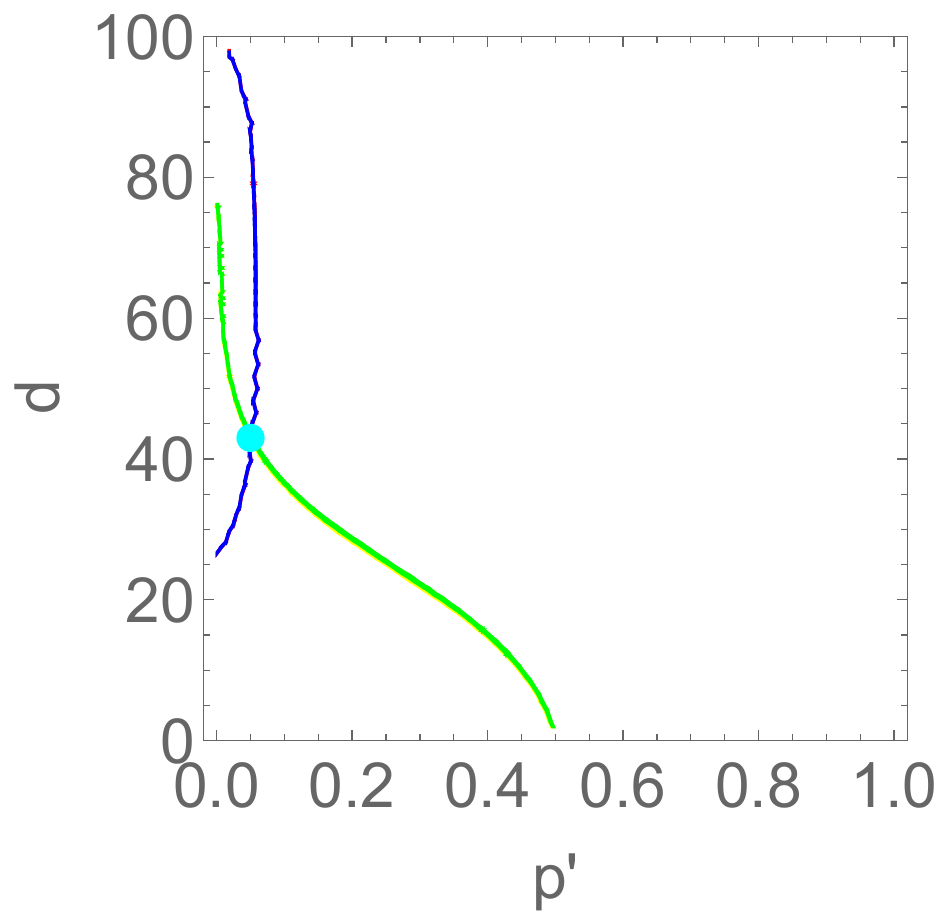}
}
\end{picture}
\caption{Experiment 9. See caption of Fig.~\ref{f:51-3}.
}
\label{f:53-9}
\end{figure}

The experiment $9$ is shown in Fig.~\ref{f:53-9}. The actual defect is in the first half of the interval with a value of $p'$ that is very small.
We are now in a case in which it is possible to fully identify the defect.
The confidence intervals for $f_\textup{ex}$ and $r_\textup{ex}$
are, respectively,
$[0.1011252,0.1019352]$ and $[3739.78,3763.11]$.
The defect is correctly identify to have $p'\in[0.049815,0.050315]$ and $d=43$, that are the pairs $(p',d)$ with integer $d$ that lay in the intersection region.

\begin{figure}[ht!]
\begin{picture}(200,145)(0,0)
\put(0,0){
  \includegraphics[width=0.25\textwidth]{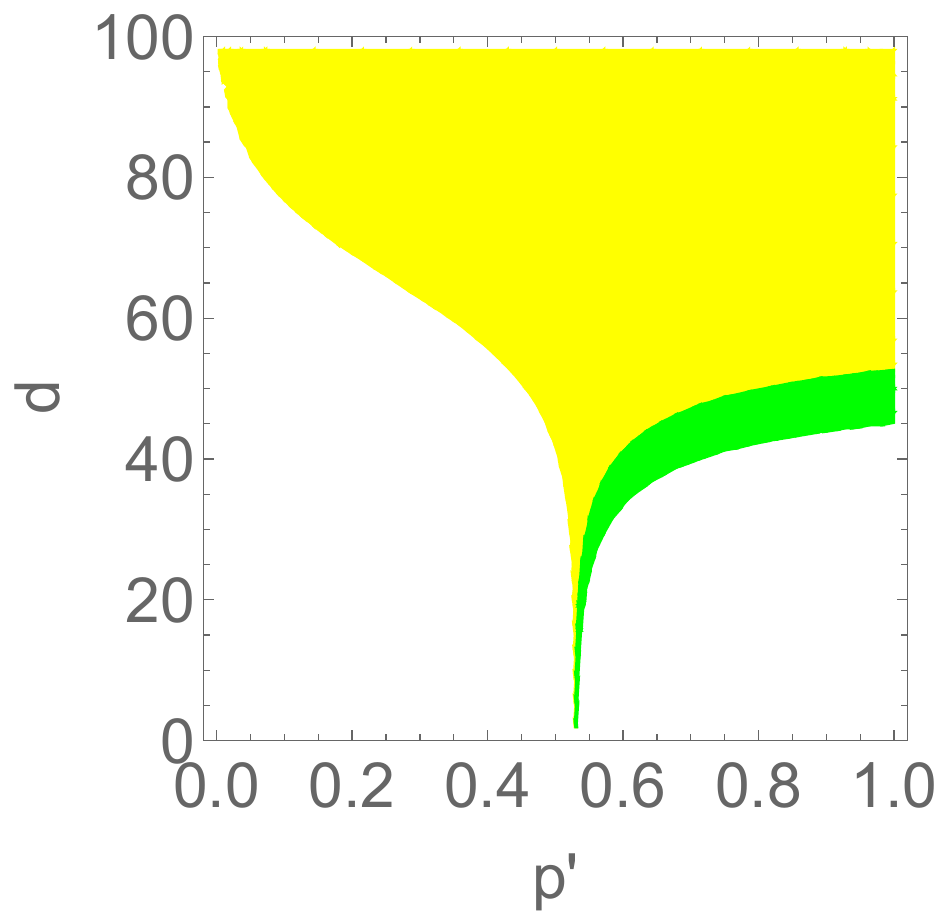}
}
\put(115,20){
  \includegraphics[width=0.11\textwidth]{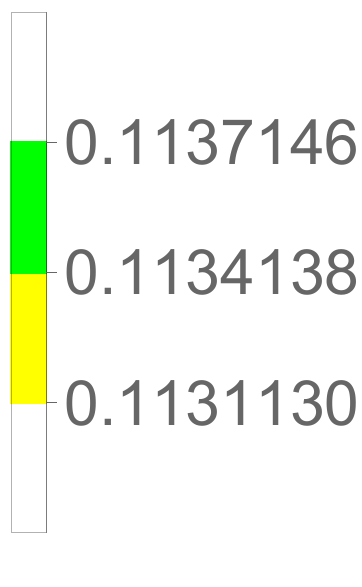}
}
\put(180,0){
  \includegraphics[width=0.25\textwidth]{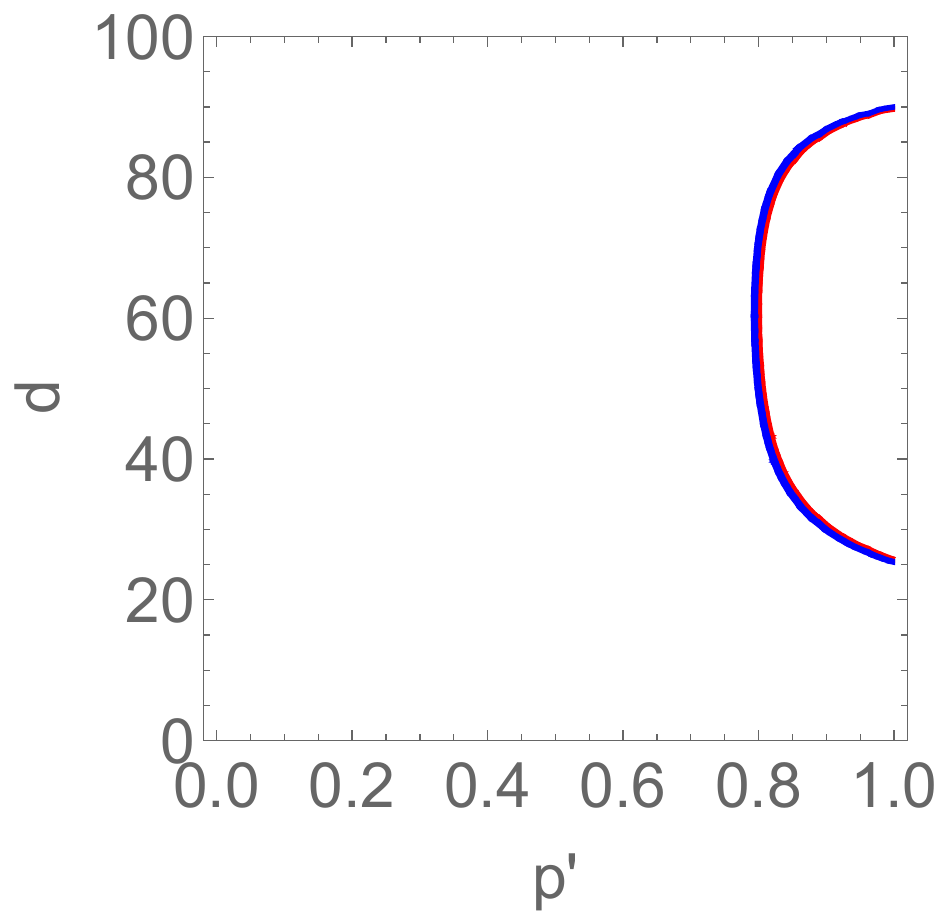}
}
\put(295,20){
  \includegraphics[width=0.10\textwidth]{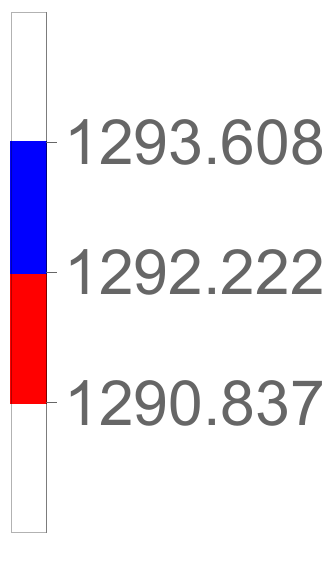}
}
\put(345,0){
  \includegraphics[width=0.25\textwidth]{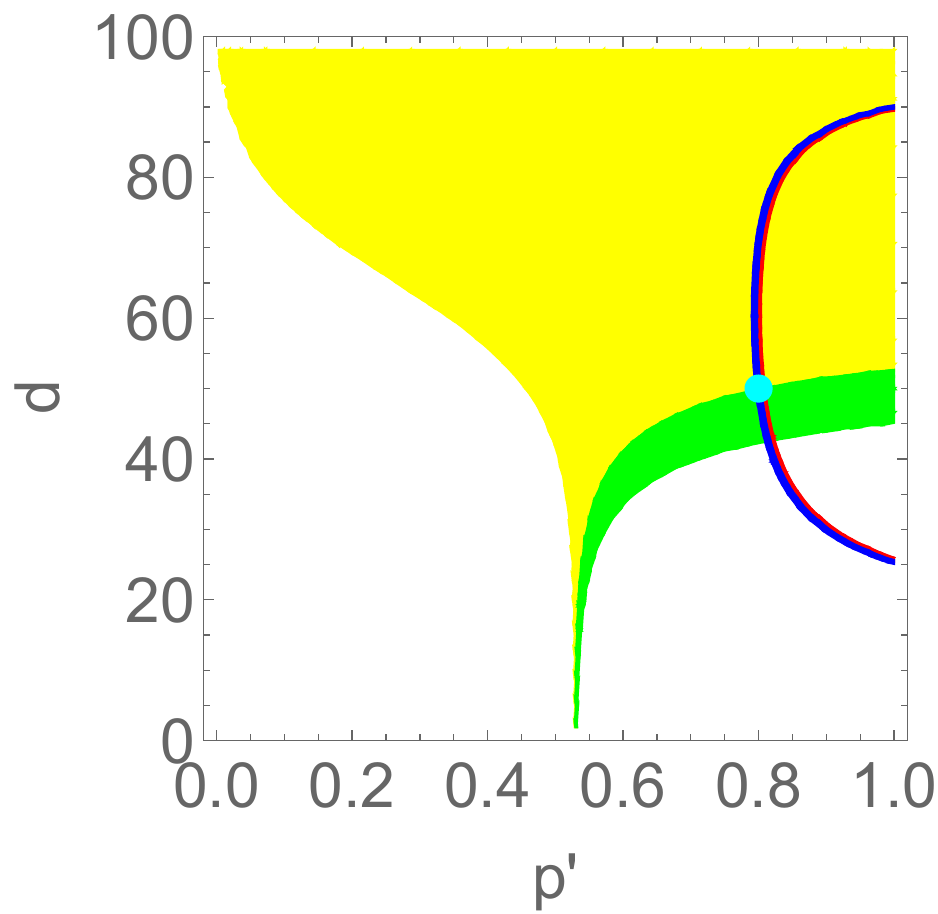}
}
\end{picture}
\caption{
Experiment $10$.
}
\label{f:53-10bis}
\end{figure}

In Fig.~\ref{f:53-10bis} we plot the results of experiment $10$.
Here, as in the case of experiments $11$--$13$, the complete identification is made difficult by the fact that the defect is far from the entrance site (the actual defect is in the middle of the lane), and the parameter $p'$ is not sufficiently high or low ($p'=0.8$ in the experiment).
In this case the information that one can easily get is that $p'>0.75$ from the residence time, and that $d>40$. Any refinement of this information would require a consistently larger number of starting particles $N$.

\begin{figure}[ht!]
\begin{picture}(200,145)(0,0)
\put(0,0){
  \includegraphics[width=0.25\textwidth]{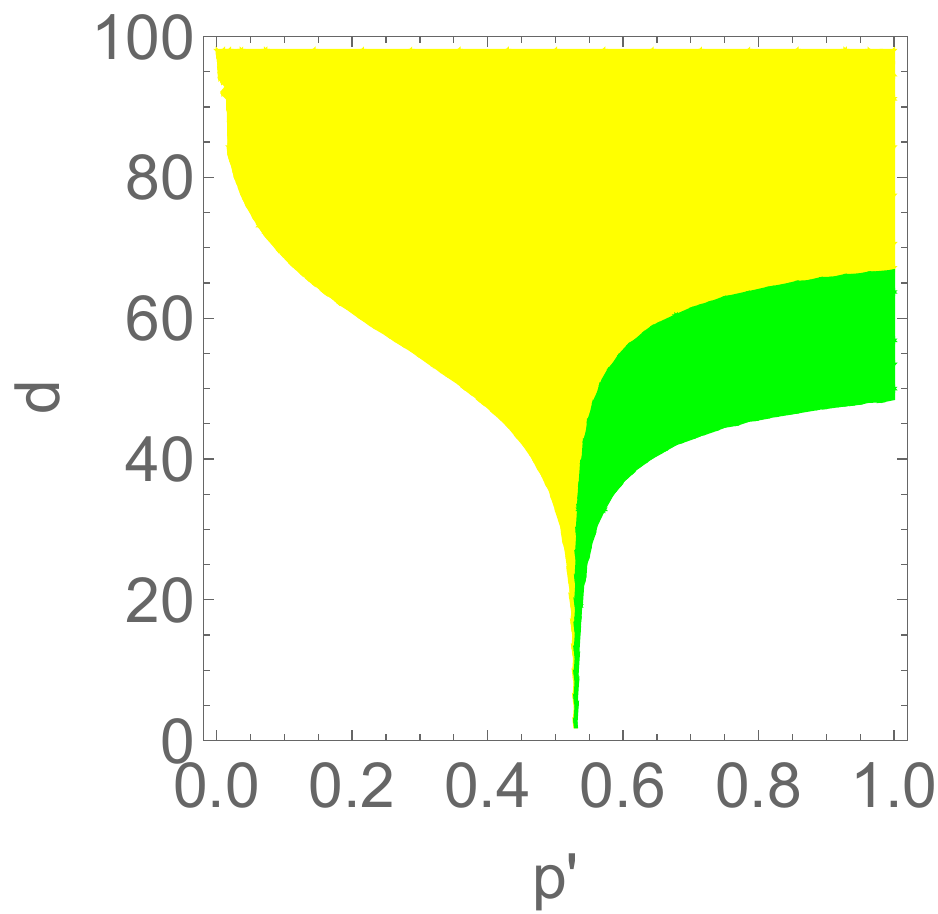}
}
\put(115,20){
  \includegraphics[width=0.11\textwidth]{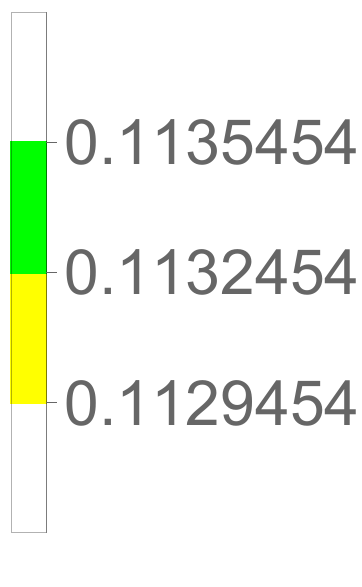}
}
\put(180,0){
  \includegraphics[width=0.25\textwidth]{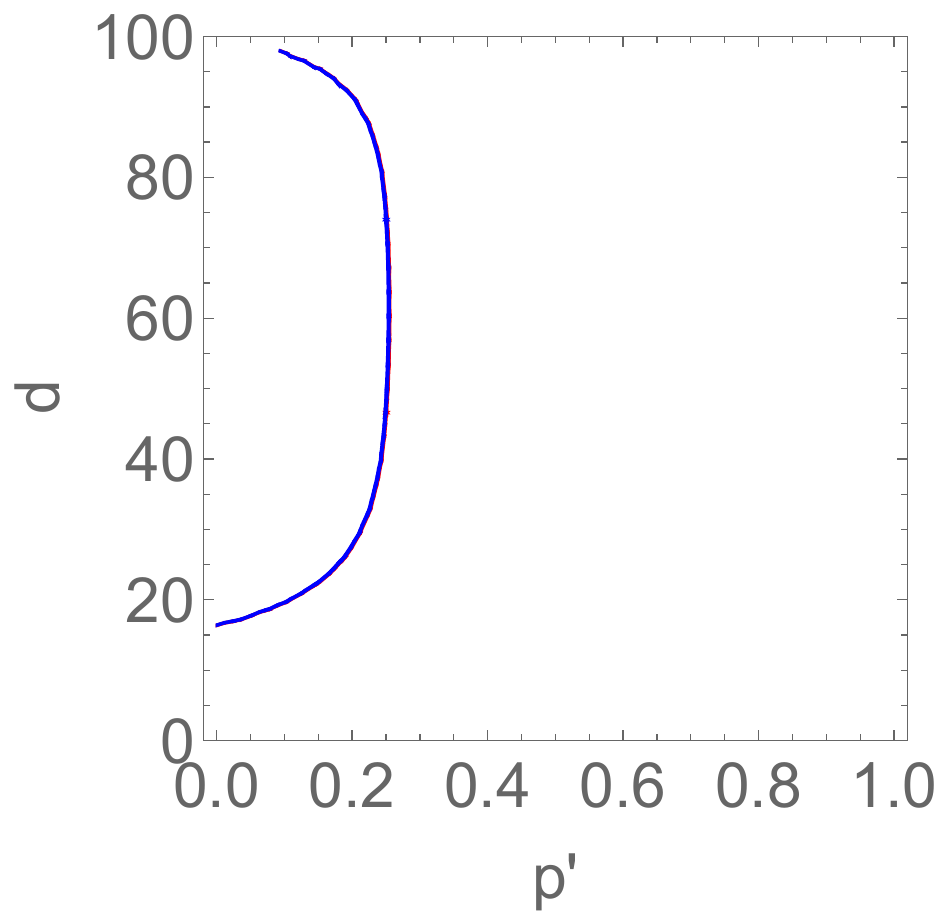}
}
\put(295,20){
  \includegraphics[width=0.10\textwidth]{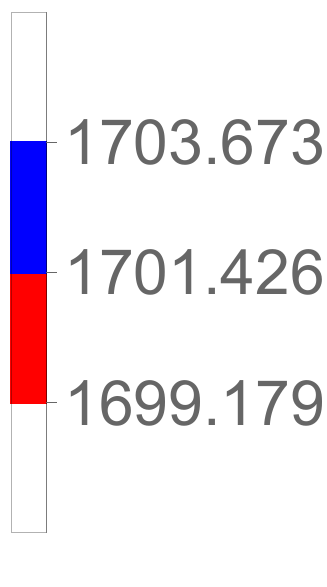}
}
\put(345,0){
  \includegraphics[width=0.25\textwidth]{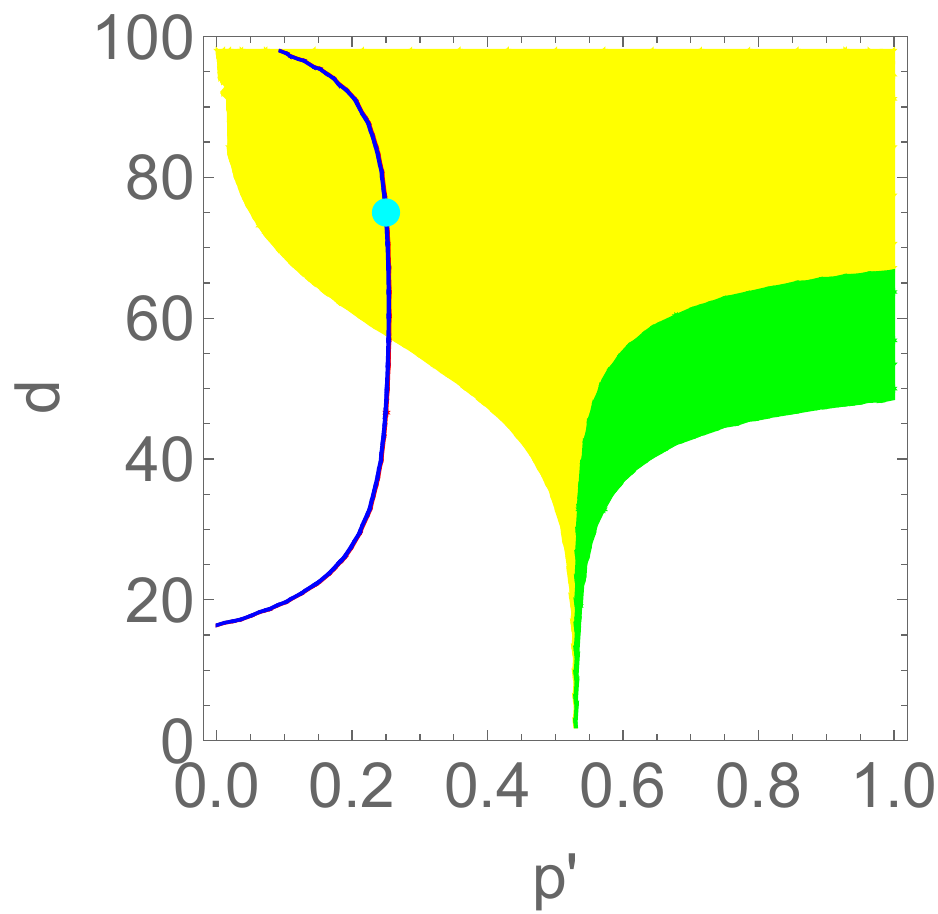}
}
\end{picture}
\caption{Experiment 11. See caption of Fig.~\ref{f:51-3}.
}
\label{f:53-11bis}
\end{figure}

\begin{figure}[ht!]
\begin{picture}(200,145)(0,0)
\put(0,0)
{
  \includegraphics[width=0.25\textwidth]{ccv-53-11bisc.pdf}
}
\put(120,15)
{
  \includegraphics[width=0.06\textwidth]{ccv-53-11bisaa.pdf}
}
\put(120,60)
{
  \includegraphics[width=0.06\textwidth]{ccv-53-11bisbb.pdf}
}
\put(150,0){
  \includegraphics[width=0.25\textwidth]{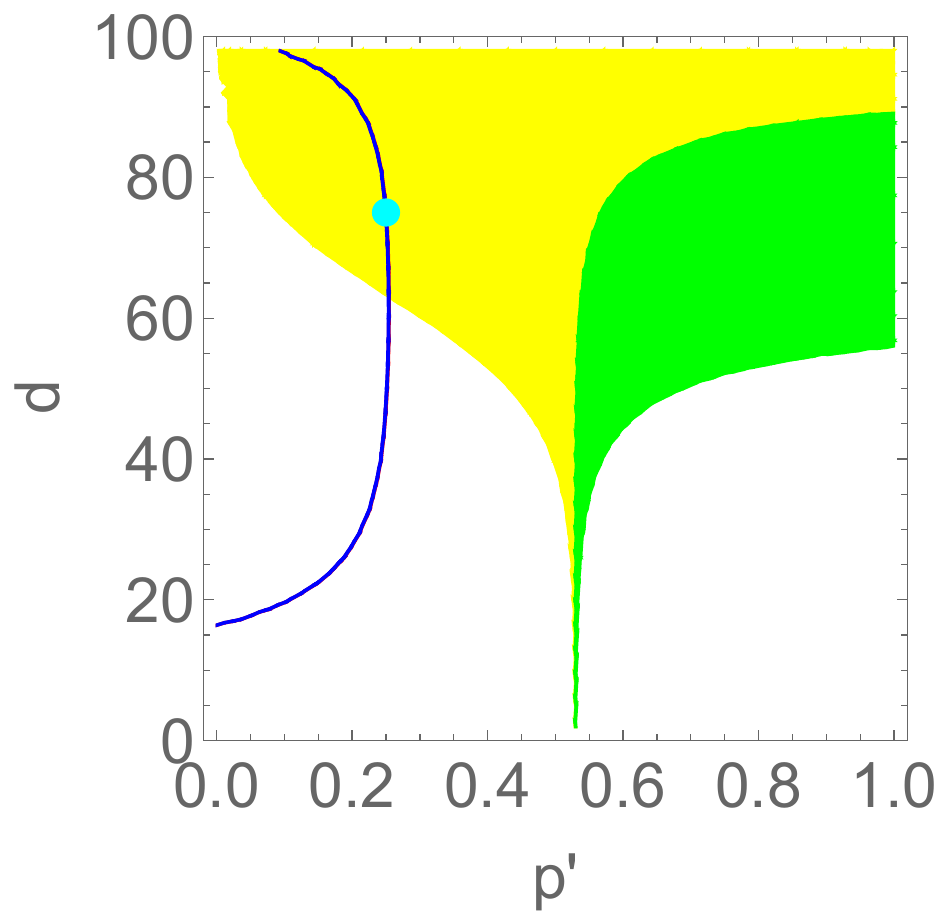}
}
\put(270,15){
  \includegraphics[width=0.06\textwidth]{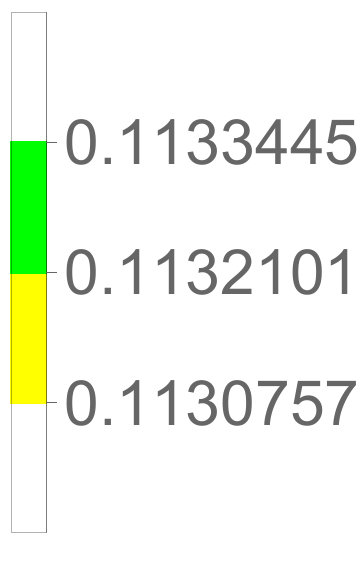}
}
\put(270,60){
  \includegraphics[width=0.06\textwidth]{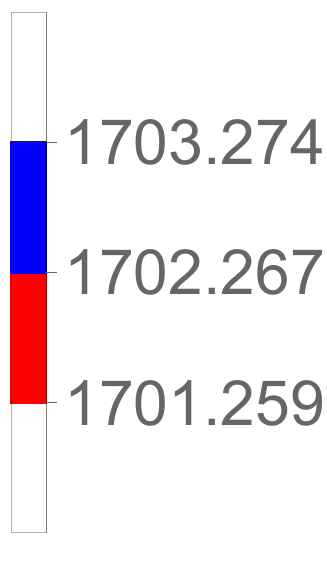}
}
\put(300,0){
  \includegraphics[width=0.25\textwidth]{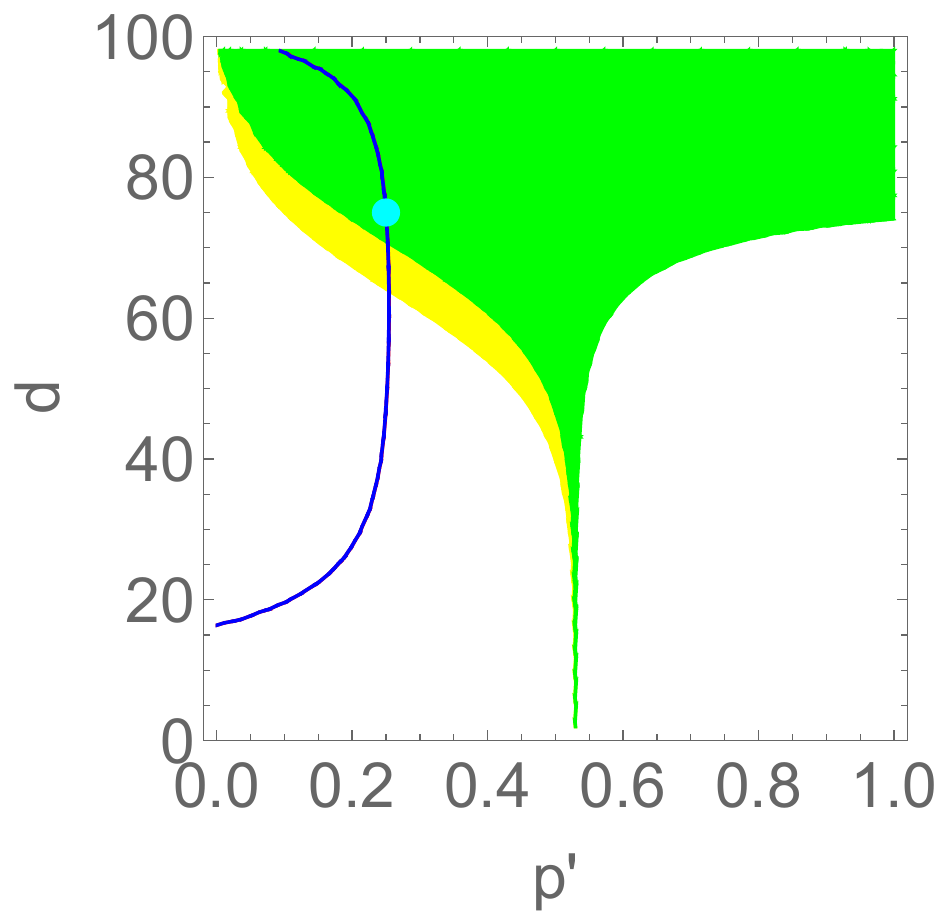}
}
\put(420,15){
  \includegraphics[width=0.06\textwidth]{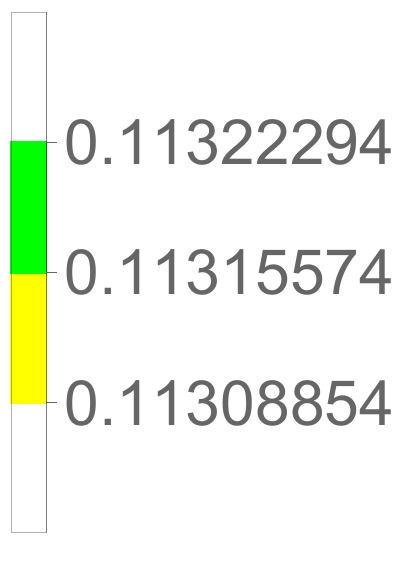}
}
\put(420,60){
  \includegraphics[width=0.06\textwidth]{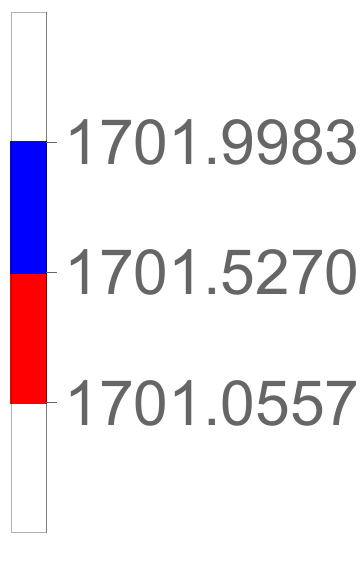}
}
\end{picture}
\caption{
From left to right: experiments $11$ -- $12$ -- $13$ respectively. The cyan point identifies the actual defect, that is the same in the three experiments. The number of starting particles is different:  $N=10^7$ in experiment $11$, $N=5\cdot 10^7$ in experiment $12$, $N=2\cdot10^8$ in experiment $13$.
}
\label{f:53-11-13bis}
\end{figure}

In Figs.~\ref{f:53-11bis}--\ref{f:53-11-13bis} we show that if the defect is not so influential on the probability to cross, the full identification of the defect will require very large $N$. 
The results of experiments $11$, $12$ and $13$ are shown in these plots.
The three experiments refer to the case of the same defect, with a different number of starting particles $N$.
It is evident that in this case the increasing of $N$ produces a very slow shrinkage of the intersection region.
Nevertheless, it soon appears clear that the defect is in the second half of the interval, and $p'$ is smaller than $p$.

\begin{figure}[ht!]
\begin{picture}(200,145)(0,0)
\put(0,0){
  \includegraphics[width=0.25\textwidth]{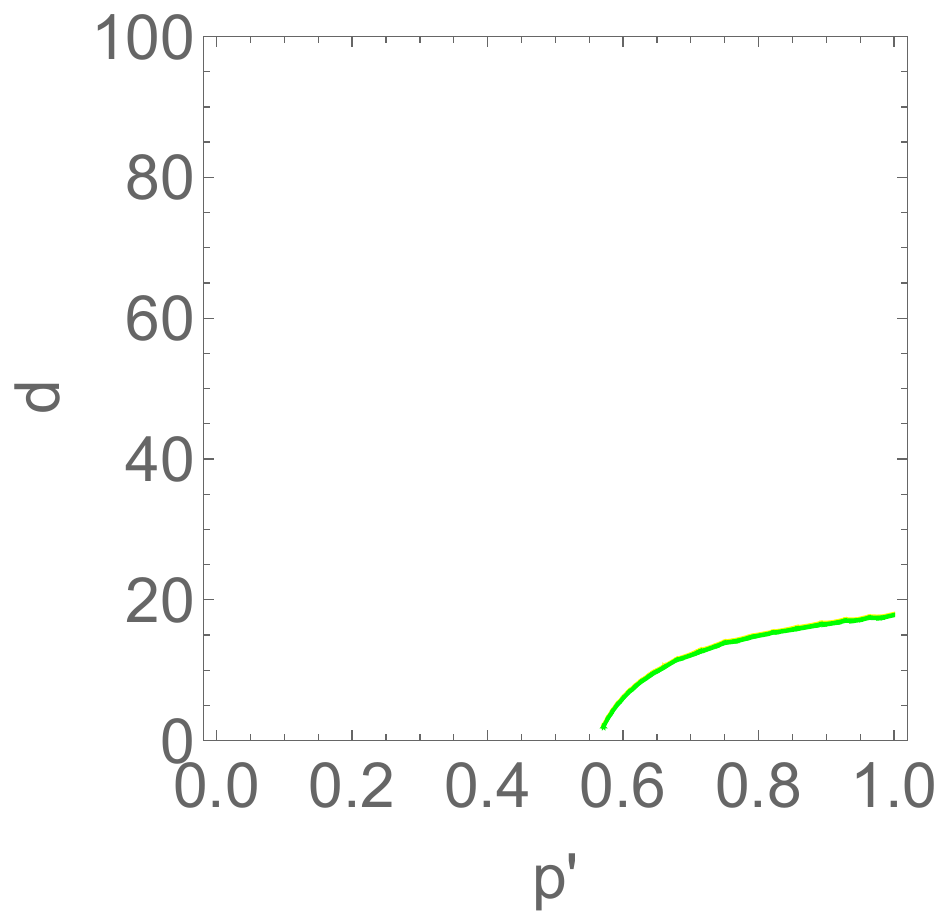}
}
\put(115,20){
  \includegraphics[width=0.11\textwidth]{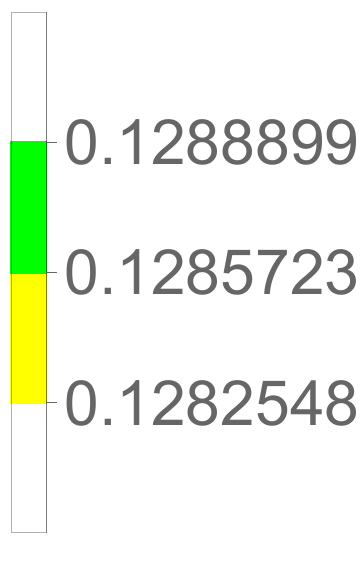}
}
\put(180,0){
  \includegraphics[width=0.25\textwidth]{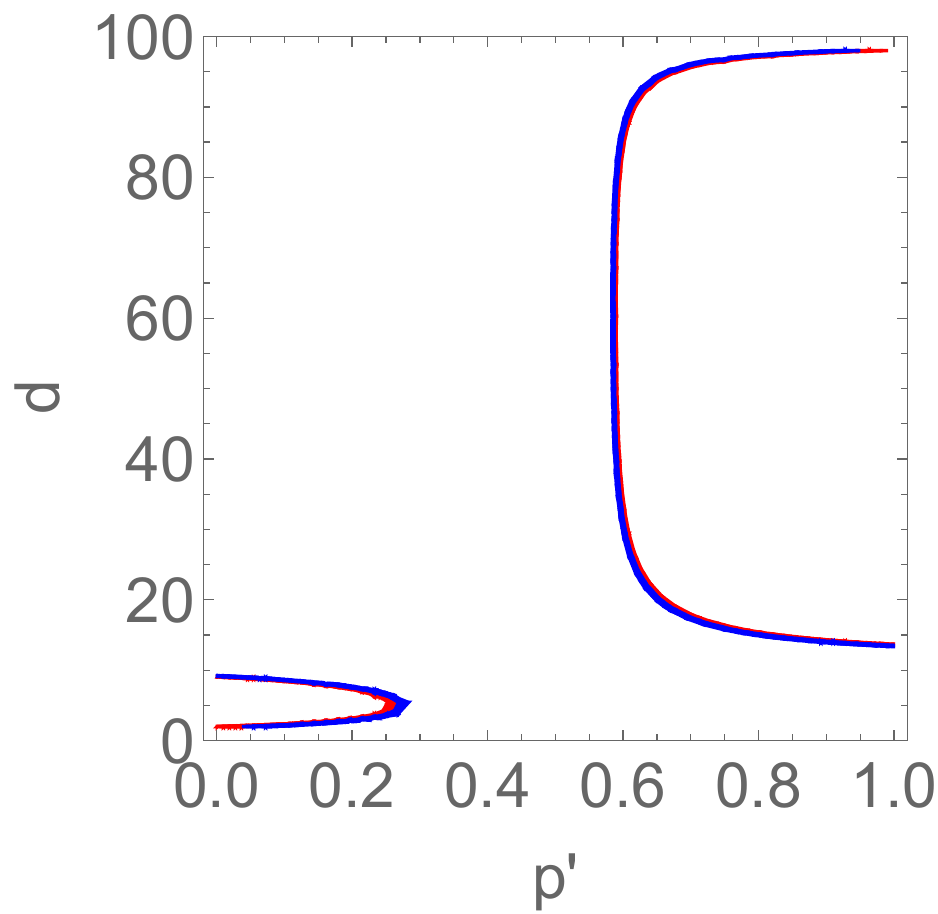}
}
\put(295,20){
  \includegraphics[width=0.10\textwidth]{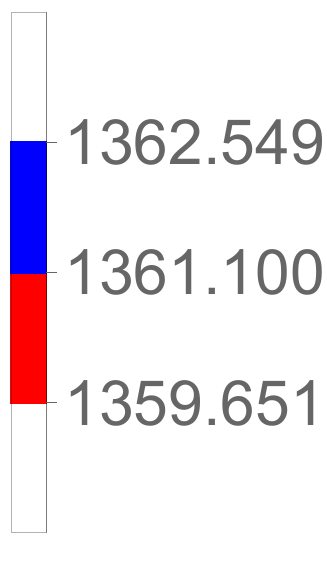}
}
\put(345,0){
  \includegraphics[width=0.25\textwidth]{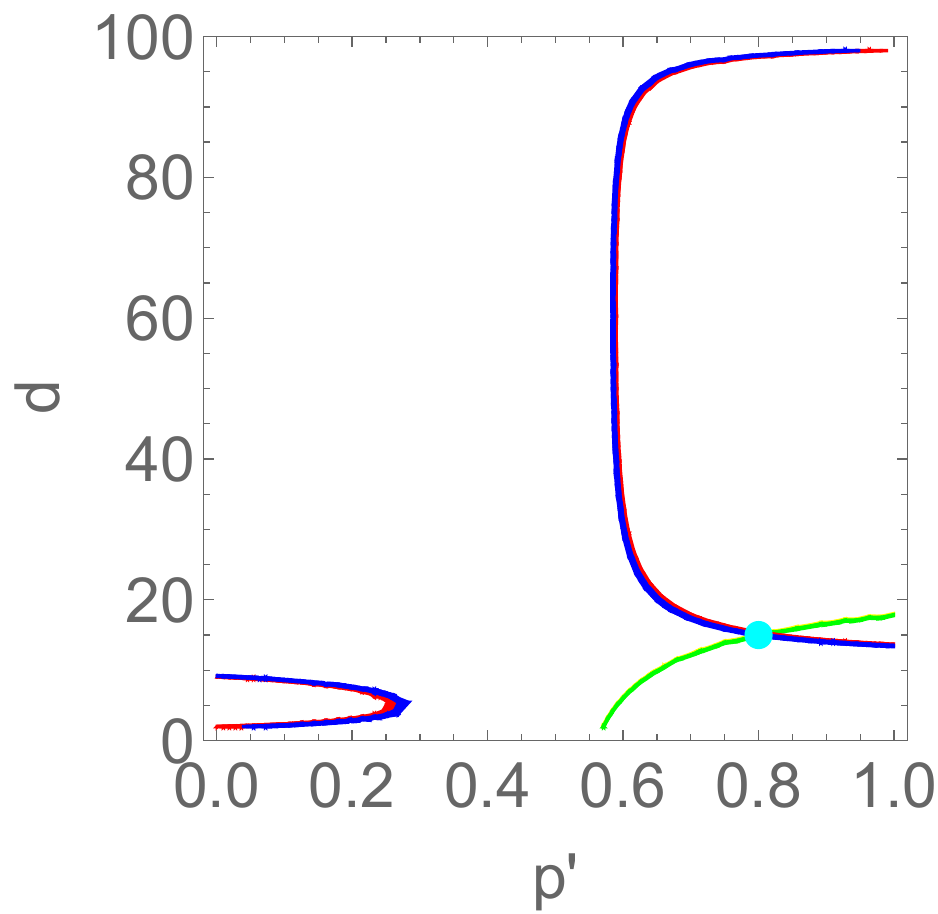}
}\end{picture}
\caption
{Experiment $14$. See caption of Fig.~\ref{f:51-3}.
}
\label{f:53-14}
\end{figure}

Instead, in Fig.~\ref{f:53-14} we show that a defect with the same $p'=0.8$ of the case of the  experiment $10$   is fully identifiable if it is sufficiently close to the starting point ($d=15$ in the experiment).
Notice that in this experiment the information about the confidence interval of the residence time alone is not sufficient to determine whether the defect is pushing the walker forward or backward, nor where the defect is located.

When $p$ increases, the possible different behaviors further reduces. 
We propose here some experiment where $p=0.55$.

\begin{figure}[ht!]
\begin{picture}(200,145)(0,0)
\put(0,0)
{
  \includegraphics[width=0.25\textwidth]{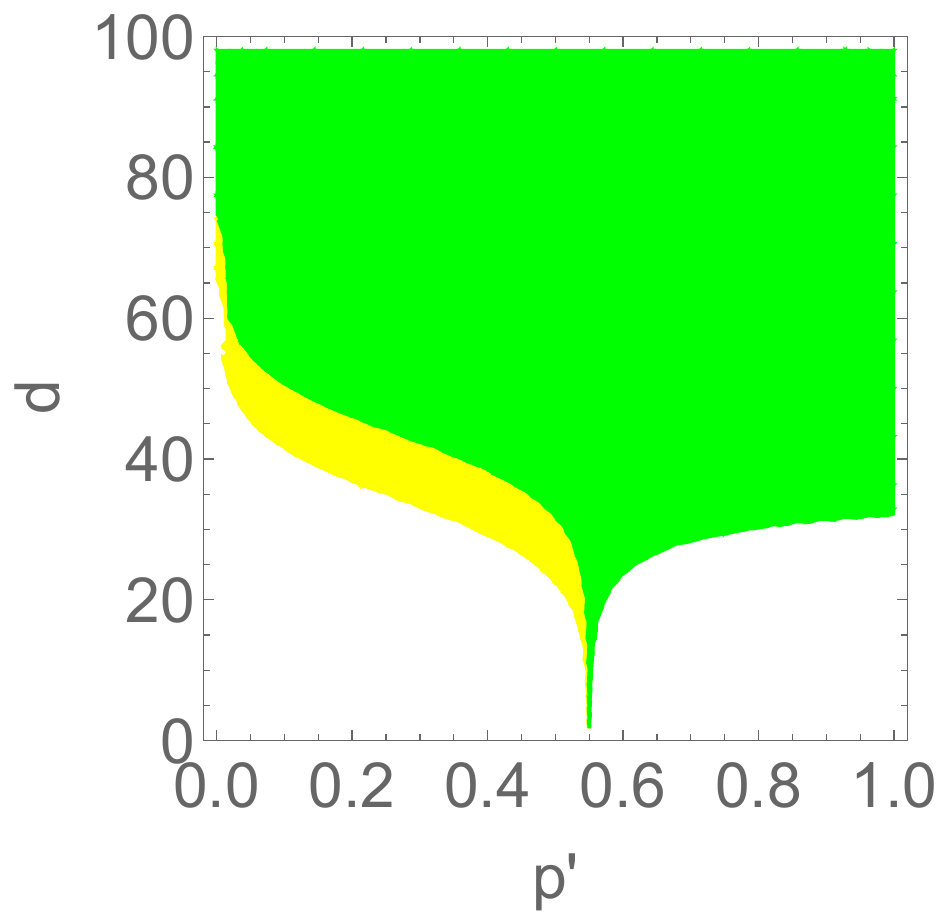}
}
\put(115,20){
  \includegraphics[width=0.11\textwidth]{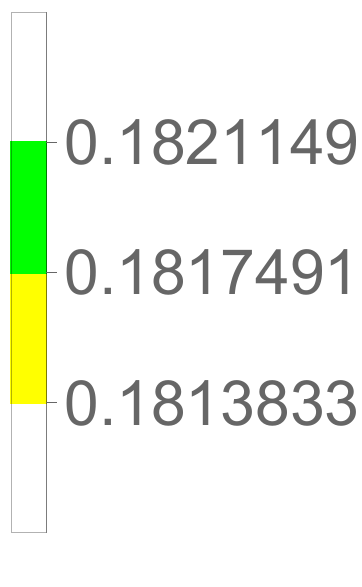}
}
\put(180,0){
  \includegraphics[width=0.25\textwidth]{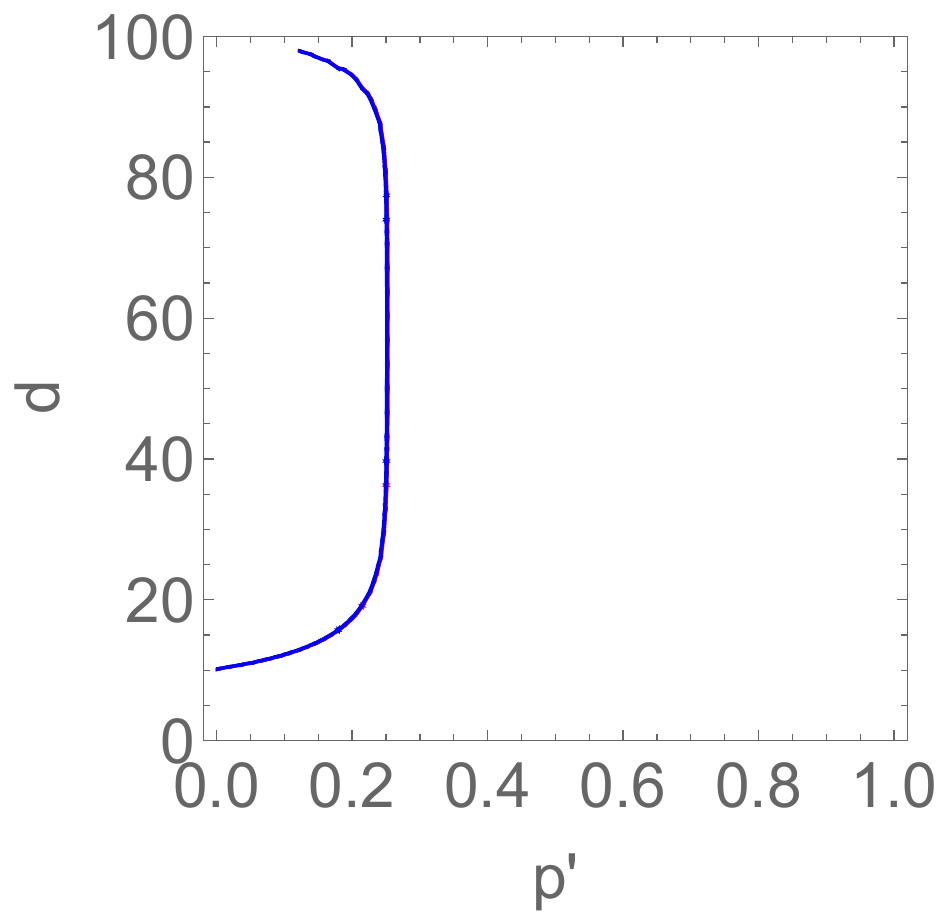}
}
\put(295,20){
  \includegraphics[width=0.10\textwidth]{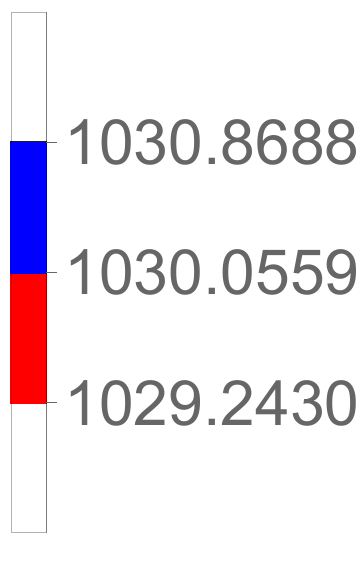}
}
\put(345,0){
  \includegraphics[width=0.25\textwidth]{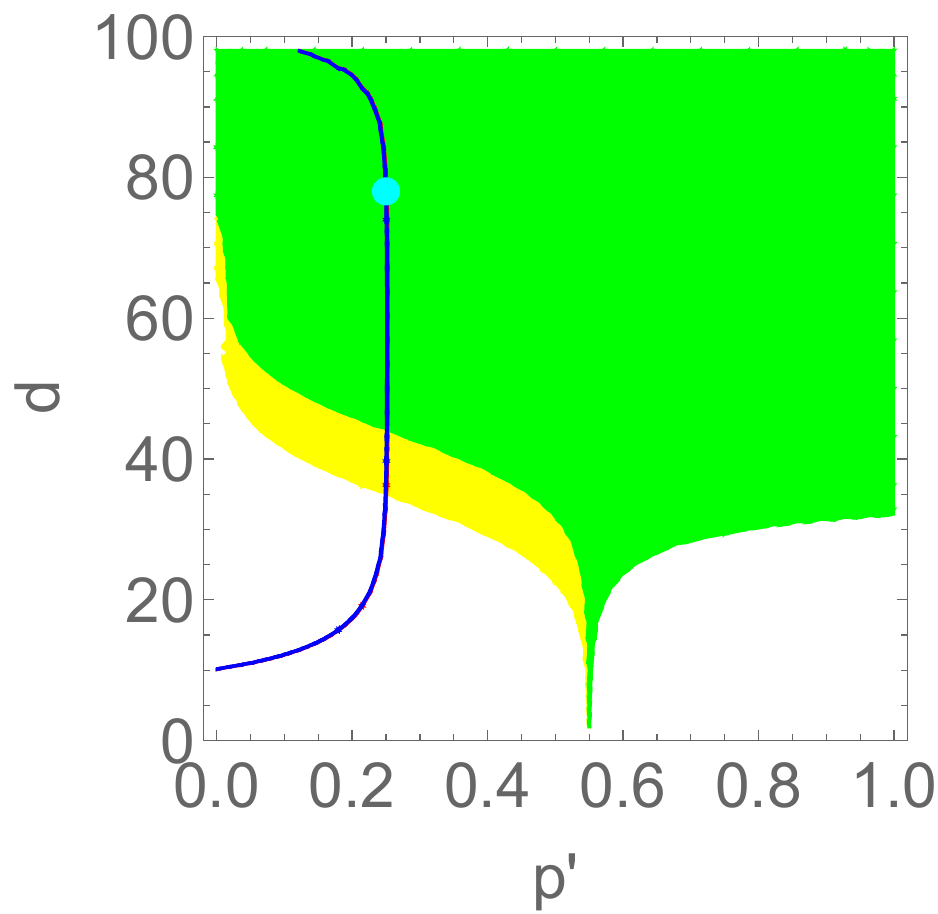}
} 
\end{picture}
\caption{
Experiment $15$. See caption of Fig.~\ref{f:51-3}.
}
\label{f:55-15bis}
\end{figure}

\begin{figure}[ht!]
\begin{picture}(200,145)(0,0)
\put(0,0)
{
  \includegraphics[width=0.25\textwidth]{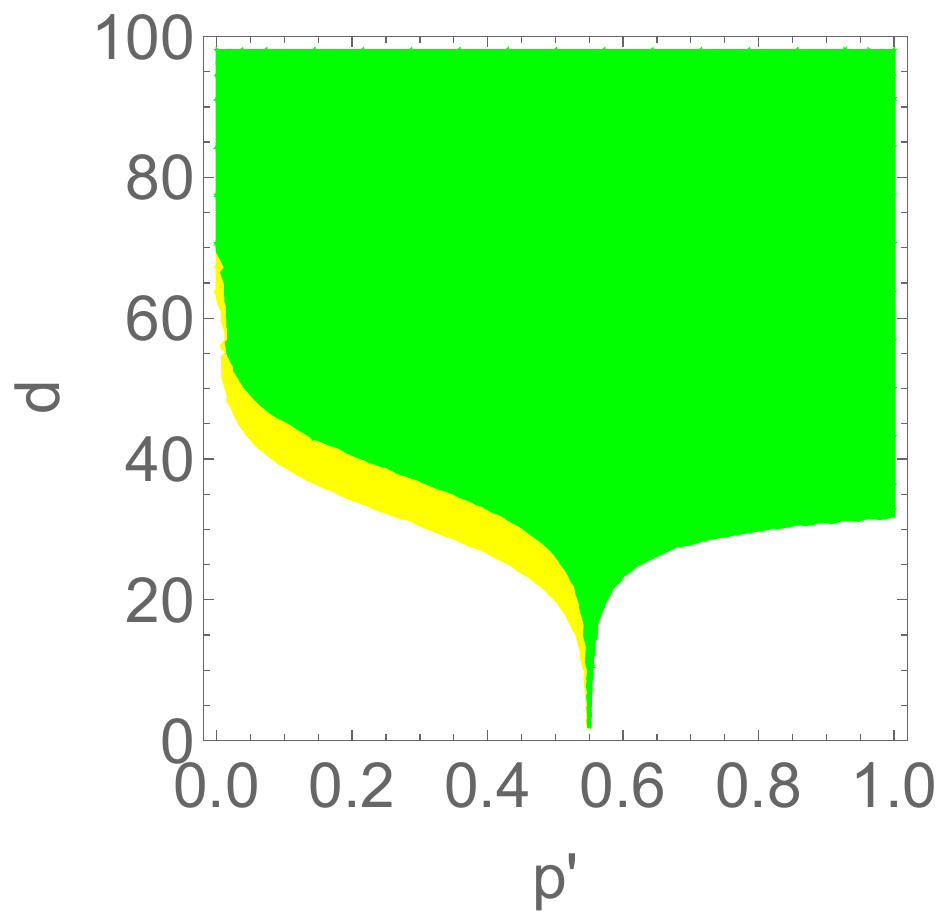}
}
\put(115,20){
  \includegraphics[width=0.11\textwidth]{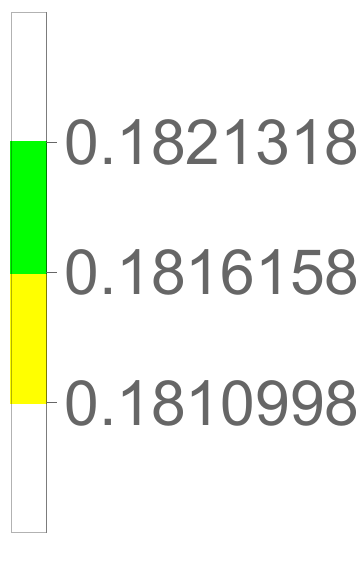}
}
\put(180,0){
  \includegraphics[width=0.25\textwidth]{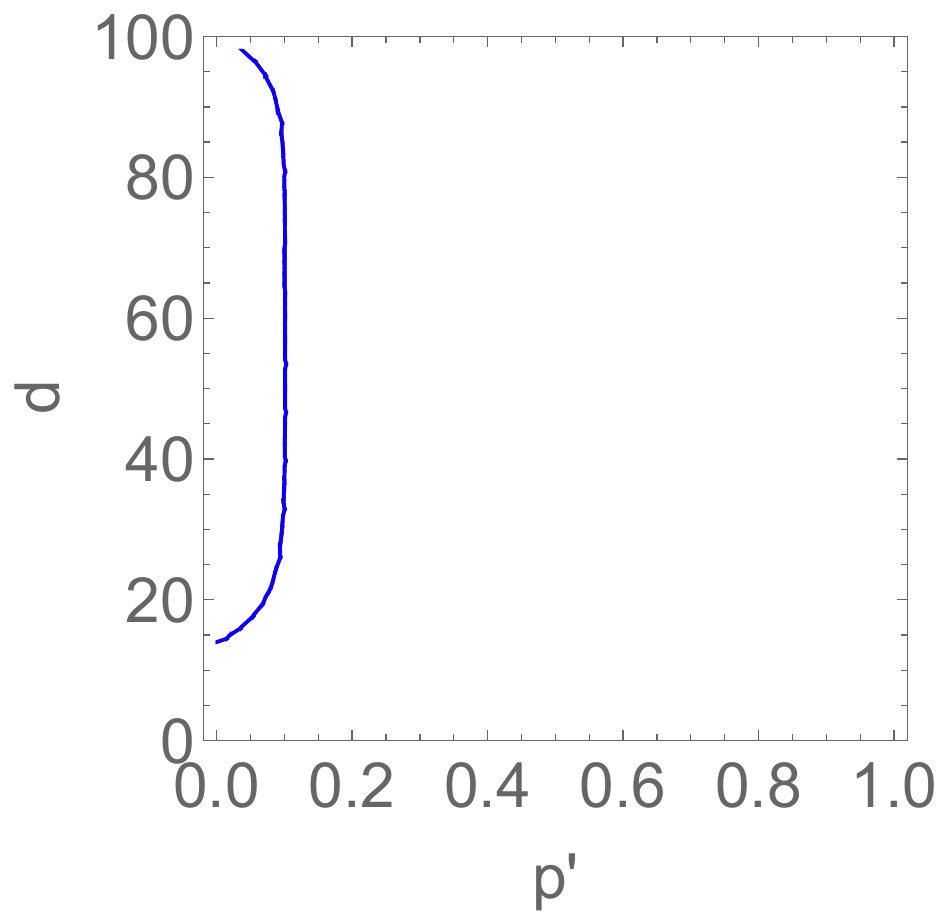}
}
\put(295,20){
  \includegraphics[width=0.10\textwidth]{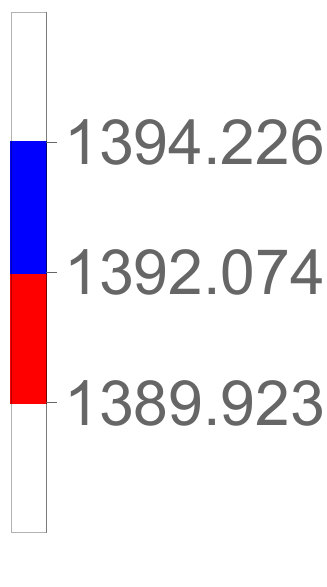}
}
\put(345,0){
  \includegraphics[width=0.25\textwidth]{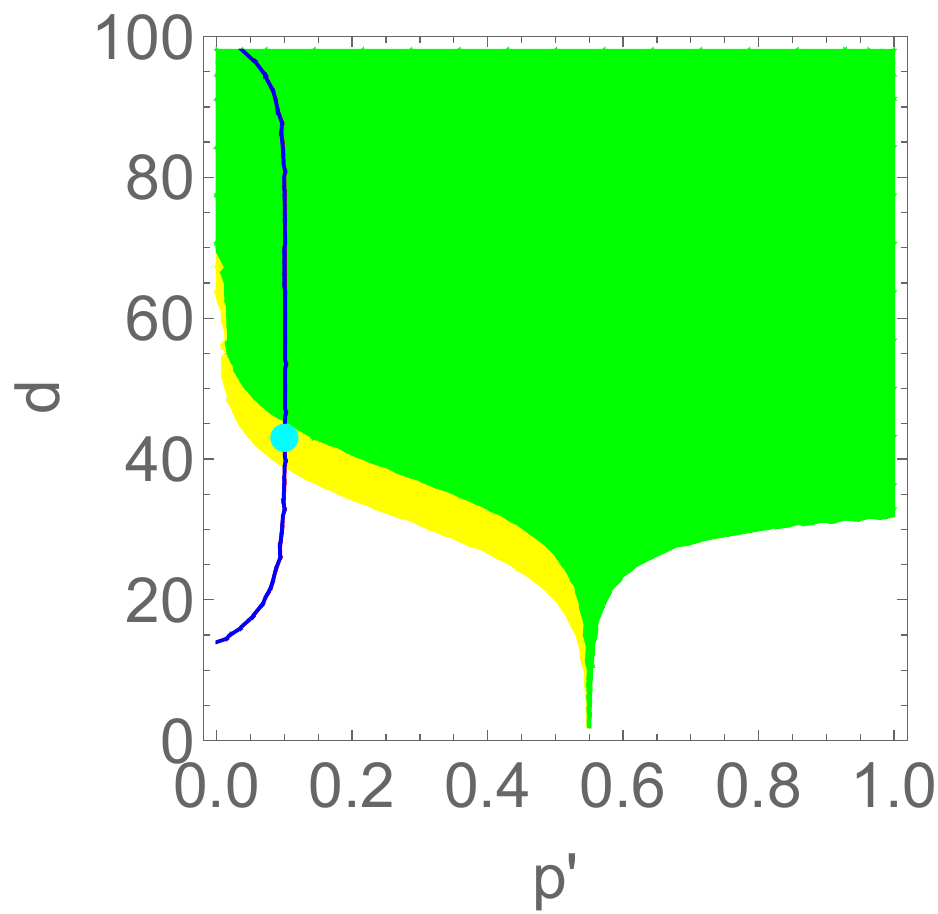}
}  
  \end{picture}
\caption{Experiment $16$. See caption of Fig.~\ref{f:51-3}.
}
\label{f:55-16bis}
\end{figure}

\begin{figure}[ht!]
\begin{picture}(200,145)(0,0)
\put(0,0)
{
  \includegraphics[width=0.25\textwidth]{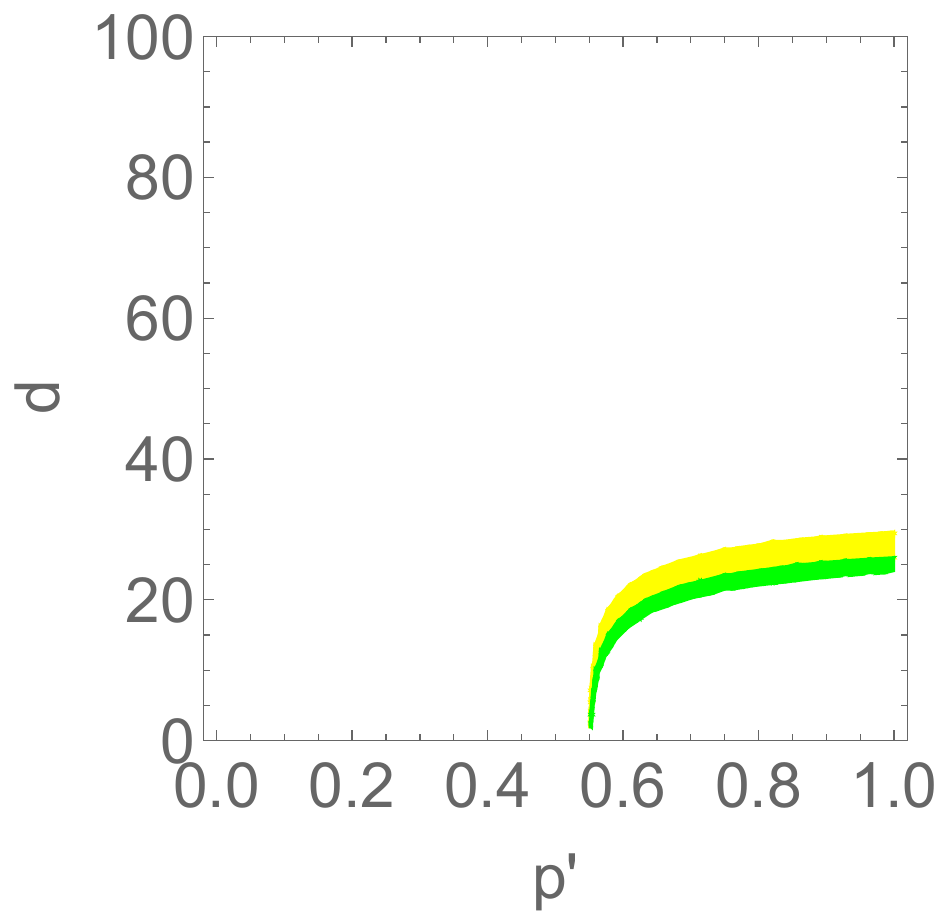}
}
\put(115,20){
  \includegraphics[width=0.11\textwidth]{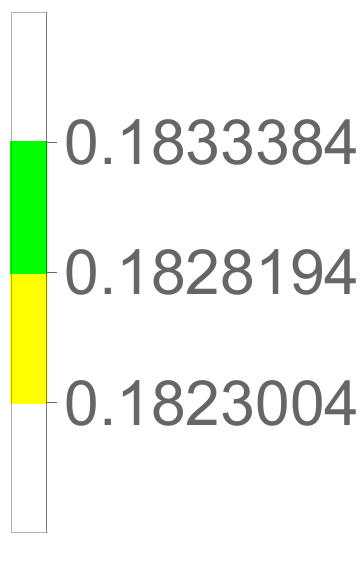}
}
\put(180,0){
  \includegraphics[width=0.25\textwidth]{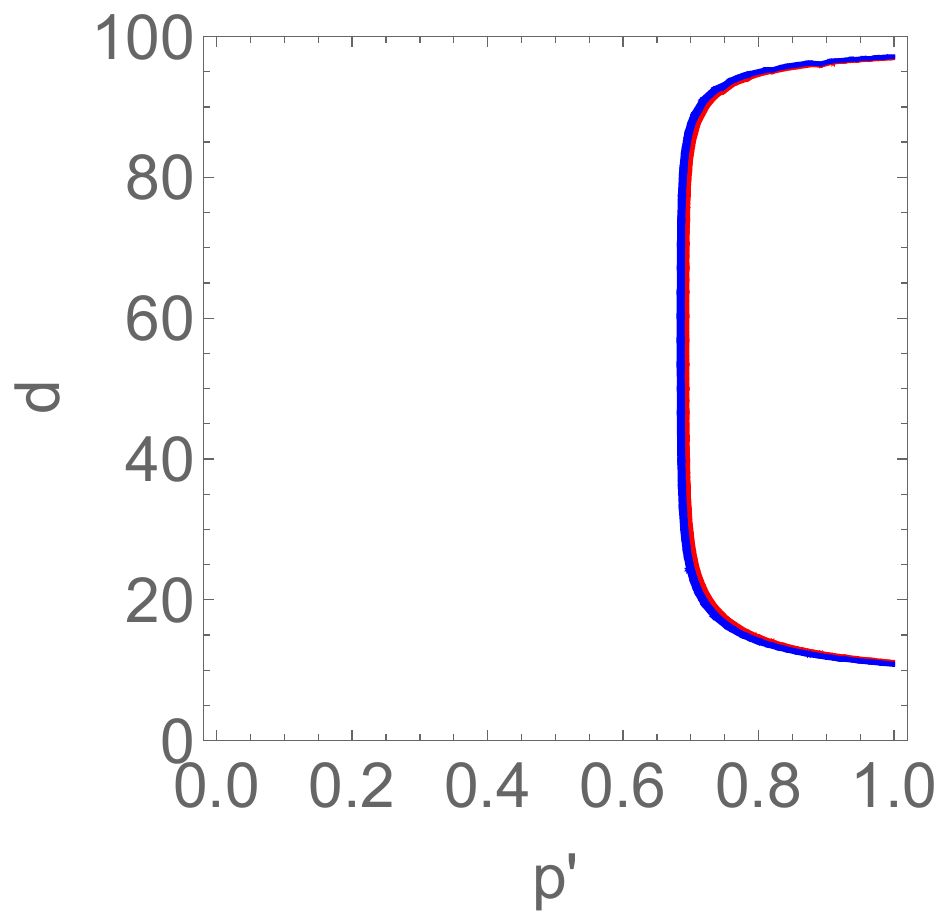}
}
\put(295,20){
  \includegraphics[width=0.10\textwidth]{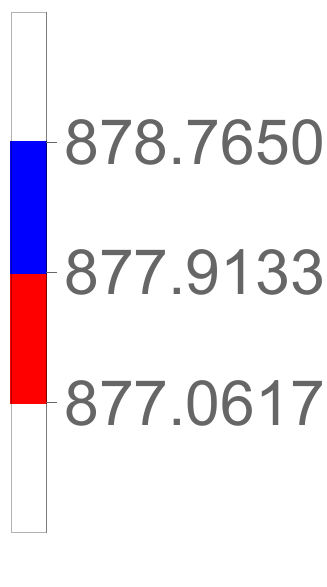}
}
\put(345,0){
  \includegraphics[width=0.25\textwidth]{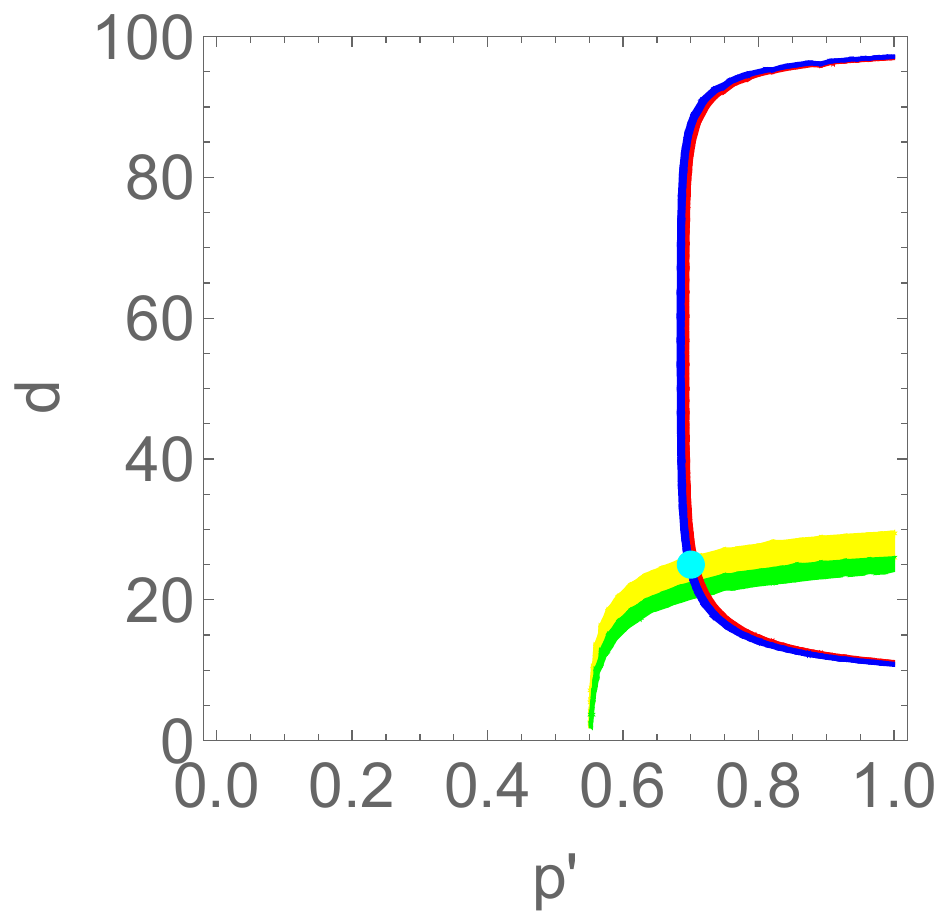}
}  
 \end{picture}
\caption{ Experiment $17$. See caption of Fig.~\ref{f:51-3}.
}
\label{f:55-17bis}
\end{figure}

In the experiments $15$, $16$ and $17$ illustrated by  Figs.~\ref{f:55-15bis}, \ref{f:55-16bis}, \ref{f:55-17bis} one can notice that the study of the region associated to the confidence interval of the experimental residence time is always useful to  identify whether the defect is favoring the walker to pass or not.
In experiments $15$ and $16$, the situation is similar to that one of experiments $10$--$13$.
The defect is not sufficiently close to the starting site $1$, so that the study of the
confidence interval associated to the experimental fraction of particles crossing is not sufficient  to fully identify the  parameters  of the defect, since many different couples $(p',d)$ are still in the intersection region.
The experiment $17$ shows instead that a defect located at $d=25$ (with $p'=0.7$ in the experiment)  produces an intersection region that restrict the possible values of the estimated $d$ to a small number, with a quite accurate estimate of the actual value of $p'$.

\section{Conclusions}
\label{s:conc}
In the framework of the 1D simple random walk, 
we have studied the effect of 
obstacles in the transport of
moving agents on a lane.
We have shown how to setup experiments to 
measure the transport properties of a lane and 
the geometrical and transport properties of possibly 
present obstacles. 

In absence of obstacles, 
we have shown that measuring the fraction of particles 
that cross the lane (experiment 1) or the 
average time taken by the crossing particles to 
cross the lane (experiment 2) is sufficient 
to deduce the transport properties of the system, that in the 
case of study is the probability for a particle to jump to its
right. To be precise, in the case of experiment 2, 
the problem is not completely identifiably, 
since the two symmetric estimates $p$ and $1-p$ are found.

In presence of defects, supposing to know $p$, we are interested 
to estimate the geometric and transport properties of 
the obstacles, that is to say, its position $d$ and 
the probability $p'$ that a particle sitting on the 
obstacle jumps to its right. 
We have shown that performing just experiment 1 or 2 is not 
sufficient to do the job, but knowing the results of both experiments 
allows to solve the problem and find an estimate of the two parameters 
$d$ and $p'$, apart from some particular case in which identifiability 
issues arise. We have discussed experiments for different 
values of the total number of moving particles $N$ and we have 
shown that the identifiability problems are weakened
when $N$ is increased.

\appendix
\renewcommand{\theequation}{\Alph{section}.\arabic{equation}}

\section{Generating function}
\label{appendice}
For the sake of completeness, we report here the construction of the
generating function $U$ introduced in
Section~\ref{s:drt} following \cite{CC19} (and references therein)
and adapting, where
necessary, the notation to the present setting.

We call $u_{n,i}$ the probability to exit the walk through the right side,
starting from $i$ and after $n$ steps.
We construct the generating function of the probability of exiting through
the right side starting from $i$, $U_i(s)=\sum_{n\geq 0} u_{n,i} s^{n}$,
see \cite{Fel,S1975,L09}. We define $t_i=\mathbb{P}_i(\rexit)$,
where the index $i$ denotes that the walk is started at the
site $i$.
Note that $t_i=U_i(1)$.
Therefore, the series defining $U_i(s)$ is totally and
thus uniformly convergent for $s\in[0,1]$.
Since the derivative of the generating function is
$U_i'(s)=\sum_{n\geq1} n u_{n,i} s^{n-1}$,
following \cite{S1975} we have that
\begin{equation}
\label{res001}
\lim_{s\to 1^-} U_i'(s)
=
\sum_{n\geq1} n u_{n,i}
=
t_i R_i,
\end{equation}
where
$R_i$ is the conditional expectation of the duration of the game given that
the random walk ends in $L$ and it is finite for any fixed
$p$, $q$, $p'$, and $q'$.

Following the approach of \cite{Fel,L09}, we find the generating function as
the solution of a system combining the equations for the generating function
in the bulks (regular sites) and on the singular site.

Recalling that on the regular sites $p$ and $q$ are, respectively,
the probabilities to jump to the right and to the left,
we find that
\begin{equation}
\label{eq:u_n,i}
u_{n+1,i}=p u_{n,i+1} + q u_{n,i-1}
\end{equation}
on the regular sites, namely, for $i=1,\dots,d-1$ and
$i=d+1,\dots,L-1$,
and boundary values
\begin{equation*}
u_{n,0}=u_{n,L}=0 \,\, \textrm{when} \,\, n\geq 1,
\quad u_{0,L}=1,\quad u_{0,i}=0  \,\, \textrm{when} \,\, i\leq L.
\end{equation*}
Thus, multiplying \eqref{eq:u_n,i} by $s^{n+1}$ and summing for
$n=0,1,2,\ldots$ we find the following equation in the bulk, i.e.,
$i=1,\dots,d-1$ and $i=d+1,\dots,L-1$,
\begin{equation}
\label{eq:U_i_bulk}
U_i(s)= ps U_{i+1}(s) + qs U_{i-1}(s),
\end{equation}
to be solved with boundary conditions
$U_L(s)=1$ and $U_0(s)=0$.
The equation for $U_d(s)$ on the defect site is
\begin{equation}
\label{eq:U_defect}
U_d(s)= p's U_{d+1}(s) + q's U_{d-1}(s),
\end{equation}
since the probabilities on the singular site are $p'$ to jump to the right and
$q'$ to the left.

It is known
\cite{Fel}
that in the bulk
of regular sites the generating function $U_i$ for fixed $s$ can be
searched in the form $\lambda^{i}(s)$, except for the case $p=q$ and $s=1$ at the same moment, where the generating function is linear, see later.
Substituting in \eqref{eq:U_i_bulk} it is found
\begin{equation}
\label{tre010}
\lambda_\pm(s)
=
\frac{1 \pm \sqrt{1 - 4pqs^2}}{2ps}
\,.
\end{equation}
The generating function in the bulk can be written as a linear combination
of terms in the form $\lambda_-^i$ and $\lambda_+^i$
for $i=0,\dots,L$.
We consider two different linear combinations in the bulk on the left and
on the right of the defect site, namely, we introduce two coefficients
on the left and two (possibly) different coefficients on the right.
Thus we find two different representation of $U_d(s)$.
Therefore, we will be able to find the unknown coefficients by
requiring that these two representation are equal for $i=d$,
that is to say
the equation at the defect \eqref{eq:U_defect} and the boundary
conditions are satisfied.
More precisely,  the generating function reads
\begin{equation}\label{tre015}
\begin{split}
&U_i(s)= G(s) \lambda_+^i{(s)} + H(s) \lambda_-^i(s), \quad i=0,1,\ldots,d;
\\
&U_j(s)= U(s) \lambda_+^j{(s)} + V(s) \lambda_-^j(s), \quad j=d,d+1,\ldots,L,
\end{split}
\end{equation}
and the coefficients $G(s)$, $H(s)$, $U(s)$, and $V(s)$ solve the system
\begin{equation}\label{tre020}
\begin{cases}
& G \lambda_+^d + H \lambda_-^d = U \lambda_+^d + V \lambda^d_-
\\ &  G \lambda_+^d + H \lambda_-^d = q' s (G \lambda_+^{d-1} + H \lambda_-^{d-1} ) + p's (U \lambda_+^{d+1} + V \lambda^{d+1}_-)
\\& 0=G+H
\\& 1 = U\lambda_+^L + V\lambda_-^L.
\end{cases}
\end{equation}
The system \eqref{tre020} has a unique solution $(G,H,U,V)$ that can be explicitly expressed in terms of  jump probabilities, $\lambda_{\pm}$ and $s$.
Thus, by substituting $\lambda_{\pm}$ given by the \eqref{tre010} in the formulas \eqref{tre015}, we find the explicit expressions of the generating function $U_i(s)$.
However, due to the length and complexity of these expressions, we prefer not to report here the solutions of the system in the general case.
Note that in the easiest case, i.e.,
if the site $d$ is regular, namely $p'=p$ and $q'=q$,
the solutions are $G=U$ and $H=V$
which reduce to the classical ones in
Feller \cite[equation (4.10) in Paragraph~XIV.4]{Fel}.

The conditional expectation $R_i$ of the duration of the game starting
from $i$ and ending in $L$ can be computed using
equation \eqref{res001}.
The limit allows us to include in this formula even the symmetric
case $p=q$, where the generating function has not the form of
a combination of powers $\lambda^i$ anymore for $s=1$.


\begin{thebibliography}{9}\label{bibliogr}

\bibitem{CPamm2017}
E.\ Cristiani and D.\ Peri.
\newblock {\em Applied Mathematical Modelling}, 45:285 -- 302, 2017.

\bibitem{Sbj1994}
M.J. Saxton.
\newblock {\em Biophysical Journal}, 66:394--401, Feb 1994.

\bibitem{HFrpp2013}
F.\ H\"ofling and T.~Franosch.
\newblock {\em Reports on Progress in Physics}, 76(4), 2013.

\bibitem{MHSbj2017}
M.A.\ Mour\~{a}o, J.B.\ Hakim, and S.~Schnell.
\newblock {\em Biophysical Journal}, 107:2761--2766, Jun 2017.

\bibitem{ESMCBjcp2014}
A.J.\ Ellery, M.J.\ Simpson, S.W.\ McCue, and R.E.\ Baker.
\newblock {\em The Journal of Chemical Physics}, 140(5):054108, 2014.

\bibitem{TLPprl2001}
K.\ To, P.\ Lai, and H.K. Pak.
\newblock {\em Phys. Rev. Lett.}, 86:71--74, Jan 2001.

\bibitem{ZGMPPpre2005}
I.\ Zuriguel, A.\ Garcimart\'{\i}n, D.\ Maza, L.A.\ Pugnaloni, and J.M. Pastor.
\newblock {\em Phys. Rev. E}, 71:051303, May 2005.

\bibitem{AMAGTOKpre2012}
F.\ Alonso--Marroquin, S.I.\ Azeezullah, S.A.\ Galindo--Torres, and L.M.
  Olsen-Kettle.
\newblock {\em Phys. Rev. E}, 85:020301, Feb 2012.

\bibitem{ZJGLAMprl2011}
I.\ Zuriguel, A.\ Janda, A.\ Garcimart\'{\i}n, C.\ Lozano, R.\ Ar\'evalo, and
  D.~Maza.
\newblock {\em Phys. Rev. Lett.}, 107:278001, Dec 2011.

\bibitem{Hrmp2001}
D.~Helbing.
\newblock {\em Rev. Mod. Phys.}, 73:1067--1141, Dec 2001.

\bibitem{HFMV2011}
D.\ Helbing, I.\ Farkas, P.\ Moln\`{a}r, and T.~Vicsek.
\newblock In M.~Schreckenberg and S.~D. Sharma, editors, {\em Pedestrian and
  Evacuation Dynamics}, pages 21--58, Berlin, 2002. Springer.

\bibitem{CMpA2013}
E.N.M.\ Cirillo and A.\ Muntean.
\newblock {\em Physica A: Statistical Mechanics and its Applications},
  392(17):3578 -- 3588, 2013.

\bibitem{MCKBud2014}
A.\ Muntean, E.N.M.\ Cirillo, O.\ Krehel, M.\ Bohm,
In ``Collective Dynamics from Bacteria to Crowds", An Excursion Through
Modeling, Analysis and Simulation Series: CISM International Centre
for Mechanical Sciences, Vol.\ 553 Muntean, Adrian, Toschi, Federico (Eds.)
2014, VII, 177 p. 29 illus, Springer, 2014.

\bibitem{CMcrm2012}
E.N.M.\ Cirillo, A.\ Muntean,
\newblock {\em Comptes Rendus Macanique} 340, 626--628, 2012.

\bibitem{CCCMm3as2018}
A.\ Ciallella, E.N.M.\ Cirillo, P.L.\ Curseu, A.\ Muntean,
\newblock {\em Mathematical Models and Methods in Applied Sciences},
https://doi.org/10.1142/S0218202518400079.

\bibitem{ABCKsjap2016}
G.\ Albi, M.\ Bongini, E.\ Cristiani, and D.\ Kalise.
\newblock {\em SIAM Journal on Applied Mathematics}, 76(4):1683--1710, 2016.

\bibitem{HFVn2000}
D.\ Helbing, I.\ Farkas, and T.~Vicsek.
\newblock {\em Nature}, 407:487--490, Sep 2000.

\bibitem{HBJWts2005}
D.\ Helbing, L.\ Buzna, A.\ Johansson, and T.\ Werner.
\newblock {\em Transportation Science}, 39(1):1--24, 2005.

\bibitem{EDLR2003}
R.\ Escobar and A.~De~La~Rosa.
\newblock In W.~Banzhaf, J.~Ziegler, T.~Christaller, P.~Dittrich, and J.T. Kim,
  editors, {\em Advances in Artificial Life, Proceedings of the 7th European
  Conference, ECAL, 2003, Dortmund, germany, September 14--17, 2003,
  Proceedings. Lecture Notes in Computer Science, vol.\ 2801.}, pages 97--106,
  Berlin, 2003. Springer.

\bibitem{JanLeb94}
S.A.\ Janowsky, J.L.\ Lebowitz,
\newblock {\em J.\ Stat.\ Phys.\/} \textbf{77}, 35--51 (1994).

\bibitem{SLM15}
B.\ Scoppola, C.\ Lancia, R.\ Mariani,
\newblock {\em J.\ Stat.\ Phys.\/} \textbf{161}, 843--858 (2015).

\bibitem{CC19}
A.\ Ciallella, E.N.M.\ Cirillo,
Conditional expectation of the duration of the classical
gambler problem with defects
{\em European Physical Journal Special Topics} \textbf{228} 111-128, 2019

\bibitem{CCSpre2018}
A.\ Ciallella, E.N.M.\ Cirillo, J.\ Sohier.
\newblock {\em Physical Review E}, 97:052116, 2018.

\bibitem{CCkrm2018}
A.\ Ciallella and E.N.M. Cirillo.
\newblock {\em Kinetic \& Related Models} \textbf{11}, 1475--1501 (2018).

\bibitem{CCMpre2016}
E.N.M.\ Cirillo, M.\ Colangeli, and A.\ Muntean.
\newblock {\em Physical Review E} \textbf{94}, 042116 (2016).

\bibitem{CCpre2017}
E.N.M.\ Cirillo, M.\ Colangeli,
\newblock {\em Phys. Rev. E} \textbf{96}, 052137, 2017.

\bibitem{CCMphysicaA2017}
E.N.M.\ Cirillo, M.\ Colangeli, A.\ Muntean,
\newblock {\em Physica A} \textbf{488}, 30--38 (2017)

\bibitem{CKMSpre2016}
E.N.M.\ Cirillo, O.\ Krehel, A.\ Muntean, and R.~van Santen.
\newblock {\em Phys. Rev. E}, 94:042115, Oct 2016.

\bibitem{CKMSSpA2016}
E.N.M.\ Cirillo, O.\ Krehel, A.\ Muntean, R.\ van Santen, and Aditya S.
\newblock {\em Physica A: Statistical Mechanics and its Applications}, 442:436
  -- 457, 2016.

\bibitem{MRVMSpre2016}
J.\ Messelink, R.\ Rens, M.\ Vahabi, F.C.\ MacKintosh, A. Sharma,
\newblock {\em Physical Review E} \textbf{93}, 01211 (2016)

\bibitem{Fel}
W.~Feller.
\newblock {\em An Introduction to Probability Theory and its Applications},
  volume~1.
\newblock John wiley \& Sons, Inc, New York -- London -- Sidney, 1968.

\bibitem{S1975}
F.\ Stern,
\newblock {\em Math.\ Mag.\/} \textbf{48}, 200--203 (1975).

\bibitem{L09}
T. Lengyel,
\newblock {\em Applied Mathematics Letters} \textbf{22}, 351--355 (2009).

\bibitem{WB77}
W.A. Beyer, M.S. Waterman,
Symmetries for conditioned ruin problems,
{\em Math. Mag.} \textbf{50} (1) (1977) 42--45.


\end{thebibliography}
\end{document}